\documentclass[acmsmall]{acmart}

\AtBeginDocument{%
  \providecommand\BibTeX{{%
    \normalfont B\kern-0.5em{\scshape i\kern-0.25em b}\kern-0.8em\TeX}}}

\setcopyright{acmcopyright}
\copyrightyear{9999}
\acmYear{9999}
\acmDOI{99.9999/9999999.9999999}

\acmJournal{TIIS}
\acmMonth{99}

\usepackage{multirow}
\usepackage{balance}       
\usepackage{graphics}      
\usepackage{color, colortbl}
\usepackage{booktabs}
\usepackage{textcomp}
\usepackage{amsmath}
\usepackage{mathtools}
\usepackage{caption}
\usepackage{subcaption}
\usepackage{geometry}
\usepackage{booktabs}
\usepackage{pdflscape}
\usepackage{longtable}
\usepackage{tablefootnote}
\usepackage{wrapfig}
\usepackage{ccicons}          
\usepackage{colortbl}
\usepackage{xcolor}
\AtBeginEnvironment{quote}{\par}
\usepackage{soul}
\usepackage{makecell}
\usepackage{lscape}
\usepackage{csquotes}
\usepackage{tabularx}
\usepackage{adjustbox}
\usepackage{arydshln} 
\newif\ifShowNotes

\global\ShowNotesfalse
\ifShowNotes
\newcommand{\FIXME}[1]{\textcolor{red}{**FIXME: #1**}}
\newcommand{\TODO}[1]{\textcolor{red}{\begingroup\raggedright TO DO: #1\\\endgroup}}
\newcommand{\NOTE}[2][gray]{\smallskip\noindent
  \colorbox{#1!30}{\parbox{.98\linewidth}{{\small\textbf{#2}}}}
}
\newcommand{\boldification}[1]{\textbf{\textcolor{blue}{\begingroup\raggedright ** #1 **\\\endgroup}}}
\newcommand{\DraftStatus}[1]{\textbf{\textcolor{black}{**Draft: #1**}}}
\else
\newcommand{\FIXME}[1]{}
\newcommand{\TODO}[1]{}
\newcommand{\NOTE}[2][gra\begin{center}
\begin{tabular}{ |c|c|c|c| } 
\hline
col1 & col2 & col3 \\
\hline
\multirow{3}{4em}{Multiple row} & cell2 & cell3 \\ 
& cell5 & cell6 \\ 
& cell8 & cell9 \\ 
\hline
\end{tabular}
\end{center}y]{}
\newcommand{\boldification}[1]{}
\newcommand{\DraftStatus}[1]{}
\fi

\newcommand{\PostGenderMag}{\textcolor{white}{\fcolorbox{PostGMGroupColor}{PostGMGroupColor}{Post-GenderMag}}}
\newcommand{\Original}{\textcolor{black}{\fcolorbox{OriginalGroupColor}{OriginalGroupColor}{Original}}}

\newcommand{\menCaption}{\textcolor{white}{\fcolorbox{MenColor}{MenColor}{men}}}

\newcommand{\womenCaption}{\textcolor{white}{\fcolorbox{WomenColor}{WomenColor}{women}}}

\definecolor{MenColor}{RGB}{32,56,100}
\definecolor{WomenColor}{RGB}{237,125,49}
\definecolor{OriginalGroupColor}{RGB}{176, 242, 248}
\definecolor{PostGMGroupColor}{RGB}{50, 205, 50}
\newcommand{\MAYBECUT}[1]{}

\begin{document}













\title{Inclusive Design of AI's Explanations: \\ Just for Those Previously Left Out, or for Everyone?}





\author{Md Montaser Hamid}
\email{hamidmd@oregonstate.edu}
\affiliation{\institution{Oregon State University}\country{USA}}
\orcid{0000-0002-5701-621X}

\author{Fatima A. Moussaoui}
\email{moussaof@oregonstate.edu}
\affiliation{\institution{Oregon State University}\country{USA}}

\author{Jimena Noa Guevara}
\email{noaguevg@oregonstate.edu}
\affiliation{\institution{Oregon State University}\country{USA}}

\author{Andrew Anderson}
\email{anderan2@oregonstate.edu}
\orcid{0000-0003-4964-6059}
\affiliation{\institution{Oregon State University}\country{USA}}

\author{Puja Agarwal}
\email{puja.agarwal@oregonstate.edu}
\affiliation{\institution{Oregon State University}\country{USA}}

\author{Jonathan Dodge}
\email{jxd6067@psu.edu}
\affiliation{\institution{The Pennsylvania State University}\country{USA}}

\author{Margaret Burnett}
\email{burnett@eecs.oregonstate.edu}
\affiliation{\institution{Oregon State University}\country{USA}}

\renewcommand{\shortauthors}{Hamid et al.}


\begin{abstract}

\textbf{Motivations:} Explainable Artificial Intelligence (XAI) systems aim to improve users' understanding of AI, but XAI research shows many cases of different explanations serving some users well and being unhelpful or worse to other users.
With non-AI systems, some software practitioners have addressed problems like this by using inclusive design approaches.
In some fortuitous cases, their improvements turned out to be ``curb-cut'' improvements---not only addressing the needs of underserved users, but also making the products better for everyone else as well.
This raises the possibility that if AI practitioners used inclusive design approaches to improve their AI product's explanations, they too might create curb-cut improvements, i.e., better explanations for everyone. \\
\textbf{Objectives:} To consider this possibility, we investigated the effects of inclusivity-driven fixes in an XAI prototype.
The prototype and fixes came from an AI product team who had adopted an inclusive design approach (GenderMag) to improve their XAI prototype.
Our objective was to investigate the curb-cut effects of the AI team's inclusivity-driven fixes on users' mental models of AI.\\
\textbf{Methods:} We ran a between-subject study with 69 participants who had no AI background. 34 participants used the original version of the XAI prototype and 35 used the version with the AI team's inclusivity fixes. 
To understand the curb-cut effects of inclusivity-driven fixes in the XAI prototype, we compared the two groups' mental model concepts scores, prediction accuracy, and inclusivity. \\
\textbf{Results:} Our investigation produced four main results.
First, it revealed several curb-cut effects of the AI team's inclusivity fixes, such as overall increased engagement with explanations and better mental model concepts scores. These improvements revealed specific inclusivity fixes with curb-cut properties.
However (second), the inclusivity fixes did not improve participants' prediction accuracy scores---instead, it appears to have harmed them.
This ``curb-fence'' effect (opposite of the curb-cut effect) on participants' prediction accuracy scores revealed the AI explanations' double-edged impact, sometimes operating for better and sometimes unexpectedly for worse.
Third, the results showed the AI team's inclusivity fixes brought significant improvements for users whose problem-solving styles had previously been underserved.
Further (fourth), the AI team's fixes reduced the gender gap by 45\%.

\end{abstract}








\maketitle

 \section{Introduction} 
 \DraftStatus{MMH: D2.5 }
 
\label{sec:introduction}  


\begin{wrapfigure}{R}{4cm}
    \centering
    \includegraphics[width=1.0\linewidth]{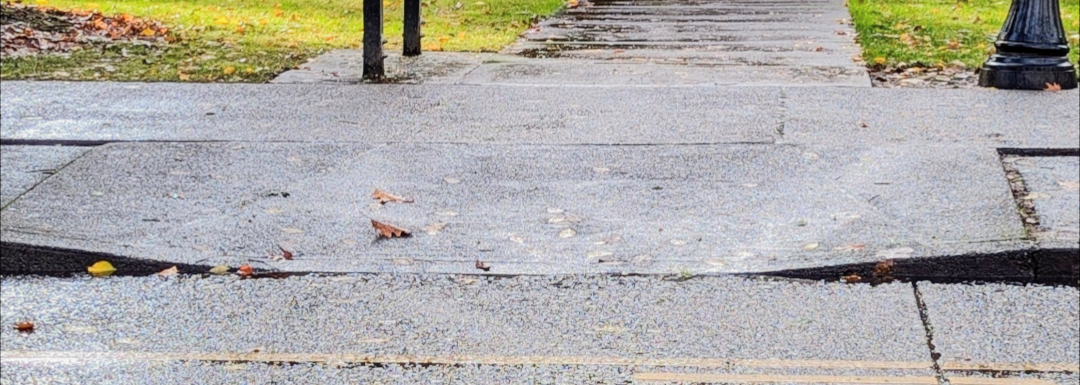}
    \caption{A sidewalk's curb cut.
    }
    \label{fig:curbCut}
\end{wrapfigure}

Consider curb cuts in sidewalks, like the one in Figure~\ref{fig:curbCut}.  
These cut-out sections of sidewalks were designed for a population who was not well-served by sidewalks before curb cuts, namely people in wheelchairs.
And they did make sidewalks better for people in wheelchairs---but also, for everybody else (e.g., parents with strollers, people pushing heavy carts, travelers with wheeling suitcases, etc.)~\cite{Blackwell16Curb}.
This property of curb cuts has been termed the \textit{curb-cut effect}: an effect of a design choice made to better serve an underserved population that also turns out to help everyone else too, even those who were already well-served~\cite{Blackwell16Curb}.

In the field of Explainable AI (XAI), many researchers have reported problems with explanations' effectiveness for diverse users in diverse situations~\cite{anderson2020mental, bhatt2020explainable, Gajos22CognitiveAI, gerlings2022explainable, rosenfeld2019explainability, wang2024roadmap}.  
This situation not only suggests (1)~the need to attend to users' diversity, but also (2)~ongoing problems generally with the effectiveness of AI explanations.  
In this paper, we consider these two problems as one, to investigate the possibilities of curb-cut effects.
Specifically, can an inclusive design approach produce explanations that not only help some underserved group(s) of users, but also are ``just plain better'' explanations for everyone?




\par

\boldification{"better" means better mental model of AI}
For our paper's XAI scope, by ``better'', we mean promoting more accurate mental models of how a user's AI system is reasoning. 
In XAI, mental models are defined as the user's understanding of how an AI system works~\cite{anderson2020mental}.
Helping users form suitable mental models is an objective of many XAI systems~\cite{kulesza2013too, nourani2021anchoring, Schraagen20Trusting}.
\par

\boldification{So we investigate curb cuts in some AI practitioners' work with an inclusivity method}
Thus, this paper investigates if inclusive design can bring curb-cut effects to XAI's explanations in the form of better mental models for everyone.
To do so, we evaluated how well two groups of end users fared in two versions of an XAI prototype. 
These versions were the work of a team of AI practitioners (Team Game) who were hoping to address inclusivity issues within their XAI prototype. 
Their prototype was a modified version of Tic-Tac-Toe games played sequentially by a red AI agent and a blue AI agent.
The prototype was supported by several kinds of explanations of the blue AI agent's behaviors. 
Prior to our study, the AI practitioners had used the GenderMag inclusivity method to find and fix inclusivity issues in their prototype~\cite{Anderson24InclusiveHAI}. 

\boldification{What's GM}
GenderMag ~\cite{Burnett16GM} is an inclusivity evaluation method that considers five ranges of users' problem-solving style types: the ranges of users' various Motivations for using technology, of their diverse Information Processing Styles, of their levels of Computer Self-Efficacy,
of their Attitudes Toward Risk in using unfamiliar technology features, and of their Learning Styles~\cite{Hamid23GM}.  
Several studies have shown GenderMag's effectiveness in non-AI systems (e.g.,~\cite{Burnett16GM, Guizani22Inlcusivity, Burnett17MS}), as we explain further in Section~\ref{sec:methodology_gendermag}.

\boldification{We ran a study to find out to what extent Team Game's inclusivity fixes were curb cuts. }
For the current paper, we ran a between-subject study where a group of 34 participants used the ``Original'' version and 35 participants used the ``Post-GenderMag'' version.
Since the prototype is attempting to explain to end users, our participants were restricted to only those with no AI background.
For the ``better for everyone'' aspect of curb-cut effects, we measured the Original participants' vs. the Post-GenderMag participants' conceptual understanding of the AI agents' behavior, their ability to predict the blue AI agent's next move, and their engagement with the system's explanations.
For the ``better for the underserved'' aspect of curb-cuts, we also disaggregated these measures by users' problem-solving style values and by their genders. 

\boldification{Our RQs were: }
Our investigation centered around these three research questions:
\begin{itemize}
    \item \textbf{RQ1:} Did the PostGenderMag participants have better \textbf{mental model concepts scores} than the Original participants?
    \item \textbf{RQ2:} Did the PostGenderMag participants have better \textbf{prediction accuracy} than the Original participants?
    \item  \textbf{RQ3:} Was the Post-GenderMag prototype \textbf{more inclusive} than the Original prototype?
\end{itemize}

\section{Background and Related Works}


\subsection{Background: GenderMag}
\DraftStatus{MMH: D2.5}
\label{sec:methodology_gendermag}

\boldification{GenderMag is a usability inspection method}
The AI practitioners team used the GenderMag inclusivity method, so our results rest on GenderMag concepts.
GenderMag (Gender-Inclusiveness Magnifier) is an inspection method to find and/or fix inclusivity bugs in problem-solving software~\cite{Hamid23GM, Burnett16GM, Guizani22Inlcusivity, Vorvoreanu19Bias}. 
The method focuses on diversity of problem-solving styles (``facets'' in GenderMag terminology) which statistically cluster by gender. 
The method uses these facets as its core, brings them to life through three faceted personas, and encapsulates the use of these facets into a systematic process through a specialized Cognitive Walkthrough (CW)~\cite{Wharton94CW}. 
Numerous software practitioners have used GenderMag to evaluate and/or fix the inclusiveness of their software products~\cite{Burnett17MS, cunningham2016supporting, Guizani22Inlcusivity, Hilderbrand20Trench,  kirillova2017gender, murphy2024gendermag, santos2023designing, shekhar2018cognitive, Vorvoreanu19Bias}.

\par
\boldification{GenderMag defines 5 facets or problem-solving styles}
The GenderMag method’s five facets capture five factors in an individual’s problem-solving approach: their Motivations for using technology, Information Processing Styles, Computer Self-Efficacy, Attitude Toward Risk in using unfamiliar technology features, and Learning Style (by Process vs. by Tinkering)~\cite{Burnett16GM}. The ``inclusivity bugs'' uncovered by GenderMag are instances where a technology product fails to fully accommodate the complete range of values associated with the five facets, leading to a disproportionate impact on individuals whose problem-solving styles are unsupported. These bugs are also considered gender-inclusivity issues because the facets reflect statistical differences among gender preferences in problem-solving styles~\cite{Burnett16GM, Guizani22Inlcusivity, Vorvoreanu19Bias, Hamid23GM}. 
GenderMag does not use peoples' gender identities to find inclusivity issues, it uses only their problem-solving styles.

\par
\boldification{GenderMag brings these facets to life with 3 research based personas}
The GenderMag method represents the range of facet values with three personas: Abi (Abigail/Abishek), Pat (Patricia/Patrick), and Tim (Timara/Timothy) (see Table~\ref{tab:facets}). 
Statistically, the problem-solving styles of Abi disproportionately skew toward women, for Tim they disproportionately skew toward men, and Pat provides a third set of values \cite{Burnett16GM, Hamid23GM}.

\begin{table}[h]
    \centering
     \caption{Three GenderMag personas and their problem-solving styles}
    \label{tab:facets}
    \begin{tabular}{p{3cm}|p{3cm}|p{3cm}|p{3cm}} 
     & \multicolumn{1}{c|}{\includegraphics[width=1.8cm, height=1.8cm]{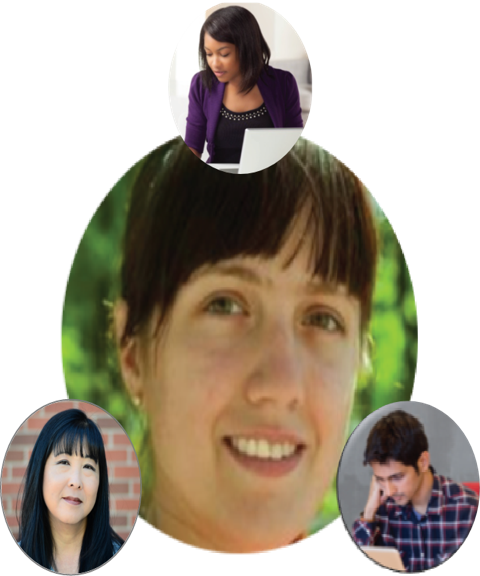}} &\multicolumn{1}{c|}{\includegraphics[width=1.8cm, height=1.8cm]{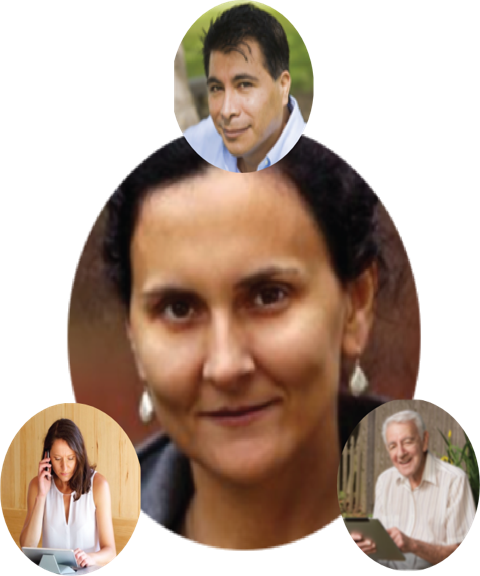}}  & \multicolumn{1}{c}{\includegraphics[width=1.8cm, height=1.8cm]{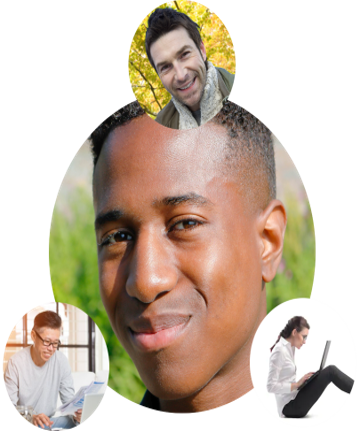}}\\  & \textbf{Abigail/Abishek (``Abi'')} & \textbf{Patricia/Patrick (``Pat'')} & \textbf{Timara/Timothy (``Tim'')} \\
    \hline
    Attitude Toward Risk &  Risk-Averse & Risk-Averse & Risk-Tolerant \\
    \hline
    Self-Efficacy &  Lower & Medium & Higher \\
    \hline
    Motivations &  Task-Oriented & Task-Oriented & Tech-Oriented \\
    \hline
    Information Processing Style &  Comprehensive & Comprehensive & Selective \\
    \hline
    Learning: by Process vs. by Tinkering &  Process-Oriented & Learns by tinkering: tinkers reflectively & Learns by tinkering (sometimes to excess) \\
    \hline
    \multicolumn{4}{l}{\footnotesize Abi and Tim have problem-solving styles at each facet's endpoints, and Pat has a mix of possible facet values.}
    \end{tabular}
\end{table}
\par

\boldification{GenderMag weaves the facets and personas together through a specialized form of cognitive walkthrough}
The GenderMag faceted personas are integrated into a systematic specialized Cognitive Walkthrough process. 
A group of evaluators adopt a specific GenderMag persona, say Abi, customize the persona’s background (optional), and systematically ``walkthrough'' each step of a use case. 
At each step, they answer three questions: (i) will <Abi> have formed this subgoal as a step to their overall goal? why?, (ii) Will <Abi> know what to do at this step? Why?, and (iii) if <Abi> does the right thing, will they know that they did the right thing and are making progress toward their goal? Why? 
If the evaluators find any answer is negative because of the persona's particular problem-solving styles, they report an inclusivity bug, provide their reasons, and note down the problem-solving styles that support their reasoning. 
When fixing the bugs, these notes help ensure that the fixes are designed around the persona's problem-solving styles that highlighted the bugs.

\par



\subsection{Related Work}
 \DraftStatus{MMH: D2.5}
\NOTE{
    Two key questions for assessing other papers to reference: \\
    Does it help improve users' mental models of AI? \\
    (Is it competition?) Who else used GenderMag-esq methods in XAI?  \\
}

\boldification{
    Theme: Improving user experience / mental models of AI with inclusivity evaluation methods--thus addressing diverse end users’ need for cognition in XAI \\
    ** below outline is out of data **
    0. (background) GenderMag is a usability inspection method and can be used to find and fix inclusivity issues.
    1. People have been working on improving mental models of AI systems leveraging different XAI techniques~\cite{Dodge19Fairness,anderson2020mental,Kulesza15Principles, Kulesza10Debug, Alipour21Improving, Brachman23Following, Schraagen20Trusting, Pierson23Comparing, nourani2021anchoring, bansal2019beyond, kulesza2013too, cai2019effects}\\
    2. Not enough to just use different techniques, need to consider human factors -> consider problem-solving styles
    3. Recent AI works have started to consider diversity in problem-solving styles, there is a gap present
    **4.  Our work is different as it examines how inclusive methods, namely GenderMag, can improve mental models of AI. \\
}

\boldification{People have been working on improving mental models of AI systems leveraging different XAI techniques (explanations). FAM: D1.8} 
XAI researchers have investigated different explanation properties' abilities to improve users' mental models and understanding of AI systems.
One group of XAI researchers examined the effect of parsimony and stochasticity of an AI’s error boundary on users' mental models of AI capabilities~\cite{bansal2019beyond}.
They found that simpler and more predictable error boundaries improved users' collaboration with AI and AI-assisted decision-making tasks~\cite{bansal2019beyond}. 
Tsai et al. showed another XAI strategy---interweaving an explanation in the conversation flow of an intelligent online symptom checker---that improved users' understanding of the diagnoses and medical recommendations~\cite{Tsai21Exploring}.
A common XAI research area involves comparing different types of explanations to understand each explanation's impacts on users' mental models of AI~\cite{Alipour21Improving, Tsai21Exploring, wang21explanations}.   
For example, Wang and Yin~\cite{wang21explanations} studied the effects of four model-agnostic explanations (feature importance, feature contribution, nearest neighbors, and counterfactual) on participants' understanding of the Machine Learning (ML) model. 
They found two effects in tasks where participants perceived themselves with higher domain expertise: (i) feature importance and counterfactual explanations increased participants' objective understanding; and (ii) all four model-agnostic explanations increased their subjective understanding~\cite{wang21explanations}.
In contrast, when participants were less knowledgeable about the task, explanations did not improve their understanding, uncertainty awareness, or trust.
In a similar study, Cai et al. found that normative explanations (i.e., showing what is ``normal'') improved user understanding and trust in an ML system, especially when the algorithm failed.
In contrast, although comparative explanations revealed algorithmic limitations, they did not always improve users' understanding of the algorithm~\cite{cai2019effects}.

\boldification{while studying different types, researchers suggested a hybrid style of explanations could be more effective in improving users' mental models}
When studying the effects of multiple explanations, XAI researchers also suggested providing more than one explanation to improve users' mental models. Anderson et al. found that participants who used multiple explanations (saliency maps and reward-decomposition bars) had better mental model scores compared to participants who saw one or no explanations~\cite{anderson2020mental}.
This result is similar to Schraagen et al.'s study where mixed explanations---combining both reasons and causes---helped participants achieve the best understanding of a self-driving car's decisions~\cite{Schraagen20Trusting}.
Dodge et al. also proposed a hybrid approach to explain ML systems to support better human-in-the-loop workflow, where global explanations would help users understand and evaluate the ML system and local explanations would help analyze individual predictions~\cite{Dodge19Fairness}.
%


\boldification{It's not enough to use different explanations for mental models, practitioners should consider factors like cognitive biases in humans, soundness and completeness of explanations and type of explanations itself for improved mental models on users. 
}

To improve users' interactions and understanding of AI, a few XAI researchers have investigated human attributes like cognitive biases, need for cognition (NFC), and user diversity. 
For example, Nourani et al.~\cite{nourani2021anchoring} investigated the effects of anchoring bias, a form of cognitive bias, on users' mental model formation in XAI systems.
They found that participants who observed system strengths first were more prone to anchor bias and made more errors, but developed a more accurate mental model of the system than the participants who noticed the weaknesses first.
In another study, Millecamp et al. examined the effects of users' NFC on their interactions, trust, and decision-making in a recommender system through the use of explanations~\cite{Millecamp20Cognitive}. 
They found that participants with high NFC actively sought and engaged with explanations to control and steer the recommendation process whereas participants with low NFC sought explanations when they encountered unexpected recommendations or when searching for specific content~\cite{Millecamp20Cognitive}. 
Similarly, Kulesza et al. studied the mental demands of explanation on users and found that explanations must balance both soundness and completeness to effectively enhance user's understanding of the AI systems, as overly simplistic or incomplete explanations can lead to cognitive overload and diminished trust~\cite{kulesza2013too}.
A recent human-centric approach in XAI involves integrating social explanations that use socio-contextual cues during the explanation process to provide actionable insights to users~\cite{Gong24SocialExplanation, Ehsan21Expanding, Liao21Human, Sun22Investigating}. 
For instance, Brachman et al. found that social explanations, which offered insights into other users' actions, were more helpful compared to system-focused explanations that provided details about the system's input processing and output generation~\cite{Brachman23Following}. 
However, the participants' mental models did not differ based on the explanation types.

\boldification{Recently, researchers have begun to consider diversity of users' problem-solving styles when using AI products. 
}
GenderMag is centrally about inclusivity, and many HCI practitioners have used GenderMag to evaluate and/or improve non-AI products' inclusivity.
Examples include using GenderMag to evaluate or improve inclusivity of Open Source Software project sites~\cite{Guizani22Inlcusivity, santos2023designing}, online courseware~\cite{Chatterjee22Inclusivity}, a system for IT professionals~\cite{Burnett17MS}, learning management platforms~\cite{Hilderbrand20Trench, shekhar2018cognitive}, education software for middle-schoolers~\cite{Burnett16GM}, and a code review tool~\cite{murphy2024gendermag}. 
A few researchers have also begun using GenderMag's problem-solving styles for measurement within AI contexts.
%
Nam et al. used the problem-solving styles, specifically Information Processing Style and Learning Style, to understand how users interacted with a Large Language Model (LLM) designed to help with code understanding~\cite{nam2024using}.
They found that participants' usage of AI features varied with their Learning Style. 
For example, participants with a process-oriented Learning Style were more likely to use the ``Followup'' AI feature than the tinkerers.
Researchers have also used problem-solving styles to measure AI products' inclusivity. 
Anderson et al. compared the inclusiveness of 16 AI products designed using Amershi et al.'s guidelines~\cite{Amershi19Guide} against versions of those products that violated those guidelines, and found that the guideline-compliant AI products were more inclusive for users with diverse problem-solving styles than those violating the guidelines~\cite{anderson2024measuring}.
They also found that supporting diverse problem-solving styles improved the demographic inclusivity of AI products.
Finally, Vorvoreanu et al. used GenderMag to evaluate and improve an AI-powered search tool~\cite{Vorvoreanu19Bias}. 
However, previous research has not investigated GenderMag, or any other inclusive design method, as a vehicle for improving users' mental models of AI systems.

\section{Methodology}
\label{sec:methodology}
\DraftStatus{MMB: top is D2.4}

\subsection{Experimental Design}
\label{sec:method:subsec:experimentalDesign}
\DraftStatus{MMH: D2.5}

To investigate our research questions, we ran a between-subjects lab study to evaluate the Post-GenderMag vs. Original versions of Team Game's XAI prototype.

\subsubsection{Participants}
\boldification{We sent out recruitment flyers through internal mailing lists at Oregon State University, as well as posting these flyers around campus in locations where students could readily see them, and this led them to an online survey for recruitment.}
To recruit participants, we distributed flyers through our university's internal mailing lists and posted them around campus. 
The recruitment emails and flyers linked to a Qualtrics survey, which collected demographic information and asked two eligibility questions to make sure our participants were at least 18 years of age and had not taken any AI courses as we wanted participants with no AI background. 
Based on these criteria, we dropped 41 responses and sent out emails to the remaining 110 eligible respondents individually. 
Ultimately, 77 respondents accepted the invitations and 69 participated in our study.
These 69 participants came from diverse backgrounds, spanning different genders, age groups, college statuses (all summarized in Table~\ref{tab:demographic}), and 39 unique college majors.

\begin{table}[h]
\centering
\caption{Summary of the 69 participants' demographics}
\hspace{1em}
\begin{minipage}{.3\linewidth}
\centering
\begin{tabular}{c|c}
\hline
\textbf{Gender} & \textbf{\#} \\
\hline
Woman & 36 \\
Man & 28 \\
Non-Binary & 2 \\
Prefer not to state & 2 \\
Woman, Transgender & 1 \\
\hline
\end{tabular}
\end{minipage}%
\hspace{2em}
\begin{minipage}{.25\linewidth}
\centering
\begin{tabular}{c|c}
\hline
\textbf{Age} & \textbf{\#} \\
\hline
18-19 & 9 \\
20-29 & 38 \\
30-39 & 17 \\
40-49 & 2 \\
50-59 & 3 \\
\hline
\end{tabular}
\end{minipage}%
\hspace{2em}
\begin{minipage}{.3\linewidth}
\centering
\begin{tabular}{c|c}
\hline
\textbf{College Status} & \textbf{\#} \\
\hline
Undergraduate & 31 \\
Graduate & 21 \\
Has graduated & 9 \\
Other & 8 \\
\hline
\end{tabular}
\end{minipage}%
\label{tab:demographic}
\end{table}

\subsubsection{Procedures}
\boldification{Once participants were assigned to sessions, based on their availability, each of those sessions were randomly assigned either the ``Before'' or ``After'' version of the prototype, and they filled out the GenderMag problem-solving style survey while they waited for the session to start}

We first assigned participants to sessions based on their availability and then randomly assigned each session to one of two versions of Team Game's prototype: the ``Original'' and the ``Post-GenderMag''.
The Original sessions included 34 participants and the Post-GenderMag sessions had 35 participants. 
Each session lasted 2 hours and had 2 to 6 participants. 
We conducted a total of 17 sessions.

\par 
\boldification{We started with the gendermag survey, gave tutorial, and then a practice game}
Upon arriving at the lab, participants filled out the GenderMag survey~\cite{Hamid23GM} (using the version in Vorvoreanu et al.~\cite{Vorvoreanu19Bias}'s supplemental document), in which participants answered Likert-scale questions about their problem-solving styles. 
Then we began the session with a 25-minute hands-on tutorial covering the game, agents, explanations, and participants' tasks. Participants then went through the first three moves of a practice game to get familiar with the prototype and the tasks. 

\par 
\boldification{Main study had 3 games and participants responded to questions after every move, after every game, and finishing all three games.}
For the main tasks, participants observed three games played by two AI agents: Agent Blue and Agent Red. Before Agent Blue made a move, participants predicted the move, provided reasoning, and rated their confidence level. After the move, they identified flaws and suggested changes. After each game, they answered two After-Action Review for AI (AAR/AI) questions~\cite{Dodge21AAR}. 
Once all three games concluded, they completed post-task questions covering agent descriptions, explanation feedback, and subjective workload assessment using NASA TLX~\cite{NASA}. Table~\ref{tab:questionnaire} lists all the questions.

While designing the questionnaire, we used some of the mental model elicitation methods recommended by Hoffman et al.~\cite{hoffman2019metrics} such as ``Prediction Task'' for measuring participants’ ability to predict, ``Task Reflection task'' for participants’ reasoning, ``Glitch Detection Task'' for participants’ ability to find flaws,  and ``Self-Explanation Task'' for participants’ understanding of AI.

\begin{table}[h]
    \centering
    \caption{The questions we asked both groups of participants}
    \begin{tabular}{p{0.9\linewidth}}
        \hline
        \textbf{Before Agent Blue made a move} \\
        \hline
        - Which square do you think the blue agent will place its next move? \\
        - Why do you think that the agent will place its move there? (A few sentences, as detailed as possible) \\
        - How confident are you that the agent will select the square that you predicted? On a scale of 1 (not confident at all) to 10 (extremely confident). \\
        \hline
        \textbf{After Agent Blue made a move} \\
        \hline
        - From looking at the explanations (right), do you think there is a flaw in the agent’s reasoning? If so, what is that flaw? (A few sentences, as detailed as possible) \\
        - What CHANGES would you make in the decisions taken by the agent? (A few sentences, as detailed as possible.) \\
        \hline
        \textbf{After each game: AAR/AI} \\
        \hline
        - What happened in this game (write any good, bad, or interesting things you’ve observed in these past moves and/or games)? (A few sentences, as detailed as possible.) \\
        - Is there anything in the explanation that helps you understand why the AI you’re assessing did the things it did? Please specify which explanation you are referring to. ($\sim$2 sentences) \\
        \hline
        \textbf{After finishing all 3 games} \\
        \hline
        - Please describe the steps in the agent's approach or method. (A few sentences, as detailed as possible.) \\
        - Explanation feedback*: i) the explanation helped in understanding Agent Blue, and ii) the explanation was sufficiently detailed. \\
        - NASA TLX. \\
        \hline
         {\footnotesize *Participants provided feedback for each of the three explanations, BestToWorst, ScoresThroughTime, and OnTheBoard. They responded on a 7-point Likert-scale from Very Strongly Disagree to Very Strongly Agree used to compute ratings.}
    \end{tabular}
    \label{tab:questionnaire}
\end{table}


\boldification{We did not use the responses for agent description as we found out that the agent was different across three games.}
From Table~\ref{tab:questionnaire}'s questions, we discarded responses from ``Please describe the steps in the agent's approach or method...'', because Agent Blue's noise levels differed. In games 1 and 3, the agent had the same noise levels, but different levels in game 2. This means participants would be describing different agents rather than one. 
To mitigate this threat, we excluded these responses from our analysis, despite Team Game's lead (who created the agent) suggesting that the noise differences didn't affect the agent's behavior much.

\subsection{Qualitative Analysis Methods}
\label{sec:methodology_conceptCodes}

\boldification{We got the data and used an affinity diagram to find the initial codeset with 35 concepts. We coded the data and got at least 80\% IRR for all four concept code groups.}

When all the study sessions were over, we used affinity diagramming to organize participant responses into concepts that the participants expressed. 
Five of the researchers independently marked participants' responses that described some aspect of their mental model of the AI agents.
We then discussed the results as a group and developed the initial concept codeset---further modifying it with input from Team Game's lead (the subject matter expert who created the AI agents).
Finally, the concept codeset had 35 concept codes divided into four groups: ``Prediction Why'', ``Strategy Should Be'', ``Flaws'', and ``How Agent Behaves''  (Appendix's Table~\ref{tab:codes_definitions}). 
Four of the researchers (R1, R2, R3, and R4) then worked in pairs of two as per Table~\ref{tab:IRR} to independently code >=20\% of each of the four concept code groups. 
Table~\ref{tab:IRR} lists the inter-rater reliability rates (IRR)~\cite{Stemler04IRR} of the agreement tests. 
Given the high IRRs, one researcher from each coder pair finished the rest of the coding for their assigned concept code group. 


\begin{table}[h]
    \centering
    \caption{Inter-rater reliability results (IRR) for each concept code group}
    \begin{tabular}{l|c|c|c}
        \hline
        \textbf{Concept Code Group} & \textbf{IRR\%} & \textbf{Test Data\%} & \textbf{Coders} \\
        \hline
        Prediction Why 
            & 82\% & 20\% & R1, R2 \\
        \hline
        Flaws 
            & 80\% & 26\% & R1, R3 \\
        \hline
        Strategy Should be 
            & 81\% & 25\% & R1, R2 \\
        \hline
        How Agent Behaves 
            & 81\% & 20\% & R1, R4 \\
        \hline
    \end{tabular}
    \label{tab:IRR}
\end{table}

\subsection{Statistical Analysis Methods}
\label{sec:method:subsec:statMethods}
\boldification{We used one-tailed t-tests for validating hypotheses on positive effects, two-tailed t-tests when effects can occur in positive or negative directions, mixed design ANOVA for considering both within and between subjects factors, and linear regression for prediction model.}
We used several statistical methods for comparing our participants.
To compare participants' mental model concepts scores and average explanation ratings we used Student's one-tailed t-tests.
One-tailed t-tests are suitable for detecting effects occurring only in one direction~\cite{Frost21tail}, whereas two-tailed tests are for both positive and negative effects.  
Considering these attributes of the two tests, we turned to the literature related to our study to investigate whether a one-tailed or two-tailed approach was more appropriate. 
For mental model concepts scores and explanation ratings, the literature points to one direction only, unanimously reporting GenderMag to improve the user experience of people with diverse problem-solving styles ~\cite{Guizani22Inlcusivity,Vorvoreanu19Bias,Burnett17MS,Hilderbrand20Trench,Blackmon03CWW,Green00CW}. 
Thus, we used one-tailed t-tests when we hypothesized that inclusivity fixes would bring positive effects.
For other comparisons, where we did not have reason to hypothesize that the effect can exist only in one direction, we used Student's two-tailed t-tests. 
For the comparisons, we used Student's one-tailed or two-tailed t-tests as appropriate when all three assumptions of independent samples t-test---1) independent observations, 2) normal distribution, and 3) equal variance were met.
The independent observation condition was always true as each group was independent and the participants used the prototype individually with no interaction between them. 
When the normality distribution failed from a Shapiro-Wilk test, we conducted a non-parametric Wilcoxon Rank-Sum test to compare participants. 
Lastly, if equal variance failed from an F-test, we conducted a Welch's t-test for the comparison.

We used mixed-design ANOVA for analyses that required investigating both within-subject and between subject-factors.
To investigate if explanation usage could predict participants' mental model concept scores, we used linear regression.



\subsection{Where the Two Versions of the Prototype Came From}
\DraftStatus{MMH: D2.5}
\label{sec:prototypes}

\boldification{As detailed in Anderson et al.'s upcoming research, between the 7th of June and the 12th of July, 2023, the researchers conducted two kinds of GenderMag sessions---find and fix sessions.}

\begin{figure}[h]
\includegraphics[scale=0.25]{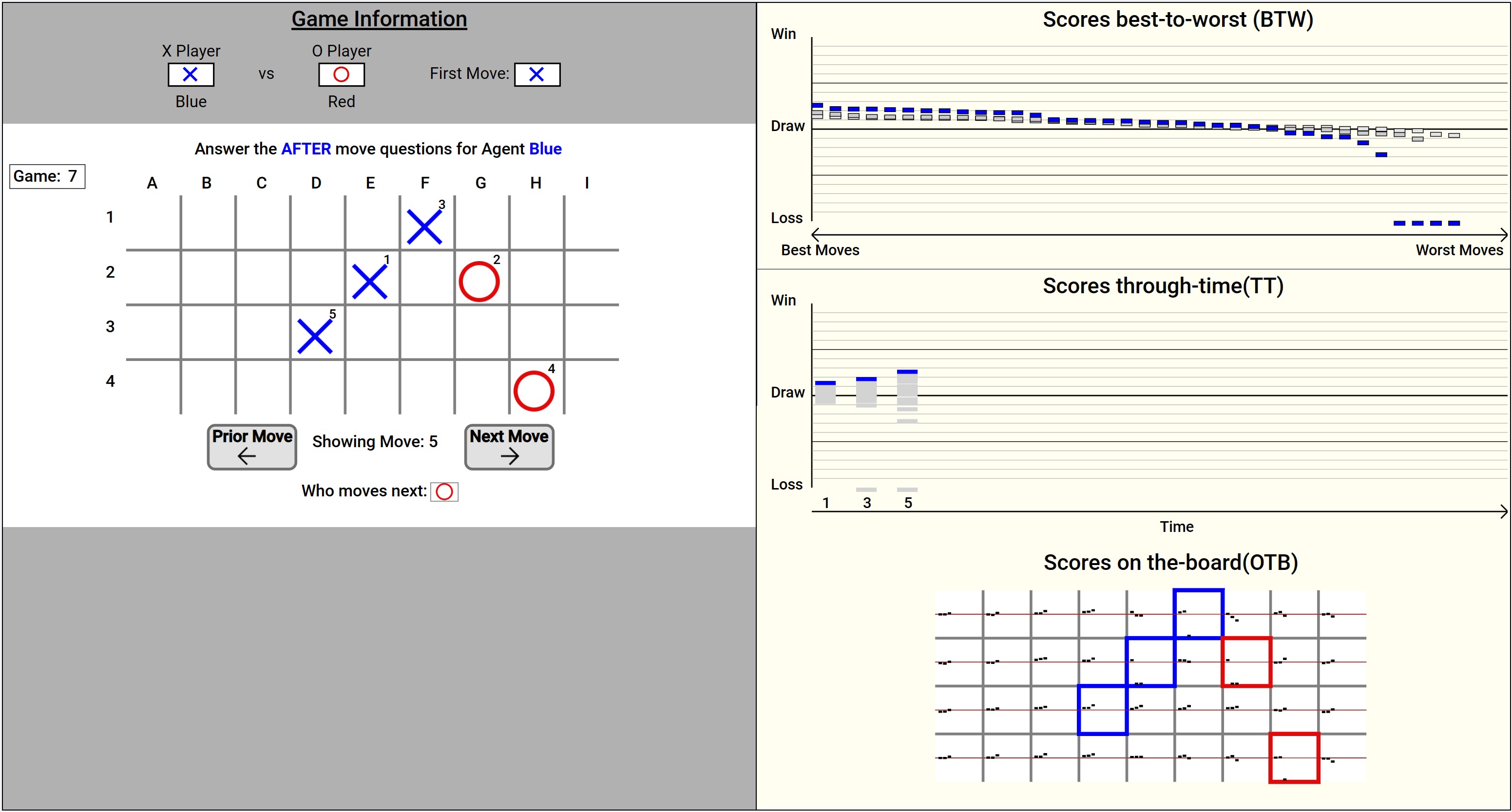}
\caption{Original prototype (we implemented via Svelte) during a game. Participants controlled the game progression with the ``Prior Move'' and the ``Next Move'' buttons. Three explanations to the right provided details on \textcolor{blue}{Agent Blue (X)}'s decisions.}
\label{figure:Pre_Prototype}
\end{figure}

Before our study, Team Game had used GenderMag to find and fix inclusivity issues in their game prototype~\cite{Anderson24InclusiveHAI}. 
This was split into ``find'' and ``fix'' sessions on the prototype (Figure~\ref{figure:Pre_Prototype}).
During these sessions, Team Game evaluated their prototype from the perspective of GenderMag's Abi persona as Abi represents the group of users who are most often left out of the technology development process. 
A snippet of what Team Game found and their proposed fixes is in Table~\ref{tab:fixes_3}. 
Table~\ref{tab:fixes_all} in the Appendix has the complete list.

%
\begin{table}[h]
\centering
\caption{Two fixes, why Team Game fixed them (problem-solving styles impacted), and the changes made for the fixes. The rest of the fixes are in the Appendix's Table~\ref{tab:fixes_all}.}
\begin{tabular}{p{2.5cm}|p{4.5cm}|p{5.8cm}}
\hline
\textbf{Fix-ID \& Fix Name} & \textbf{Fixed-Why + Problem Solving Styles (PSS)} & \textbf{Changes Made for the Fixes} \\
\hline
Fix-2: Exact values and Exact actions  & ``She is not clear on what she should do.....will not take this step.'' PSS: Risk, Learn, Info, SE, Motiv. & 
\hspace{-2mm} -- Showed exact win\%, loss\%, and draw\% for every rectangle in the explanations using tooltips.

\hspace{-2mm} -- Added tooltips that informed about the exact actions offered by the next and the prior move buttons. 
 \\
\hline
Fix-7: Top 5 Moves & ``...but they may not be sure how to reason with this information.'' PSS: Risk, Learn, Info, Motiv. & 
\hspace{-2mm} -- Added ``Top 5'' moves with their score, win, loss, and draw percentages.\\


\hline
\end{tabular}
\label{tab:fixes_3}
\end{table}


\begin{figure}[h]
 \includegraphics[scale=0.3]{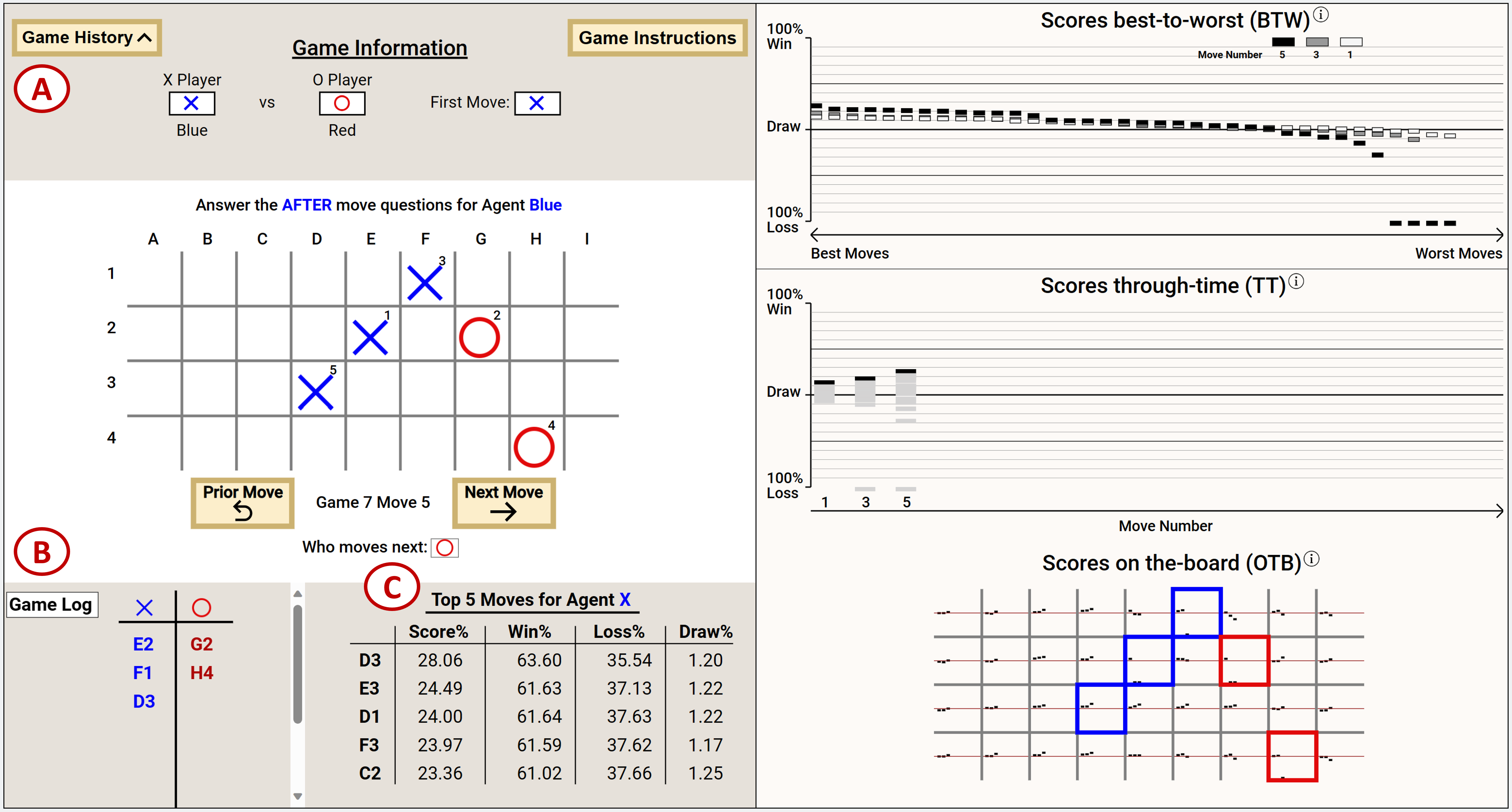}
\caption{Post-GenderMag prototype during a game at the same state as the Original prototype as in Figure~\ref{figure:Pre_Prototype}. Some fixes resulted in new additions such as the (A) Game History, (B) Game Log, and (C) Top 5 Moves. The Appendix's Table~\ref{tab:fixes_all} enumerates all differences between the Post-GenderMag version and the Original version.}
\label{figure:Post_Prototype}
\end{figure}

\boldification{This gave us two versions of the prototype to answer our research questions, a ``Before GenderMag''and ``After GenderMag'' version.}
We call the original version ``Original'' and the redesigned version ``Post-GenderMag'' (as its changes came from the GenderMag-driven walkthroughs). With these two versions, we aimed to address our research questions about
GenderMag-driven inclusivity fixes' impacts on participants' mental model concepts scores, prediction accuracy, and inclusivity.


\subsection{Game Overview and  Prototype}
\DraftStatus{FAM: D 2.8}
\label{sec:methodology_gameOverview}
\boldification{A team of researchers  (Team Game) approached us, interested in making their AI-powered system more inclusive.}

\boldification{They were working in the sequential decision domain of MNK games, based off of Dodge et al.'s original prototype~\cite{Dodge22Rank}. Here's an overview of how MNK games work.} 

Team Game was working in the sequential decision domain of MNK games, using a version of Dodge et al.'s  prototype~\cite{Dodge22Rank}.
MNK games are an abstraction of Tic-Tac-Toe where M and N represent the dimensions of the gameboard ($M \times N$) and the goal is to get a sequence of length K. 
To enable Team Game to make changes in a lightweight way, we simulated the original prototype presented in Dodge et al.~\cite{Dodge22Rank} using a web-based platform using Svelte (JavaScript framework), shown in Figure~\ref{figure:Pre_Prototype}.
In this MNK game, the gameboard---on the left half of the prototype---has 9 columns and 4 rows (36 squares) where the winning condition is getting 4 in-a-row (9-4-4).
Two AI agents, Agent Blue (X) and Agent Red (O) play the game and use a convolutional neural network to predict outcome tuples $O = (Win\%, Loss\%,  Draw\%)$ for each square on the gameboard before making a move.
The agents then assign a ``Score'' for each square using the function $Win\% - Loss\%$ and make moves based on the scores. To the right of the gameboard, the prototype provides three explanations for Agent Blue's move but not for Agent Red.
The prototype used the explanations presented in Dodge et al.~\cite{Dodge22Rank}, namely Scores Best-To-Worst (BestToWorst), Scores Through Time (ScoresThroughTime), and Scores On-the-Board (OnTheBoard).

\subsubsection{Explanation Type: Scores Best-to-Worst (BestToWorst)}

\begin{figure}[h]
\includegraphics[scale=0.5]{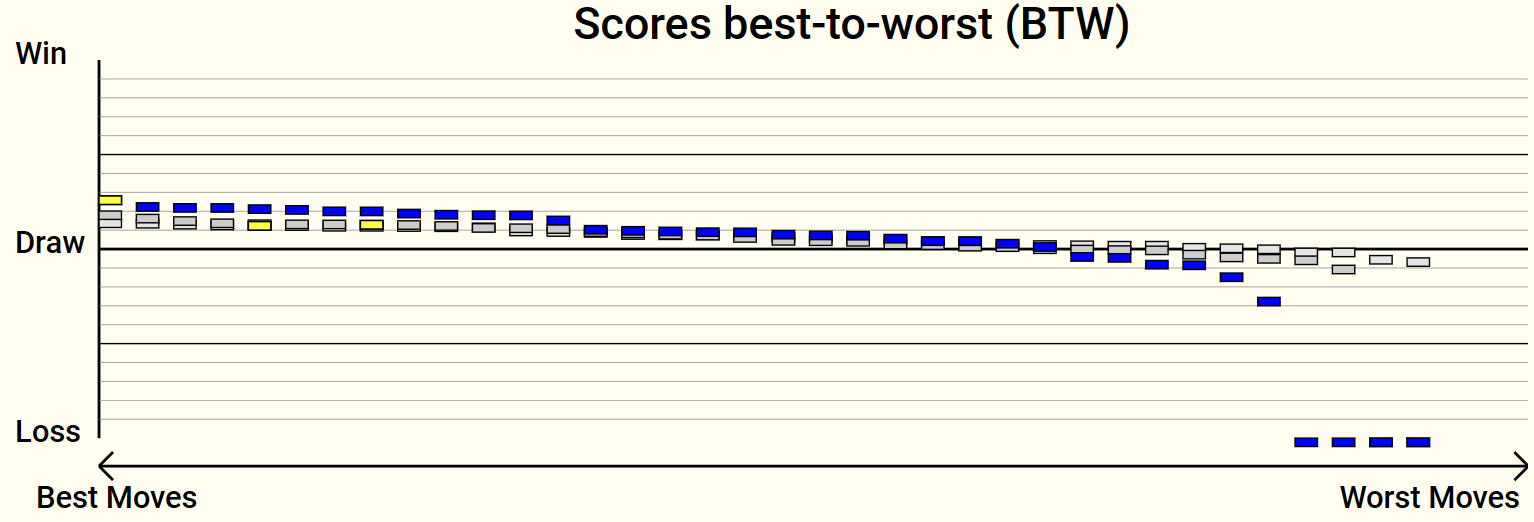}
\caption{BestToWorst in the Original prototype. Agent Blue has made three moves so there are three series of scores. The most recent highest scoring square (blue series) is highlighted along with its past scores.}
\label{figure:Pre_BestToWorst}
\end{figure}

\boldification{The BestToWorst explanation takes all 36 of the AI's predicted moves (one for each square) and arranges them from the highest value on the left to the lowest value on the right}
The Scores Best-to-Worst (BestToWorst) explanation presents all of Agent Blue's predicted scores for the 36 gameboard squares ranked from the best move to the worst move on a graph (Figure~\ref{figure:Pre_BestToWorst}). 
The X-axis orders scores in descending order with the best moves (higher score) on the left. 
The Y-axis shows the score value from -100\% to 100\%.
Agent Blue assigns the lowest scores for illegal moves (i.e., moves that are already taken). 
With each move by Agent Blue, a new series of scores appears in a dark color in the graph, and the prior set of scores turns lighter in color. 
Hovering a mouse cursor over a score in the most recent data series highlights its corresponding past scores and gameboard square.
Similarly, hovering over a gameboard square highlights its corresponding scores in the explanations.

\subsubsection{Explanation Type: Scores Through Time (ScoresThroughTime)}

\boldification{The TT explanation shows scores' changes throughout the game duration, whenever Agent Blue makes a move. This is intended to provide participants with a way to gauge the response to the evolving game through time.}

\begin{figure}[h]
\includegraphics[scale=0.45]{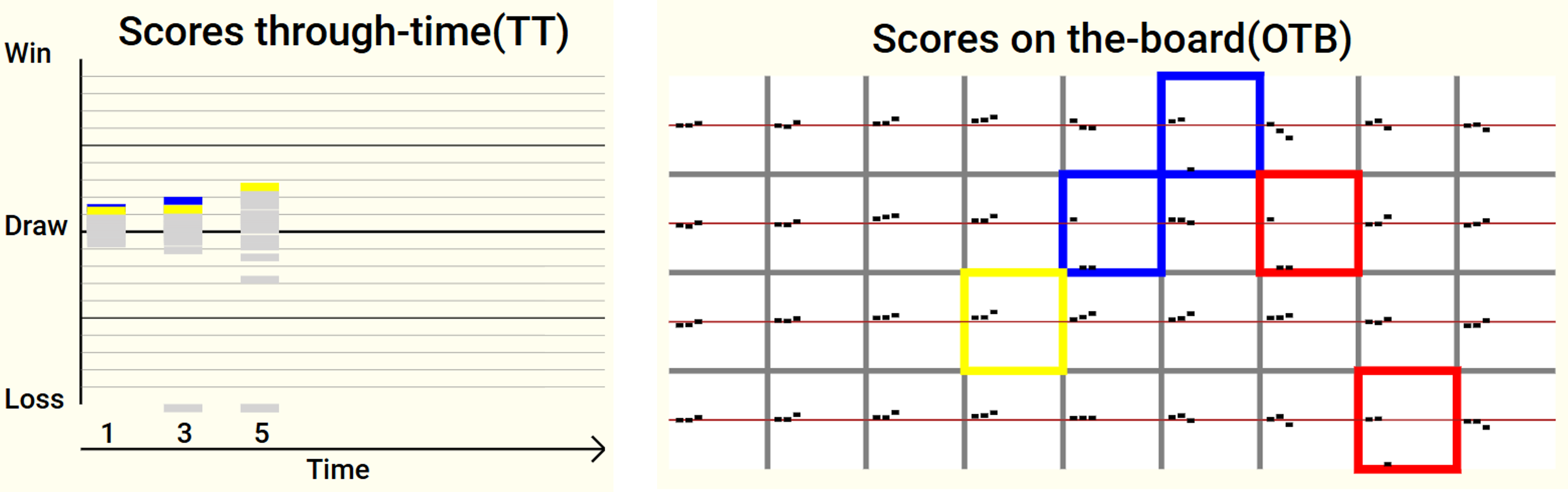}
\caption{Left: ScoresThroughTime in the Original prototype. Agent Blue has played 3 moves so there are 3 columns of scores and the number under each column represents the move number. \\
Right: OnTheBoard in the Original prototype. The square highlighted in yellow is the most recent move by Agent Blue which has a high score.}
\label{figure:Pre_ScoresThroughTime_OnTheBoard}
\end{figure}

The Scores Through Time (ScoresThroughTime) explanation shows the changes in score distribution over the course of the game (Figure~\ref{figure:Pre_ScoresThroughTime_OnTheBoard}: Left).
The X-axis is time (decision points) and the Y-axis is the score.
Each time Agent Blue makes a move,  a new column appears in ScoresThroughTime with the agent's predicted scores for all 36 gameboard squares.
Each column organizes the scores from the highest at the top (marked in a darker color) to the lowest at the bottom. 
ScoresThroughTime has the same highlight interactions as BestToWorst.


\subsubsection{Explanation Type: Scores On-the-Board (OnTheBoard)}

\boldification{The OnTheBoard explanation shows the scores through time but at the level of granularity of each of the squares, where each little dot shows the AI's evaluation of that square through time.}
The last explanation, Scores On-the-Board (OnTheBoard), shows a mini-version of the ScoresThroughTime chart for each square on the board  (Figure~\ref{figure:Pre_ScoresThroughTime_OnTheBoard}: Right). 
Each dot shows Agent Blue's predicted score for that square over different moves.
When Agent Blue makes a move, OnTheBoard adds a new data point to each square's chart.
If a square is occupied, it has a highlighted border in the same color (blue or red) as the occupying agent.
Hovering over a square in the gameboard highlights the corresponding chart in the explanation and vice versa.

\subsection{Measuring Mental Model Concepts for this Game}
\DraftStatus{MMB: top is D3 (because it's empty)}

\subsubsection{Mental Model Concepts Rubric}
\label{sec:methodology:subsec:mentalModelRubric}
\DraftStatus{MMH: D3}
\boldification{Team Lead at PSU picked the codeset from our initial list of codes for measuring the mental model. They have reasoned why they should go with this set of codes. These selected codes center around the basic concepts the designers of the system wanted the agent to abide by.}

We needed a mental model concepts rubric to calculate the mental model concepts score of each participant. From the four concept code groups of the codeset, Team Game's lead, who had created the AI agents, selected from the list in the Appendix's Table~\ref{tab:codes_definitions}, ``Flaws'' and ``How Agent Behaves'' for the mental model concepts rubric (Table~\ref{tab:ConceptDefinition}).

\boldification{Team Game's lead picked flaws as it provides how good participants were in identifying and detailing flaws.}
The agent creator pointed out that to identify flaws, participants needed to understand the scoring system, the agent's decision-making process, and the gameboard. They also mentioned both local and global understanding of the Agents' moves were necessary for detecting flaws.
For these reasons, the agent creator picked the three ``Flaws'' concepts (top three rows in Table~\ref{tab:ConceptDefinition}) that reflected Agent Blue's decision-making ``Flaws'' and assigned positive weights for these concepts in the mental model concepts rubric.
\par

\boldification{Team Game's lead picked these codes because they provided insights on participants' overall understanding of the AI agents}
The agent creator also selected the seven ``How Agent Behaves'' concepts (Table~\ref{tab:ConceptDefinition}) to capture participants' perceptions of the AI agents' behaviors, including agents' priorities, strengths, and weaknesses. According to the agent creator, these concepts would cover the required understanding to evaluate a participant's comprehensive mental model of the AI agents, justifying the concepts' integration into the mental model concepts rubric. The agent creator also pointed out that mentioning the correct concepts (top three rows in Table~\ref{tab:ConceptDefinition}'s ``How Agent Behaves'' section) indicated participants' better understanding of the AI agents. In contrast, the INcorrect concepts (bottom four rows) indicated a lack of understanding. Thus, the agent creator assigned positive weights for the correct concepts and negative weights for the incorrect ones in the mental model concepts rubric. 


\begin{table}[h]
    \centering
    \caption{Concepts used in the mental model concepts rubric and their weights/weight ranges. \textbf{$+$} sign
 indicates correct concepts with positive weights and \textbf{$-$} sign indicates incorrect concepts with negative weights.}
    \includegraphics[trim={0 11.2cm 0 2.5cm}, clip, width=0.99\linewidth]{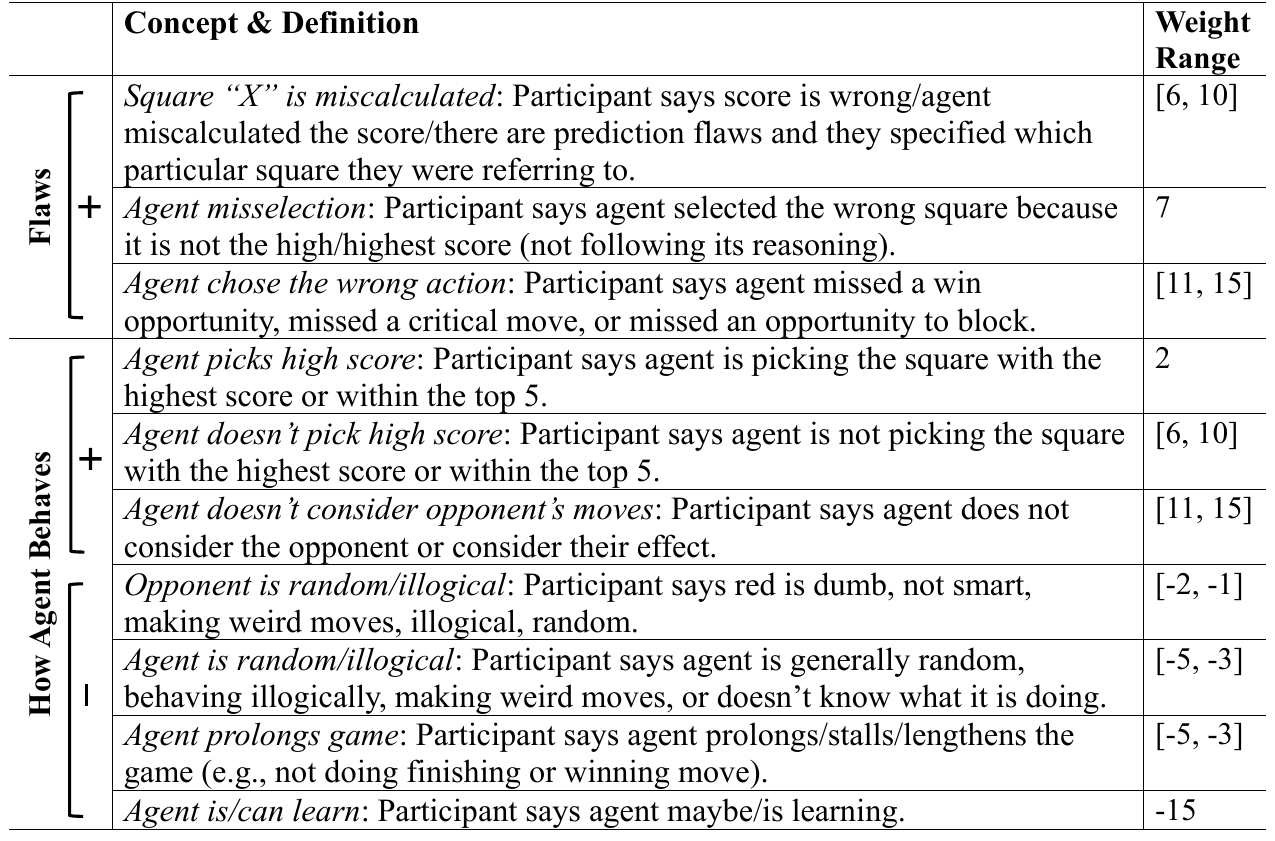}
    \label{tab:ConceptDefinition}
\end{table}

\subsubsection{Assignment of Weights: Finding the Weight Ranges}
\DraftStatus{MMH: D2.5}

\boldification{Team Game's lead and our team then worked on the possible weights of the selected codes and assigned the weights. We reasoned why some codes should have more weights, and some less}

Team Game's lead finalized specific weights for certain concepts and a range of weights for others (Table~\ref{tab:ConceptDefinition}). A concept with a range of weights indicates that any values within the range would be appropriate for the mental model concepts rubric, as defined by Team Game's lead. Our team also provided feedback on possible weights when Team Game's lead made the decisions. 
From these, we derived Table~\ref{tab:4criteria_weights}'s final weights as follows.



\begin{table}[h]
        \caption{Criteria for weight selection derived from Table~\ref{tab:ConceptDefinition}'s weight ranges. HAB = How Agent Behaves.}
        \begin{tabular}{l|c|c|c}
            \hline
            \textbf{Criterion} & \textbf{Flaws} & \textbf{HAB (Correct)} & \textbf{HAB (INcorrect)} \\
            \hline
            Criterion 1: Max within constraints & Middle & Max & Max \\
            \hline
            Criterion 2: Min within constraints & Min & Max & Min\\
            \hline
            Criterion 3: INcorrects' effect check & Min & Max & Middle \\
            \hline
            Criterion 4: Flaws' effect check & Min & Max & Max\\
            \hline
        \end{tabular}
        \label{tab:4criteria_weights}
\end{table}


\subsubsection{Assignment of Weights: Finalizing the weights}
\DraftStatus{MMB: D2.5}
\label{sec:3.5.3}

\boldification{To finalize the exact weights from the ranges, we came up with four criteria.}

To finalize the weights, we came up with four weighting strategies, described via Table~\ref{tab:4criteria_weights}'s four criteria. These criteria analyzed the possible effects of the two concept code groups within the mental model concepts rubric: (i) Flaws, (ii) How Agent Behaves (Correct), and How Agent Behaves (INcorrect). 
To check the effects of these groups, we considered different combinations of weights (min, middle, max). 

For Criteria 1 and 2, we assigned all the concepts with maximum/middle and minimum weights within their ranges respectively. 
Criterion 3 varies from Criterion 2 only in INcorrect concepts' weights to check if ``INcorrect'' concepts make much difference within the mental model concepts rubric. 
Lastly, Criterion 4 varies from Criterion 1 only in Flaws concepts' weights to check if ``Flaws'' concepts make much difference. 
As an example of the Min, Middle, and Max weights is the concept ``Square `X' is miscalculated,'' which has a weight range of [6, 10].
Within this range, we considered its min weight as 7, middle weight as 8, and max weight as 10.


\boldification{Rationale behind our weight selection criteria}
The reasonings behind the weights in the columns of Table~\ref{tab:4criteria_weights} were as follows.
\par
For the \textbf{Flaws} concepts, the agent creator emphasized the importance of detecting the flaws as it shows a participant’s understanding of the agent’s gameplay. However, the agent creator also noted that detecting the flaws was not very difficult.
Thus, we assigned low weights to the flaws' concepts with ``min'' weights in three criteria, ``middle'' weights in one, and ``max'' weights in none.

\par
Next, the agent creator concluded that \textbf{How Agent Behaves (correct)} concepts provide critical insights into the participant’s overall understanding of the agent's behavior and recommended assigning higher weights for these concepts.
So, we assigned the maximum possible weights within these concepts' ranges in all four criteria.

\par
Lastly, the agent creator wanted the \textbf{How Agent Behaves (INcorrect)} concepts to account for participants' incorrect understanding of the AI agents.
However, they didn't specify any priorities for these concepts.  
So, we tested all three---min, middle, and max weights for these concepts within the four criteria.

Table~\ref{tab:exact_weights_4Criteria} shows all the concepts' exact weights that we derived from Table~\ref{tab:4criteria_weights}'s four criteria.

\begin{table}[htbp]
    \centering
    \caption{Exact concept weights used in the four criteria. All are within the ranges specified in Table~\ref{tab:ConceptDefinition}, and were chosen as per the strategies in Table~\ref{tab:4criteria_weights}. C-1 = Criterion 1, C-2 = Criterion 2 etc.}
    \begin{tabular}{l|c|c|c|c}
        \hline
        \textbf{Concept} & \textbf{C-1} & \textbf{C-2} & \textbf{C-3} & \textbf{C-4} \\
        \hline
        Square ``X'' is miscalculated & 8 & 7 & 7 & 7 \\
        \hline
        Agent misselection & 7 & 7 & 7 & 7 \\
        \hline
        Agent chose the wrong action & 13 & 11 & 11 & 11 \\
        \hline
        Agent picks high score & 2 & 2 & 2 & 2 \\
        \hline
        Agent doesn't pick high score & 10 & 10 & 10 & 10 \\
        \hline
        Agent doesn't consider opponent's moves & 15 & 15 & 15 & 15 \\
        \hline
        Opponent is random/illogical & -2 & -1 & -1 & -2 \\
        \hline
        Agent is random/illogical & -5 & -3 & -4 & -5 \\
        \hline
        Agent prolongs game & -5 & -3 & -4 & -5 \\
        \hline
        Agent is/can learn & -15 & -15 & -15 & -15 \\
        \hline
    \end{tabular}
    \label{tab:exact_weights_4Criteria}
\end{table}



\section{RQ1 Results: Which Participants Had Better Mental Model Concepts Scores, and Why?}
\DraftStatus{MMB: Top is D2.5}
\label{results_sec:mentalModelScores}

Curb-cut effects are a pair of effects---a positive ``better for everyone'' effect and a ``better for an underserved group'' effect.
We begin by considering the ``better for everyone'' question.
\par 

\begin{figure}[h]
\includegraphics[scale=0.35]{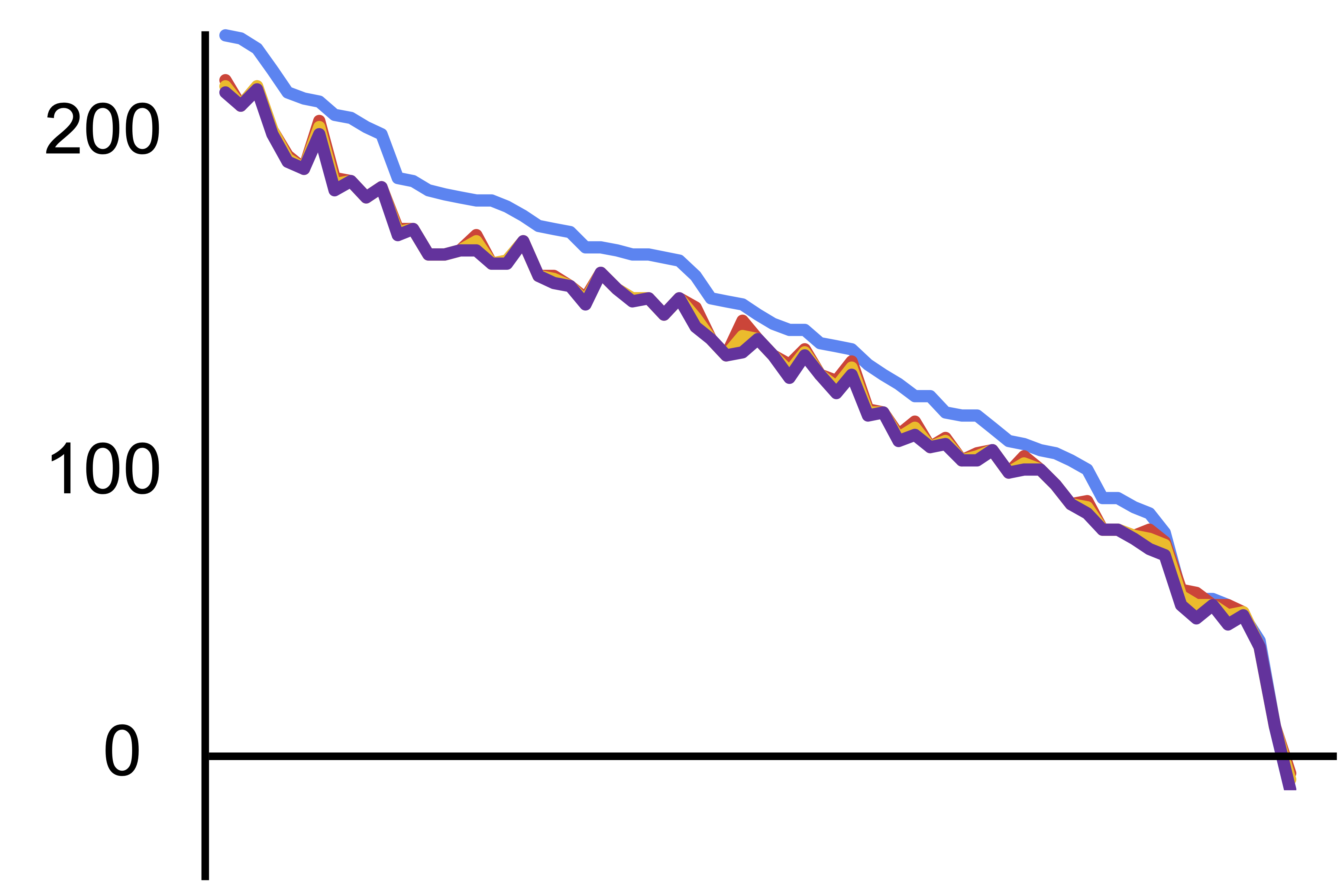}
\caption{Mental model concepts scores (y-axis) for each participant (x-axis) using \textcolor{blue}{Criterion 1 (blue)} vs. \textcolor{red}{Criterion 2 (red)} vs. \textcolor{orange}{Criterion 3 (orange)} vs. \textcolor{purple}{Criterion 4 (purple)}.  
Participants are ordered by rank (highest to lowest mental model concepts scores as per \textcolor{blue}{Criterion 1}). 
Criteria \textcolor{red}{2}, \textcolor{orange}{3}, and \textcolor{purple}{4} rank participants almost identically, and all three rank them similarly to \textcolor{blue}{Criterion 1}.
}
\label{figure:mentalmodelscore4criteria}
\end{figure}

\boldification{Which criterion to use? Turns out they produce similar results: Top-ranked participants are top-ranked across 4 criteria. Same goes for lower-ranked participants}
To answer RQ1's question about participants' mental model concepts, we needed to calculate each participant's mental model concepts scores using the concept weights and concept counts\nolinebreak\footnote{The concept count is the total number of times we identified a particular concept in a participant's responses.},~
using one of the four criteria from Table~\ref{tab:4criteria_weights}---but which one?
Fortunately, it turned out that the choice among these four criteria made little difference: the four criteria generated similar participant rankings (Figure~\ref{figure:mentalmodelscore4criteria}). 
For example, 9/10 of the top ten participants by Criterion 1 were also the top ten participants of Criteria~2, 3, and~4.  
Likewise, 10/10 of the bottom ten were identical in all four criteria. 
Indeed, Criteria 2, 3, and 4 gave almost identical ranking results, as their almost identical overlapping lines show (Figure~\ref{figure:mentalmodelscore4criteria}).

\par

\boldification{We rejected criteria 1 because of its differences from the other three criteria, which closely agreed with one another. Of the remaining three, we picked Criterion 4 because it had Min values for Flaws and Max values for How Agent Behaves.}

In choosing one among such similarly behaving  criteria, we rejected Criterion 1 because its line graph was a little different from the other three criteria. 
Of the remaining three, we selected \textbf{Criterion 4: Flaws' effect check} for two reasons.
First, it avoids overemphasizing the ``Flaws'' concepts.
This was important because ``Flaws'' responses came throughout the games' progressing stages, when participants' mental models were still evolving, so sometimes it did not reflect participants' final understanding. 
Second, Criterion 4 emphasizes ``How Agent Behaves'' concepts, which came from the end-of-game AAR/AI questions and thus represented participants' final mental models.
From here on, the term ``mental model concepts scores'' will refer to the scores calculated using Criterion 4.


\subsection{Which Participants Had Better Mental Model Concepts Scores?}
\DraftStatus{MMB: D3}
\label{subsubsec:PostGMSignficantMMScore}

\boldification{We used t-tests. previous research showed that gendermag fixes had positive impacts on participants. So we ran a one-tailed greater t-test to check if Post-GenderMag had better mental model concepts scores}

To compare the two participant groups' mental model concepts scores, we used a one-tailed t-test, following the statistical reasoning explained in Section \ref{sec:method:subsec:statMethods}.
The results showed that Post-GenderMag participants had significantly higher (better) mental model concepts scores than the Original participants ($t(67)=1.6788, p=.0489, d=.4043$; Post-GenderMag participants' $mean = 129.06$; Original participants' $mean = 109.12$).





\subsection{Why?---Three Concepts}
\DraftStatus{MMB: D2.5}
\label{subsubsec:WhyPostGMBetter}



\par 
 


\boldification{why Post-GM had better mm score? to find out we compared the 2 groups' acquisition of mental model concepts to identify the key differences.}
To find out why Post-GenderMag participants had higher mental model concepts scores than the Original participants, we compared each group's responses about the mental model concepts defined in Section~\ref{sec:methodology:subsec:mentalModelRubric} to identify which ones differentiated the Post-GenderMag participants from the Original participants.

\begin{figure}[h]
\includegraphics[scale=0.5]{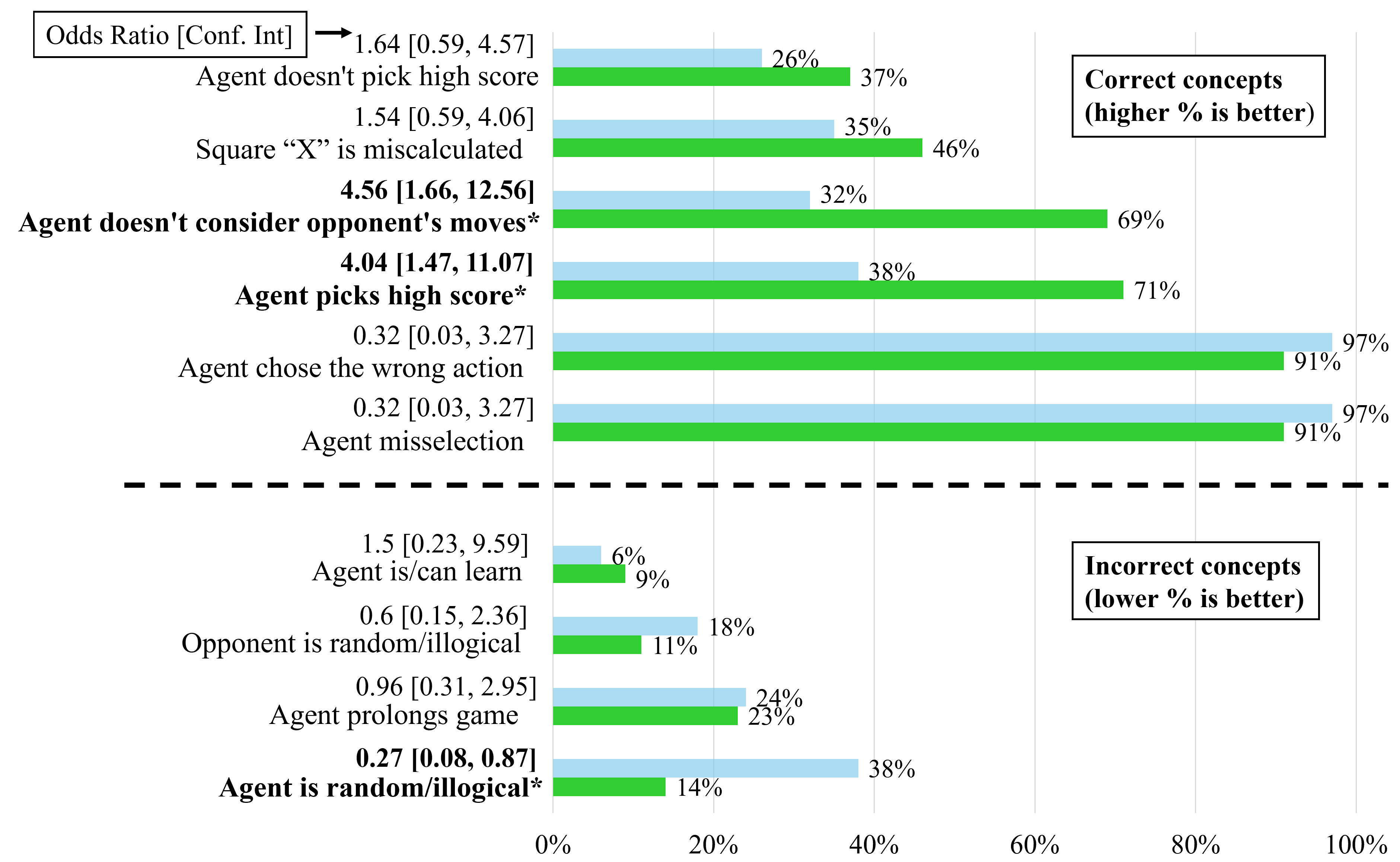}
\caption{Percentage of \Original (n=34) and \PostGenderMag (n=35) participants whose responses included $>=$1 statement about that concept.  Correct concepts (above the dashed line) increase the mental model concepts score, and incorrect concepts decrease it. \\
Odds Ratios: The concept is more likely ($>$ 1), equally likely ($=$ 1), or less likely ($<$ 1) to occur in the Post-GenderMag group.
\textbf{*Bold} concepts' Odds Ratios are statistically significant. 
}
\label{figure:MentalModelPercentageOddsRatio}
\end{figure}

\boldification{Figure below illustrates the differences in concept acquisition.}
Figure~\ref{figure:MentalModelPercentageOddsRatio} shows the results: concepts above the dashed line were correct statements about how this particular AI worked, and concepts below the dashed line were incorrect statements. 
(Recall the complete definitions of each concept from Table~\ref{tab:ConceptDefinition}.)

\boldification{3 concepts stood out with big differences in percentages and high odd ratios.}
Three concepts particularly differentiated the Post-GenderMag participants' remarks from those of the  Original participants: ``Agent picks high score'', ``Agent doesn't consider opponent's moves,'' and ``Agent is random/illogical''.
To measure how much each concept differentiated the two groups, we calculated the odds of Post-GenderMag participants vs. Original participants expressing these concepts; the odds ratios are shown in Figure~\ref{figure:MentalModelPercentageOddsRatio}.
For example, as the figure shows, Post-GenderMag participants were 4.56 times more likely to correctly say that the ``Agent doesn’t consider opponent’s moves''; 4.04 times more likely to correctly say that the ``Agent picks high score''; and 3.7 ($1/0.27$) times \textit{less} likely to \textit{in}correctly say that the ``Agent is random/illogical'' than the Original participants.

\boldification{the odds ratio of these 3 concepts were also significant}
In fact, these three concepts were significant differentiators of Post-GenderMag participants' vs. Original participants' responses about their mental model concepts, at the $\alpha=.05$ level.
Calculating significance of an odds ratio for $\alpha=.05$ requires computing its $95\%$ confidence interval; if the confidence interval does not include 1, then the odds ratio is significant.
Figure~\ref{figure:MentalModelPercentageOddsRatio} highlights the three significant concepts by these calculations in boldface and marks them with asterisks (*), which we will term the \textit{``differentiating''} concepts.


\subsection{The Fixes Behind the Differentiating Concepts}
\DraftStatus{MMB: top is D3}
\label{subsubsec:GMFixTiedMentalModel}

\par 
\boldification{We then investigated how the fixes are related to these 3 concepts}
These significant differences in participants' acquisition of the three differentiating concepts raise the question of what \textit{particular fixes} were tied to these particular differences.
We again turned to Post-GenderMag participants' responses to see what they had to say about this.

\begin{figure}[h]
\includegraphics[scale=0.5]{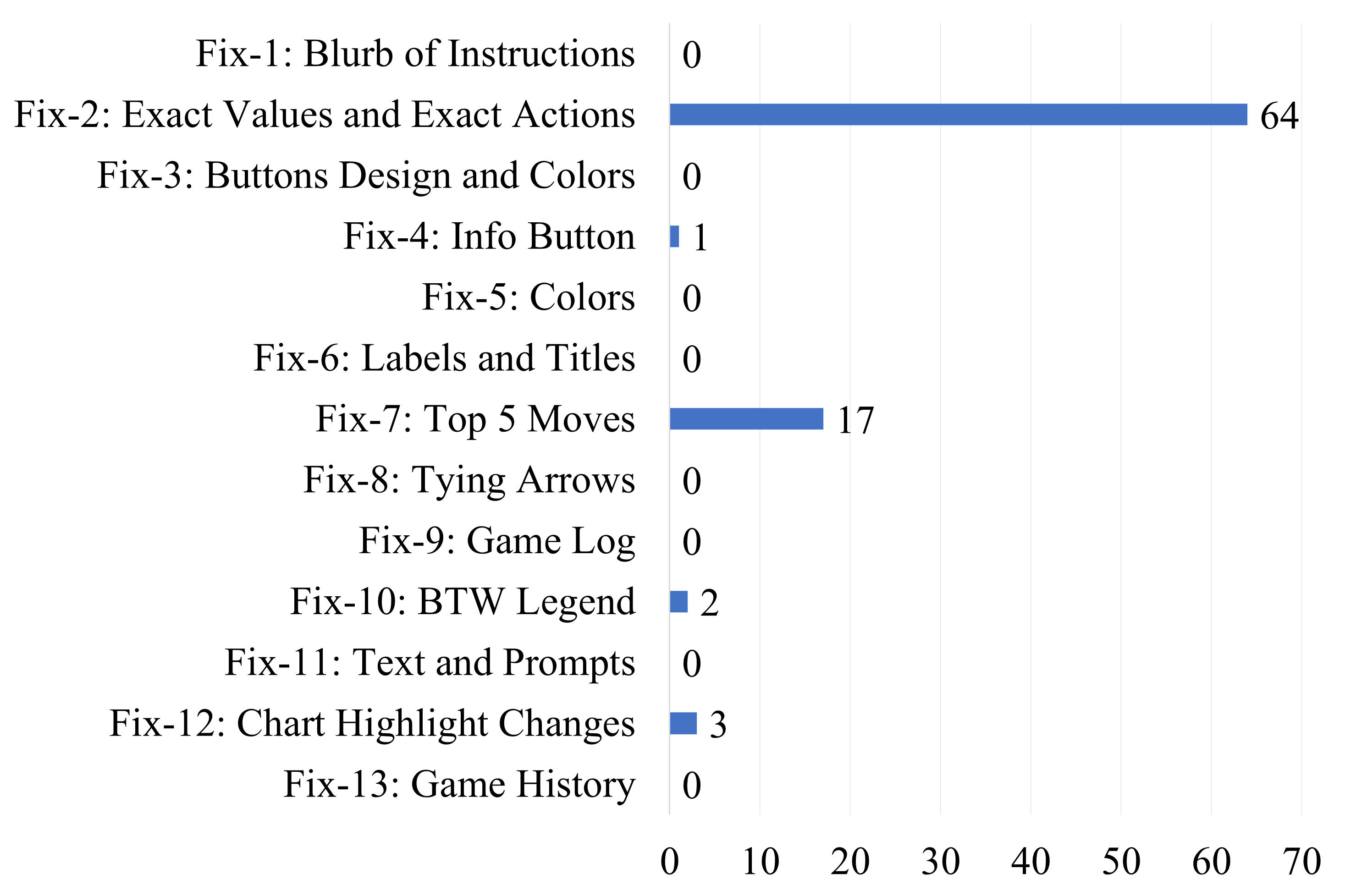}
\caption{Total number of times Post-GenderMag participants referred to each fix. Another quantitative way (not shown) Fix-2 and Fix-7 stood out  was that 26/35 (74\%) and 12/35 (34\%) individual participants referred to Fix-2 and Fix-7, respectively.} 
\label{figure:fix_count}
\end{figure}

\boldification{We did so by investigating the explanation comments (since the fixes are mostly about expls, and lo! -- 2 fixes stood out, namely Fix-2 and Fix-7}
For this purpose, we considered only responses to questions about explanations (explanation feedback in Table~\ref{tab:questionnaire}), because 9 of Team Game's 13 fixes were changes to explanations. 
Within those responses, two coders independently coded Post-GenderMag participants' references to each fix.
Their inter-rater reliability (IRR) was 83\% with 20\% of the data.
Given this acceptable level of consensus~\cite{Stemler04IRR}, the coders then split up the coding of the rest of the data.
Figure~\ref{figure:fix_count} shows the resulting counts for each fix. 
As the figure shows, two fixes stood out---Fix-2 and Fix-7.

\boldification{These 2 most referred fixes also show their influence on Post-GenderMag participants' mental model concepts scores. Post-GenderMag participants who mentioned at least one of these two fixes had significantly higher mental model concepts scores than the participants who did not mention either of the fixes.}

\begin{figure}[h]
\includegraphics[scale=0.47]{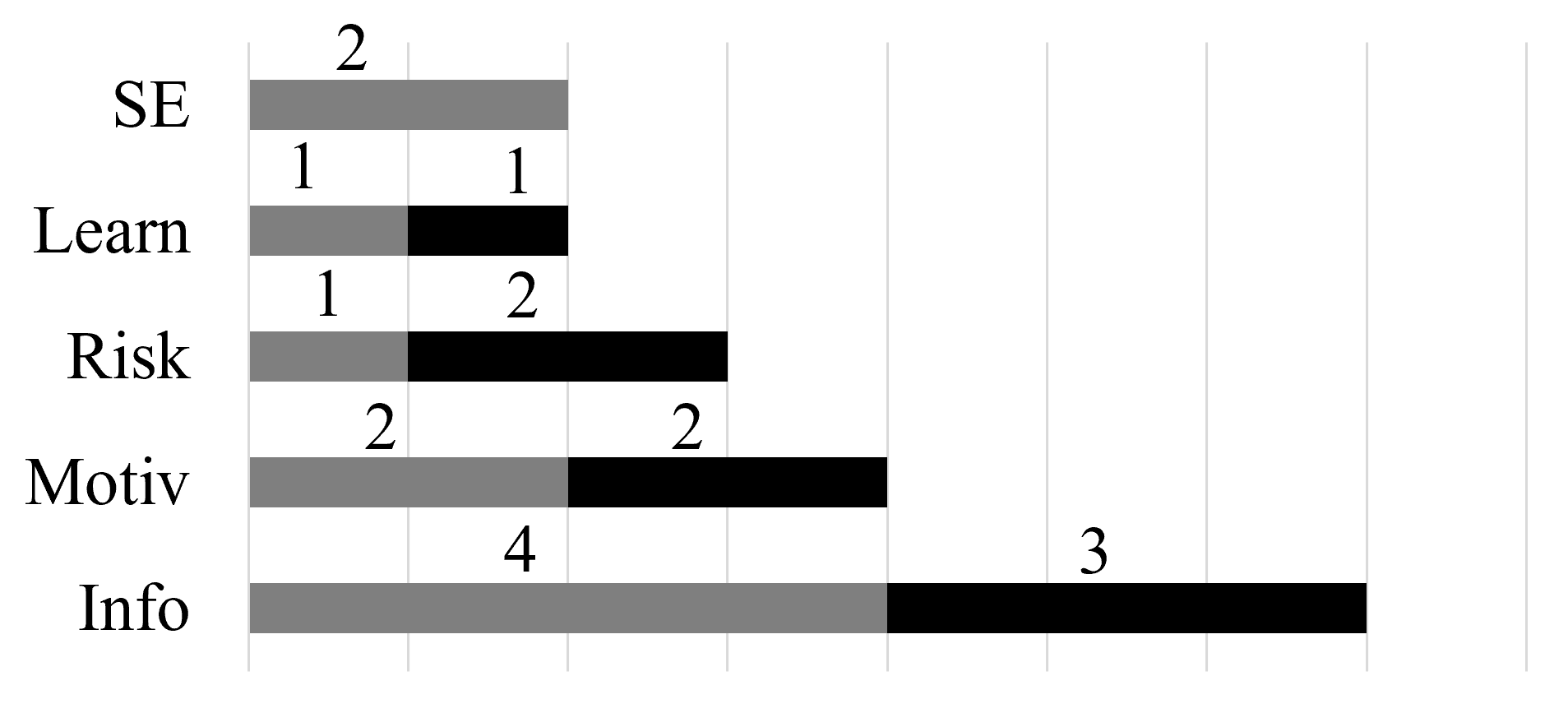}
\caption{Number of times Team Game mentioned the GenderMag problem-solving styles during the GenderMag sessions that led to the development of \colorbox{gray}{\textcolor{white}{Fix-2}} and \colorbox{black}{\textcolor{white}{Fix-7}}.}
\label{figure:Fix_2_7_Facets}
\end{figure}

Recall that these fixes were triggered by Team Game's earlier analysis of the ``inclusivity bugs'' in their Original prototype, and were then designed by Team Game to resolve the bugs in whatever way they saw fit~\cite{Anderson24InclusiveHAI}. 
During that inclusivity analysis, Team Game had noticed several instances of leaving out users whose Information Processing Style was the ``comprehensive'' style; i.e., those who like to gather a sizable batch of relevant information before proceeding with a problem-solving action or decision.
In fact, as Figure~\ref{figure:Fix_2_7_Facets} shows, Team Game talked about this problem-solving style more than any other during their discussions leading to these fixes. 
Ultimately, they devised Fix-2 and Fix-7 to add more information in order to be more inclusive to users with comprehensive Information Processing Styles, who had not been explicitly considered in the Original prototype.

The statistical evidence of Fix-2's and Fix-7's joint impact on participants' mental model concepts scores is very strong.
A two-tailed Student's t-test revealed that the
27 Post-GenderMag participants who referred to at least one of Fix-2 and Fix-7 had significantly higher mental model concepts scores ($mean=143.67$) than the 8 Post-GenderMag participants ($mean=79.75$) who did not refer to these fixes, $t(33)=4.051, p=.00029, d=1.6306$.\footnotemark

\footnotetext{Although one group had 8 participants we still conducted a Student's t-test as all three assumptions of independent samples t-test were met: (1) the observations in each group were independent of each other as each participant used the prototype individually and there was no interaction between them, (2) a Shapiro-Wilk normality test with $W$ statistics closer to 1 (0.95798 and 0.96535) indicating that both groups may be normally distributed and $p>.05$ (.33 and .86) indicating that the data does not deviate significantly from normality, and (3) a Levene's test with $p=.55$ indicating there is no significant difference in variances between the groups.}

\subsubsection{Fix-2: Exact win/loss\% available}
\DraftStatus{MMB: D2.5}

\begin{figure}[h]
\includegraphics[scale=0.62]{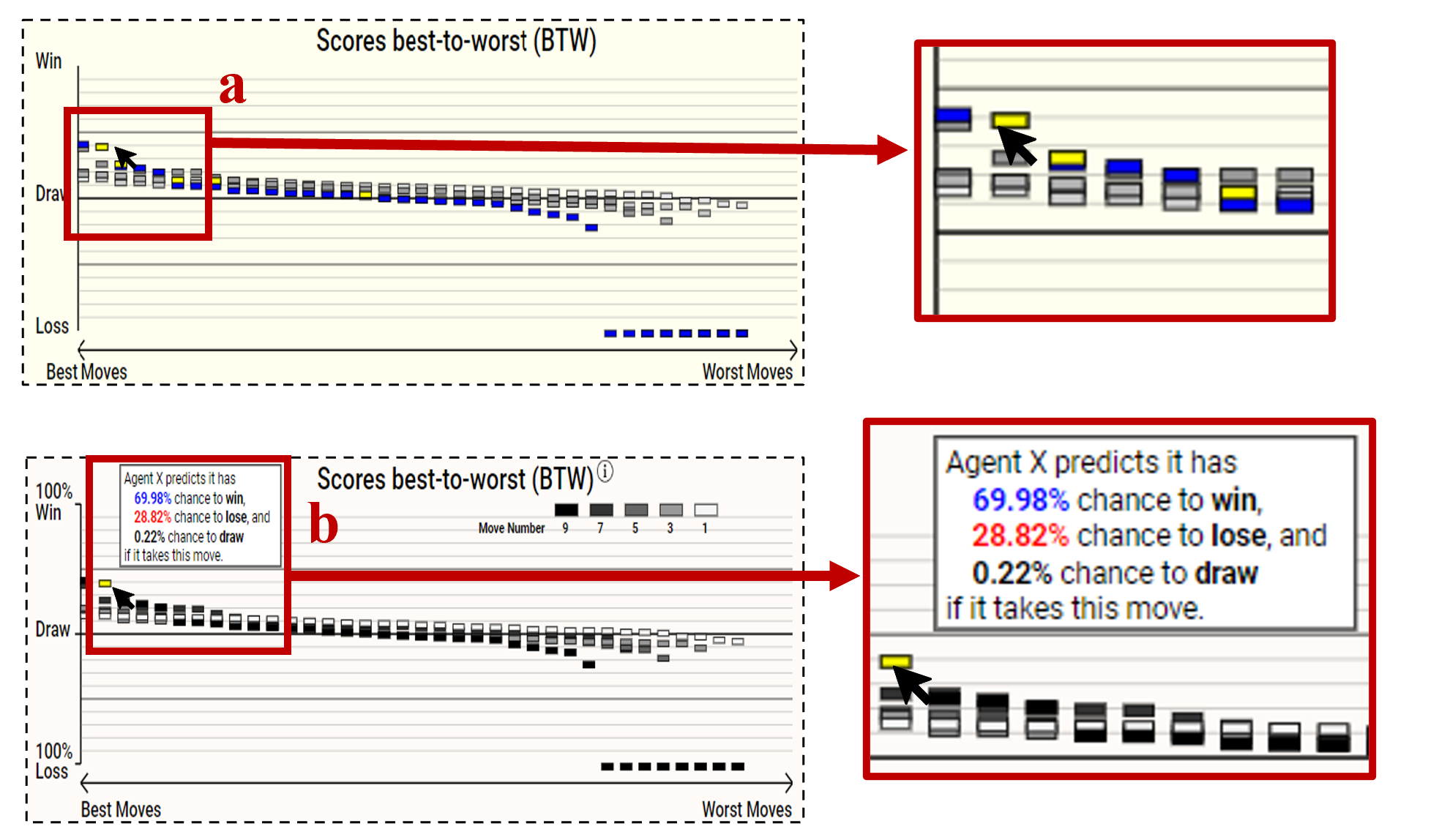}
\caption{Top: (a) Before Fix-2 (Original version). Hovering over a rectangle from the latest data series highlights the rectangles representing the same gameboard square in every data series\protect\footnotemark  ~from the previous decision points, but does \textit{not} show the exact probabilities and does not show which rectangles came from which data series. \\
Bottom: (b) After Fix-2 (Post-GenderMag version). Hovering over the rectangle now explicitly shows the win\%, loss\%, and draw\% of the gameboard square represented by that rectangle.}
\label{figure:Fix2Tooltip}
\end{figure}

\boldification{Information regarding the agent's decision-making process was incomplete, so Team Game fixed it by adding a tooltip that showed a breakdown of the decision with win, loss, and draw percentages}
To increase support across the range of diverse users' Information Processing Styles while also addressing diverse Learning Styles, Motivations, and Attitudes Towards Risk, in Fix-2 (Figure~\ref{figure:Fix2Tooltip}: Bottom) Team Game created a detailed breakdown of Agent Blue's decisions, available via a mouse hover.
%
Specifically, whereas in the Original version (Figure~\ref{figure:Fix2Tooltip}: Top), participants could only guess at an agent's exact scoring (win\%-loss\%) using the graph's placement of the tiny rectangle of interest, 
Fix-2 enabled participants to hover over an explanation's rectangle to see the exact win\%, loss\%, and draw\% if Agent Blue picked the move represented by this rectangle (Figure~\ref{figure:Fix2Tooltip}: Bottom).
Note how this feature supports a wide range of Information Processing Styles: comprehensive information processors can hover to see it, and selective information processors can simply not hover.
Perhaps this is why it was so well-received: 74\% of the Post-GenderMag participants referred to it at some point during the study.

\boldification{Showing the win/loss helped Post-GM group in noticing that the agent picks high score.}

The ability to view these exact win/loss/draw \%s may explain why Post-GenderMag participants were $4.04$ times more likely to notice ``Agent picks high score''. 
As one Post-GenderMag participant wrote:

\begin{quote}
P06-PostGM: \textit{The AI did appear to attempt to select the squares with the highest probability of winning/highest score <Available via Fix-2>.....}
\end{quote}

P06-PostGM continued, speculating about how a particular square's exact win/loss/draw \%s affected the probability of winning.  
Such a specific estimate was made possible by Fix-2's explicit display of win/loss probabilities:
\begin{quote}
P06-PostGM (continuing): \textit{.....even though I don't understand how the probability of winning in move 7 with square F1 wasn't 100\%.}
\end{quote}


\footnotetext{Every time Agent Blue makes a move, the BestToWorst explanation generates a data series containing 36 small rectangles, 1 rectangle for each gameboard square. Where the rectangle lies vertically tells how good the agent thinks moving that square would be. Multiple data series show this info for different decision points.}

 \par
 \boldification{similarly, exact win/loss/draw \%s may have helped Post-GM in understanding that the agent does not consider its opponent' move}
Showing the agent's predicted win/loss\% from Fix-2 may have also influenced why Post-GenderMag participants were $4.56$ times more likely to express the concept ``Agent doesn’t consider opponent’s moves'' compared to the Original participants.
This concept is correct---the agent emphasizes its \textit{own} calculations rather than considering what the opponent agent is doing, as one Post-GenderMag participant pointed out: 
\begin{quote}
    P04-PostGM: \textit{The AI explanations said the moves had high/highest chances of winning<Fix-2 showed Win\%>, but it didn't seem to account for the moves made by red....}
\end{quote}

\par
\boldification{lastly, as Fix-2 provided more info on the agent, Post-GM avoided saying that the agent is random.}
Fix-2's emphasis on the agent's own calculations also may have helped most Post-GenderMag participants avoid an error that the Original participants made: thinking that the ``Agent is random/illogical'', which is not true. 
Compared to Post-GenderMag participants, Original participants were 3.7 ($1/0.27$) times more likely to incorrectly assume that Agent Blue was random/illogical.
In contrast, Post-GenderMag participants' responses reflected an understanding of the logic behind Agent Blue's decisions:
\begin{quote}
    P02-PostGM: \textit{BTW <BestToWorst> allowed me to assess all possibilities the AI was considering \& how it percieved it's win \% <Available with Fix-2>.}
\end{quote}

\begin{quote}
P68-PostGM: \textit{
...Chose A2 because of the win\% and loss\% <Available with Fix-2>. The AI is following the best move for the entire game.}
\end{quote}

\subsubsection{Fix-7: Add move rankings for comparison}
\DraftStatus{MMB: D 2.5}
\label{subsec:fix7}

\boldification{In addition to Fix-2, Team Game felt the need for a feature for comparing the top-ranked squares. So they made Fix-7 that included a top 5 moves table 
}

\begin{figure}[h]
\includegraphics[scale=0.6]{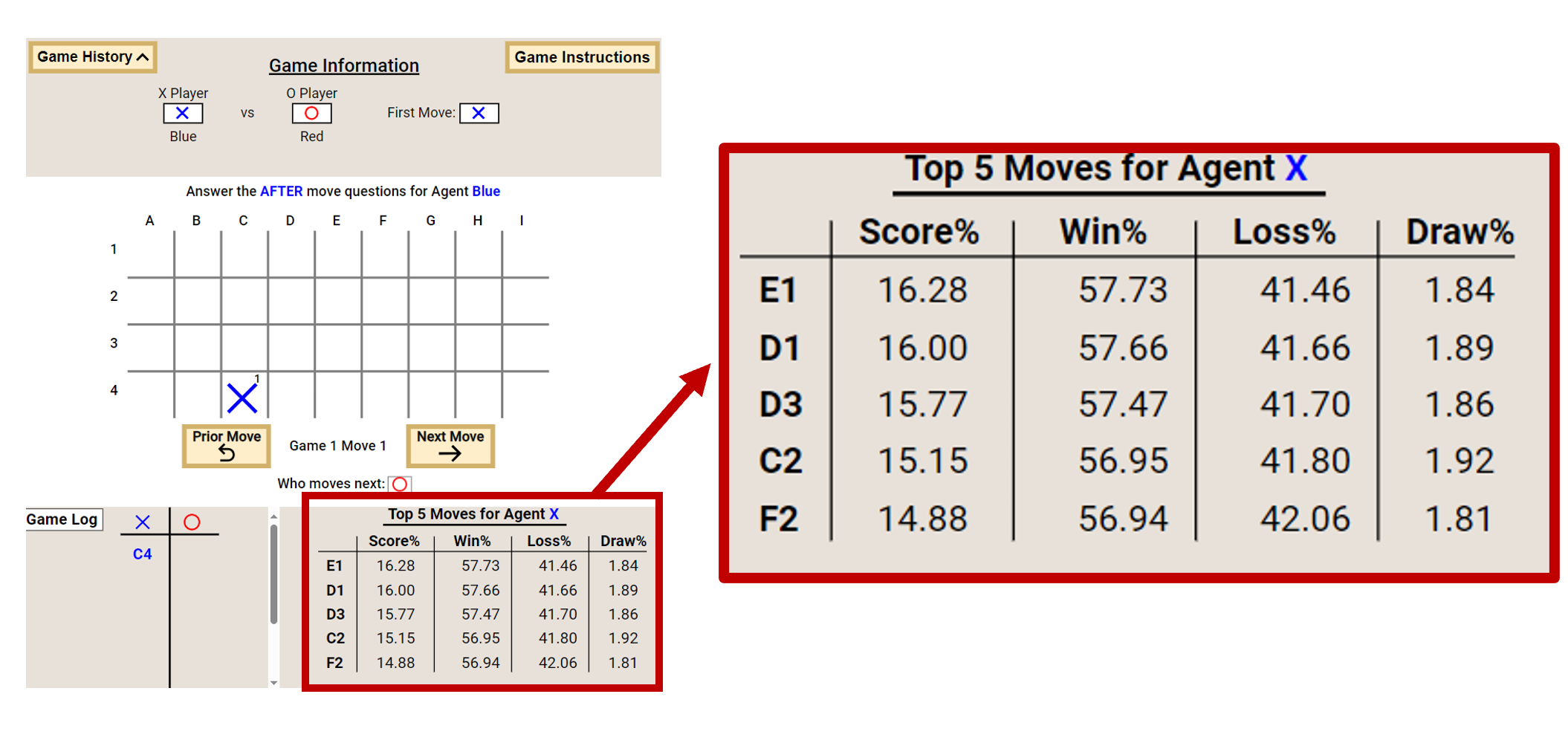}
\caption{Fix-7. Post-GenderMag version showed the top 5 moves (red-bordered callout shows at readable size) with their score\%, win\%, loss\%, and draw\%. }
\label{figure:top5moves}
\end{figure}

Finding similarities and differences is a central component of many cognitive processes 
~\cite{markman1996differences}, and Team Game decided that supporting this ability would be particularly helpful to comprehensive information processors.
Thus, with Fix-7, Team Game added a new explanation ``Top 5 Moves'' (Figure~\ref{figure:top5moves}), which enabled participants to compare the scores and win/loss/draw \%s of the top contenders for Agent Blue's prior move.

Seeing the exact scores and comparing the top-ranked moves could be the reasons why more Post-GenderMag than Original participants said that ``Agent picks high score'':

\begin{quote}
P07-PostGM: \textit{The ``scores best-to-worst'' and the ``top 5 moves'' <Available with Fix-7> help explain it. It shows that it's <Agent Blue> picking squares based on their scores.}
\end{quote}

Participant P39-PostGM also explained how the top 5 moves helped them see the agent's behavior as being purposeful, not random/illogical:
\begin{quote}
P39-PostGM: \textit{Yes - it is clear that X was trying to place moves based on top score (Top5 Moves) <Only Post-GenderMag version explicitly showed the top 5 moves>. It worked in the end.}
\end{quote}

Note a nuance of Fix-7: it follows a different inclusivity strategy than Fix-2 did.  
With Fix-2, participants who leaned toward a comprehensive Information Processing Style had to mouse hover to retrieve the additional information Fix-2 provided.  
Thus, participants who did not want more information were not ``force-fed'' it. 
In contrast, Fix-7's additional information was always on-screen.
This still does not force everyone to read it---arguably, a user can always ignore it---but its constant presence communicates an encouragement for everyone to read it. 
In Section~\ref{subsec:predDiffer_Why}, we will return to ways Fix-7 may have affected participants' mental models.



\subsection{The Post-GenderMag Participants' Mental Models and Curb Cuts}
\DraftStatus{MMB: ??}
\label{subsec:MMscoresAndCurbCuts}

\boldification{As Intro said, the Curb Cut effect is better sidewalks for everyone.  This section suggests that general effect: higher MM scores overall}

In the Introduction, we alluded to the ``curb cut effect,'' which occurs when using inclusive design to improve an environment for underserved users helps everyone---including those who were not previously underserved.  
The results just presented suggests that this is exactly what happened in this study.
Team Game's inclusivity fixes to the Post-GenderMag prototype's explanations, which they designed to make their explanations more inclusive for ``Abi-like'' users (Section~\ref{sec:methodology_gendermag}), actually led to significantly higher overall mental model concepts scores for the entire Post-GenderMag group than in the Original group. 

\boldification{Post-GM were also more engaged with the explanations}
The inclusivity fixes' effectiveness are also suggested by the two groups' explanation usage counts.
A two-tailed Student's t-test revealed that Post-GenderMag participants tended to have a higher explanation usage count than the Original participants (Post-GenderMag $mean=18.8$, Original $mean=15.26$), a marginally significant difference ($t(67)=1.9327, p=.0575, d=.4653$).

\boldification{However, there's more than 1 way to measure MMs.  We'll try a different way in the next section.}

However, the mental model concepts scoring used in this section is not the only way to measure mental models.
We next consider what a different measure of participants' mental models reveals.

\section{RQ2 Results: Prediction Accuracy and Why}
\label{results_sec:predictionPerformance}


\subsection{The Prediction Accuracy Results}
\DraftStatus{MMB: D2.4}
\label{results_subsec:predictionResults}

\boldification{The previous section's way was conceptual, but another way to calc MMs it thru ability to predict what the agent will do next}

The mental model concepts score used in the previous section measured participants' mental models in terms of \textit{concepts}  of the AI's behavior that participants understood.
In this section, we consider an alternative measure of mental models, namely participants' ability to \textit{predict} the AI's next behavior. 
Whereas the mental model concepts score measured participants' conceptual understanding of the AI's behaviors, measuring a participant's prediction accuracy measures a participant's ability to actually mimic what the AI does, namely choose the same next behavior.

\boldification{We calculate participants' prediction accuracy scores (PredError) as follows.}

We calculate participants' prediction accuracy using the absolute value of Dodge et al.'s method of calculating \textit{``Loss in Value'' per prediction}~\cite{koujalgi2024PredictionAccuracy}. 
Loss in value is a measure of how erroneous a prediction is; in this paper, we term it  \textit{PredError}.
Its calculation is:

\begin{equation}
\begin{aligned}
\text{PredError} = &|\text{ Score of the square selected by the agent (correct answer)} \\
& - \text{Score of the square predicted by the participant}|
\end{aligned}
\end{equation}
where \textit{Score} is the difference between the AI agent's predicted $Win\%$ and $Loss\%$. 
We calculated participants' PredError for 17 predictions across all three games and their average PredError. A lower PredError indicates better prediction accuracy.


\boldification{method of comparison: We used Mixed-design ANOVA as we had both within-subject and between-subject factors}
To compare participants' prediction accuracy, we used mixed-design ANOVA for investigating both within-subject (to use each of a participant's 17 predictions as a data point) and between-subject (to compare the Original group's predictions vs. the Post-GenderMag group's predictions). 
A mixed-design ANOVA can account for both the repeated measures within each participant and the differences between the two groups.

\boldification{Surprising results. A marginally significant main effect of group where the mean loss values showed Original better than Post-GenderMag. }
To our surprise, contrary to the mental model concepts scores, the Post-GenderMag group's PredError scores were not better than the Original group's---in fact, the opposite may have been the case.
ANOVA revealed that the Original group had lower PredError scores with a marginally significant difference (ANOVA, main effect of group: $F(1,997)=2.9, p=.0889$; Original $mean=0.1133$, Post-GenderMag $mean=0.1264$).

\boldification{A clue to this surprise---it depends on the task.}

The ANOVA analysis also gave a clue as to a possible reason for this surprising result. 
It showed that the PredError between the Post-GenderMag and the Original groups varied significantly depending on the prediction tasks (ANOVA, interaction effect: $F(16,997)=1.716, p=.0386$).\footnote{The main effect of prediction tasks on PredError was not significant overall (when participants' groups were not considered), $F(16,997)=0.511, p=.9429$.}



\begin{figure}[h]
\includegraphics[scale=0.55]{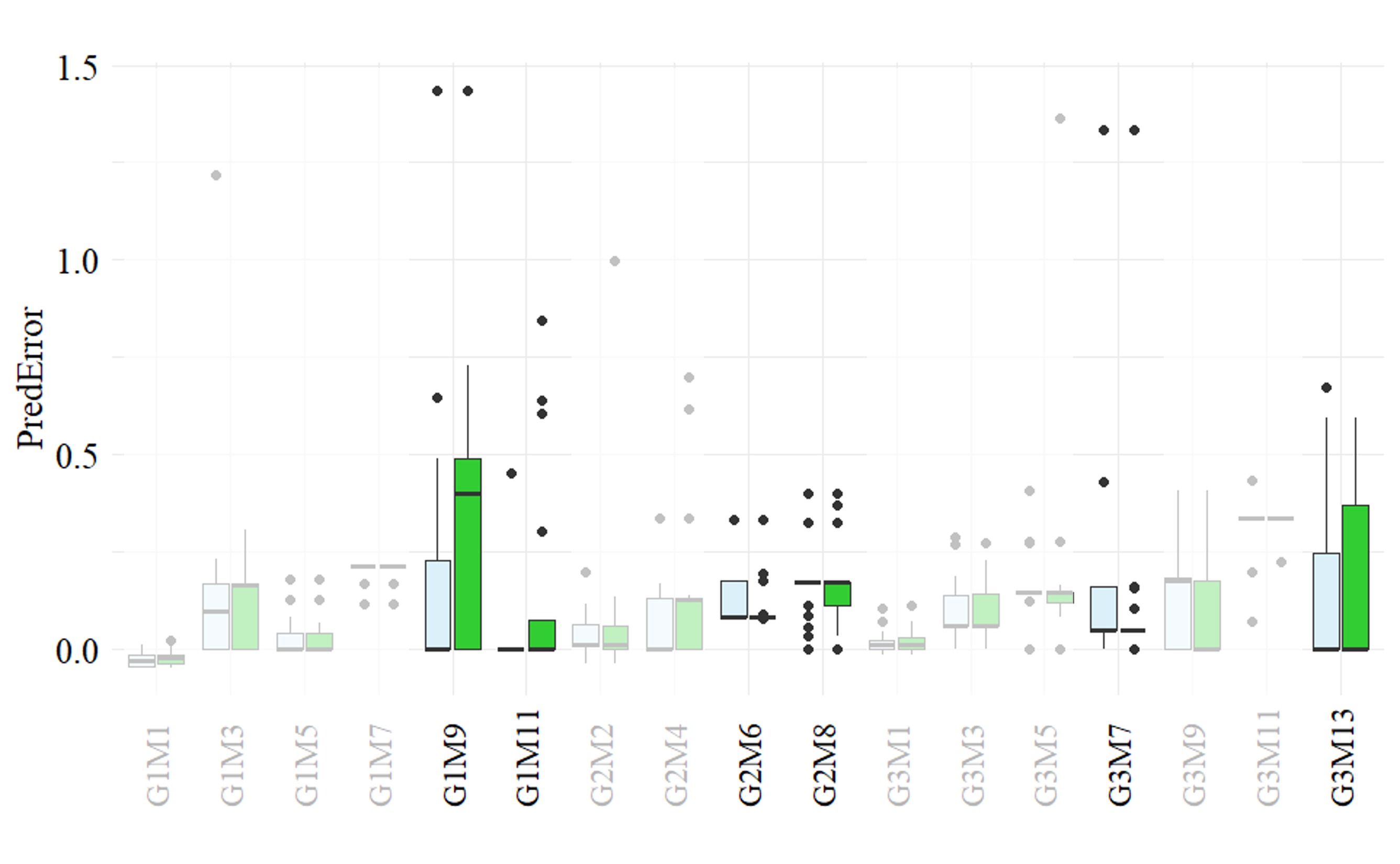}
\caption{PredError for 17 prediction tasks across three games. G=Game, M=Move. 
For the 11 faded prediction tasks, participants in both the \PostGenderMag group and the \Original group made almost identical predictions. But this was not true for the 6 prediction tasks not faded---G1M9, G1M11, G2M6, G2M8, G3M7, and G3M13. (Lower is better, because low PredError indicates high prediction accuracy.)} 
\label{figure:PredError_All}
\end{figure}

Details of the two groups' PredError scores by task are shown in Figure~\ref{figure:PredError_All}. 
Note that the two groups had almost identical PredErrors in 11 of these tasks (faded colors), with visibly different box sizes/positions in only 6 tasks (not faded).

\subsection{Why the Difference? Explanations As Both Friend and Foe}
\DraftStatus{MMB: D 2.4}
\label{subsec:predDiffer_Why}
\boldification{Q:Why? A:Explanations. We think that for 3 reasons...}

The ``why'' points toward the explanations themselves.
One indicator is that almost all differences between the two prototypes were fixes to explanations.
Second, it turns out that explanations really mattered to participants---overall, participants made numerous mentions of explanations in their responses (mean: 17 mentions; median: 18 mentions).
Third, it turns out that explanation usage significantly predicted participants' mental model concepts scores (linear regression, coefficient $=2.899, p=.0001$).
\begin{align}
\mu(&\text{Mental Model Concepts Score} | \text{Explanation Usage})= \beta_{0} + \beta_{1} \times \text{Explanation Usage}
\end{align}
where explanation usage is the frequency of each participant's responses meeting the definition of code \textit{``Participant used prior explanation''} (Appendix's Table~\ref{tab:codes_definitions}).

\boldification{Hmmm, are they paying too much heed to some kind of explanation?}

Given that the explanations helped participants' mental model concepts scores, it seems logical that they would also help participants' PredError scores.
But what if, in some cases, participants followed explanations ``too closely''?

\boldification{is it the "Top 5 Moves explanation"? Here's why it could be that.} 

Our participant data enabled considering this question for one of the explanation differences between the two prototypes---the addition of the ``Top 5 Moves'' explanation (Fix-7) into the Post-GenderMag prototype. 
Two aspects of the ``Top 5 Moves'' explanation are pertinent to our question of why participants' mental model concepts scores were better in the presence of this explanation but their prediction accuracy scores were worse.
The first aspect is that this explanation's explicit inclusion of the Agent's actual scores emphasizes them.  
The concept of the Agent emphasizing its scores, which was part of the mental model concepts scoring (recall Table~\ref{tab:4criteria_weights}), was one that was significantly more salient to the Post-GenderMag participants (as shown earlier in Figure~\ref{figure:MentalModelPercentageOddsRatio}).
This evidence suggests that the ``Top 5 Moves'' explanation impacted Post-GenderMag participants' mental model concepts scores positively.

The second pertinent aspect of ``Top 5 Moves'' is its emphasis on Agent Blue's \textit{previous move's} top five. 
If a participant attended too much to that information, it could influence their prediction of Agent Blue's next move either positively or negatively.
It would be positive if the game remains on a similar trajectory as on the previous move, since those top 5 are likely to be good candidates for the next move.  
However, it would be negative if Agent Red's move markedly changed the game's situation, because the previous top 5 would not be likely to be good candidates in the present situation.

\boldification{To find out about participants' focusing heavily on the Top 5 Moves, we investigated which group Predicted one of the previous top 5. Siginificant main effect of group indicates that Post-GM referred to 'top 5' more than their Original counterparts.}
To investigate how much impact ``Top 5 Moves'' had on the Post-GenderMag participants' predictions, we used a mixed design ANOVA to test which group's predictions included more of the previous move's top 5 list.
Specifically, if a participant's prediction was in the previous move's top 5 list, we counted it as 1; if a participant suggested multiple moves with at least one in the top 5, we counted it as a fraction (1/moves); otherwise we counted it as 0 (zero). 
We did this for 14 out of 17 prediction tasks (all three games), excluding only the three prediction tasks occurring before any explanations had been shown.
The results showed a significant main effect of the group, indicating that the Post-GenderMag participants predicted an entry just listed in the Top 5 Moves ($mean=3.854$) significantly more often than the Original participants ($mean=3.502$), $F(1,801)=6.145, p=.0134$.%
\footnote{A marginally significant effect further suggested that whether participants predicted from the entries listed in the previous move's Top 5 depended on the particular prediction task, $F(13,801)=1.592, p=.0818$.~
The interaction effect between PredError and the group was not significant.} 




\boldification{So, Explanation was their friend: Post-GM did what we wanted- predicted from top 5(yay). }

What this shows is that the Post-GenderMag participants did exactly what Team Game's Fix-7 explanation encouraged---basing their predictions on the Top 5 Moves explanation. 
This strategy was effective (the explanation was their ``friend'') when the agent's top 5 from the previous move helped in predicting the next move.  
This situation occurred in Figure~\ref{figure:PredError_All}'s G2M6 (Game 2 Move 6), G2M8, and G3M7.

\boldification{For example, in G2M6...}
For example, in the move before Game 2 Move 6, one of the top 5 moves had been moving to G2, which Post-GenderMag participants saw on display when they were predicting the agent's upcoming Move 6.
It turned out that Post-GenderMag participants were about 9\% more likely than Original participants to predict move G2, which suggests that they were influenced by the Top 5 Moves explanation.
And, because G2 remained a pretty good choice for Move 6 (still highly ranked by the agent), this choice helped Post-GenderMag participants' prediction accuracy, as Figure~\ref{figure:PredError_All} shows.

\begin{figure}[h]
    \includegraphics[scale=0.45]{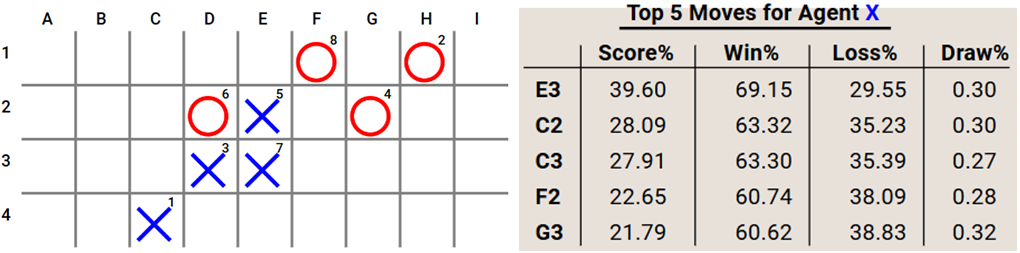}
    \caption{Gameboard state and the Top 5 Moves that participants saw before predicting Game 1, Move 9. Here the explanation ``Top 5 Moves'' turned out to be a foe for Post-GenderMag participants. The correct prediction is F3 but they picked mostly from the Top 5 Moves.}
    \label{figure:G1M8_GB_Top5_Foe}
\end{figure}

\boldification{But sometimes, following the explanations was a bad idea. We have examples from the games.}
However, when the Top 5 Moves for the previous move did not include an advantageous choice for the upcoming move, the explanation acted as a ``foe''.
One such example was Game 1 Move 9 (G1M9 in  Figure~\ref{figure:PredError_All}).
For this move, Post-GenderMag participants' PredError was higher/worse, perhaps because Post-GenderMag participants followed the Top 5 Moves more than the Original participants: 
10/35 Post-GenderMag participants based their predictions on the Top 5 Moves compared to only 5/34 Original participants.
But the correct choice for this move, F3, was not in Top 5 Moves display, because the game situation had changed too much.
Thus, the Top 5 Moves explanation from the last move was misleading (Figure~\ref{figure:G1M8_GB_Top5_Foe}) so much that only 8/35 Post-GenderMag participants predicted ``F3'' compared to 20/34 Original participants.

\boldification{conclusion: explanations were correct but relying on them for predicting future did not work out for Post-GM and affected their prediction accuracy}

Thus, although the fixed explanations were helpful for Post-GenderMag participants' mental model concepts scores, using them to predict the next move sometimes worked against them.
Our results can be seen as a nuanced case of overreliance---relying on the explanations to build conceptual understanding turned out to be appropriate, but relying on them to predict what the agent will do next turned out not to be generally appropriate.



\section{RQ3 Results: Inclusivity}
\DraftStatus{**?}
\label{results_sec:inclusivity}

\boldification{Curb cut effects have to show 2 things: everyone and underserved -- this section addresses underserved. }
For curb-cut effects to be present, the effects need to be both ``better for the underserved'' population and ``better for everyone else too.''
The previous two sections investigated the latter effect, and found that, although the Post-GenderMag group did not achieve better prediction accuracy, the entire Post-GenderMag group did benefit with higher mental model conceptual understanding. 
This section considers the ``underserved'' effects, i.e., whether the underserved population also benefited with higher mental model conceptual understanding.

\subsection{Inclusivity for Abi-Like Problem-Solvers}
\label{sec:results:subsubsec:InclusivityforAbi}

\boldification{Here, the underserved are the Abi's, so RQ3's question is about the Abis.}
In this study, ``Abi-like'' users are the underserved population, because that is the population Team Game focused on supporting better with their inclusivity bug fixes.
RQ3 asks whether the Post-GenderMag prototype provided better inclusivity for \textit{these} particular users.


\boldification{Here's how we went after RQ3.}
To investigate this question, we statistically tested mental model concept scores using the statistical reasoning explained in Section~\ref{sec:method:subsec:statMethods}. 
We compared the mostly Abi-like subset of the Post-GenderMag group (participants with 3, 4, or 5 Abi-like problem-solving styles as per their questionnaire responses) vs. the mostly Abi-like subset of the Original group.
We hypothesized that GenderMag-driven inclusivity fixes would best support the mostly Abi-like Post-GenderMag group's mental model concepts scores.

\boldification{And yaay -- it did benefit the Abi's MMscores.}
The results showed that Team Game's explanation fixes did improve mental model scores for Abi-like participants.
Specifically, Abi-like Post-GenderMag participants  achieved significantly higher mental model concepts scores than the Abi-like Original participants  (Student's one-tailed t-test: $t(36)=1.9441, p=.0299, d=.4653$, Post-GenderMag $mean=141.10$, Original $mean=112.83$).

\begin{figure}[h]
    \centering 
    \begin{subfigure}[b]{0.45\textwidth}
        \hspace{-0.5cm} 
        \includegraphics[scale=0.42]{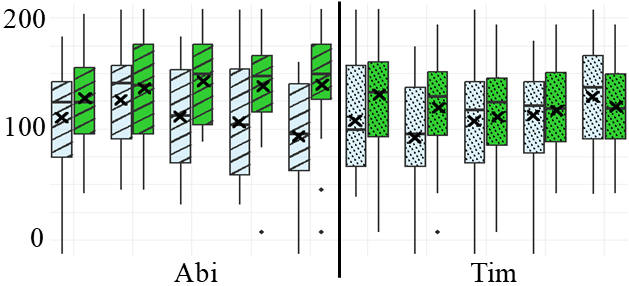}
        \caption{Mental Model Concepts Score}
        \label{figure:MMScoreShrinked}
    \end{subfigure}
    \quad
    \begin{subfigure}[b]{0.45\textwidth}
    \hspace{0.2cm}
        \includegraphics[scale=0.42]{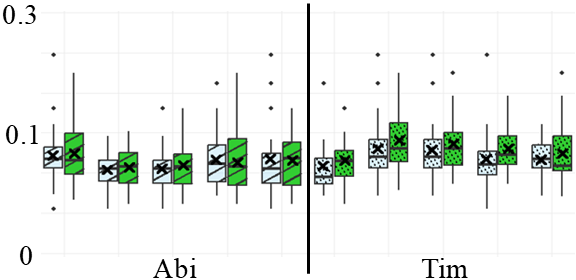}
        \caption{PredError}
        \label{figure:PredErrorShrinked}
    \end{subfigure}
    \caption{(a): \PostGenderMag Abi (striped) and Tim (dotted) participants had higher (better) mental model concepts scores than their \Original counterparts. The five problem-solving styles (from left to right): Info, Learn, Motiv, Risk, and SE (same order in adjacent plots). 
    (b): \PostGenderMag Abi's and Tim's PredError were higher (worse) but the difference from the Original group was not significant. x: mean, ---: median.}
    \label{figure:facet-by-facetAnalysis}
\end{figure}

\boldification{Problem-solving styles showed positive signs for Post-GM participants.}
Figure~\ref{figure:facet-by-facetAnalysis}(a) shows a close-up view of these improvements, one problem-solving style at a time. 
At this level of detail, inferential statistics are not viable, but the raw numbers are still illustrative.
As the figure shows, for participants with each of the Abi-like problem-solving values, the Post-GenderMag participants' average mental model concepts scores were always higher than the Original participants'.
That the Abi-like gains were visibly higher than the Tim-like gains for every one of the problem-solving style types confirms the ``especially for underserved'' aspect of curb-cut effects.
That this was also true to a lesser extent for all but one of the Tim-like problem-solving values confirms the ``for everyone else'' aspect of these inclusivity fixes' curb-cut effects.

\boldification{But the PredLoss stuff was still a disaster. }
On the other hand, as Figure~\ref{figure:facet-by-facetAnalysis}(b) shows, prediction errors abounded with the new fixes, and with no advantageous effects for Abi-like participants or for anyone.
This is unsurprising, given the results of Section~\ref{results_sec:predictionPerformance}. 
We will return to prediction errors in Section~\ref{sec:discussion}.


\subsection{Inclusivity by Gender}
\DraftStatus{**?**}
\label{subsec:inclusivitybyGender}

\boldification{Remember, GM is about both problem-solving and genders. So for RQ3 on gender, we'll assume women to be the underserved population.}
Investigating curb-cut effects through the lens of gender inclusivity requires identifying the underserved gender.  
Recall that the GenderMag inclusivity method has a dual focus: it aims to address problem-solving inclusivity in order to also address gender inclusivity.
Thus, because Team Game used GenderMag's Abi persona, which was designed to capture problem-solving styles more common among women than other people, Abi-like users being the underserved population also suggests women to be an underserved population.

\boldification{We begin with 2 genders...}
We begin with the two genders who made up about 93\% of our study population, thereby allowing inferential statistics---women and men. 
For participants in the LGBTQIA* community%
\footnote{LGBTQIA* used based on Scheuerman et al.’s living document~\cite{schmidt2020calibrating}.}, 
we then provide a non-statistical representation of their data. 

\subsubsection{Statistical Gender Results}

\begin{figure}[h]
    \includegraphics[scale=0.55]{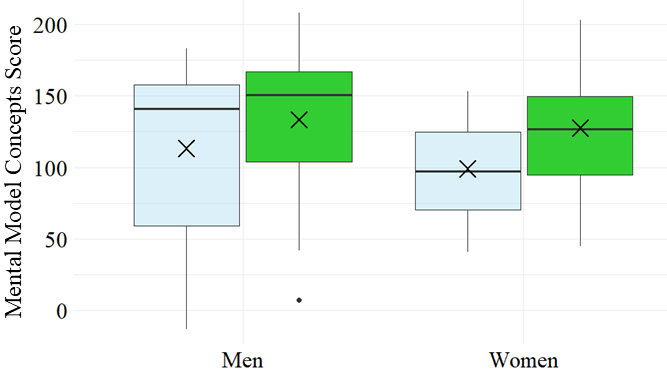}
    \caption{\PostGenderMag men's and women's mental model concepts scores were higher than their \Original counterparts. The difference was significant for women. x: mean, ---: median.}
    \label{figure:MMScore_Gender}
\end{figure}

\boldification{The gender-based analysis showed significant gains.}
As Figure~\ref{figure:MMScore_Gender} illustrates, Team Game's fixes were especially beneficial to women's mental model concepts scores.
Using the statistical process explained in Section~\ref{sec:method:subsec:statMethods}, Post-GenderMag women (i.e., Post-GenderMag participants whose self-reported gender was ``Woman'') had significantly higher mental model concepts scores than Original women (Student's one-tailed t-test: $t(34)= 2.2537, p=.0154, d=.7512$, Post-GenderMag women $mean=127.5$, Original women $mean=99.06$).
Post-GenderMag men's mental model concepts scores also tended to be higher than Original men, but the difference was not significant.
As for PredError scores, comparisons by gender did not reveal any significant differences.
Thus, as with the problem-solver inclusivity results above, the gender perspective also shows both aspects of the curb-cut effect being present---the Post-GenderMag prototype was more inclusive for women and made it slightly better for men as well for their mental model concepts. 

\boldification{Post-GenderMag women's inclusivity gains might be related to their significantly higher ratings for explanations and explanation usage count}

Post-GenderMag women's explanation ratings and explanation usage count are consistent with these results.
Their average explanation ratings were significantly higher than those of Original women (Student's one-tailed t-test: $t(34)= 2.0594, p=.0236, d=.6865$, Post-GenderMag women $mean=5.19$, Original women $mean=4.63$).
Also, Post-GenderMag women's explanation usage count was significantly higher than that of Original women (Student's two-tailed t-test: $t(34)= 2.4814, p=.0182, d=.8271$, Post-GenderMag women $mean=20.06$, Original women $mean=13.11$). 
Explanation comparisons for the Post-GenderMag men vs. Original men did not reveal any significant differences.
The Post-GenderMag women's higher ratings and usage, plus the fact that most differences between the prototypes were explanation fixes, suggest that their superior mental model concepts scores were  due to Team Game's explanation fixes.

\begin{figure}[h]
    \includegraphics[scale=0.5]{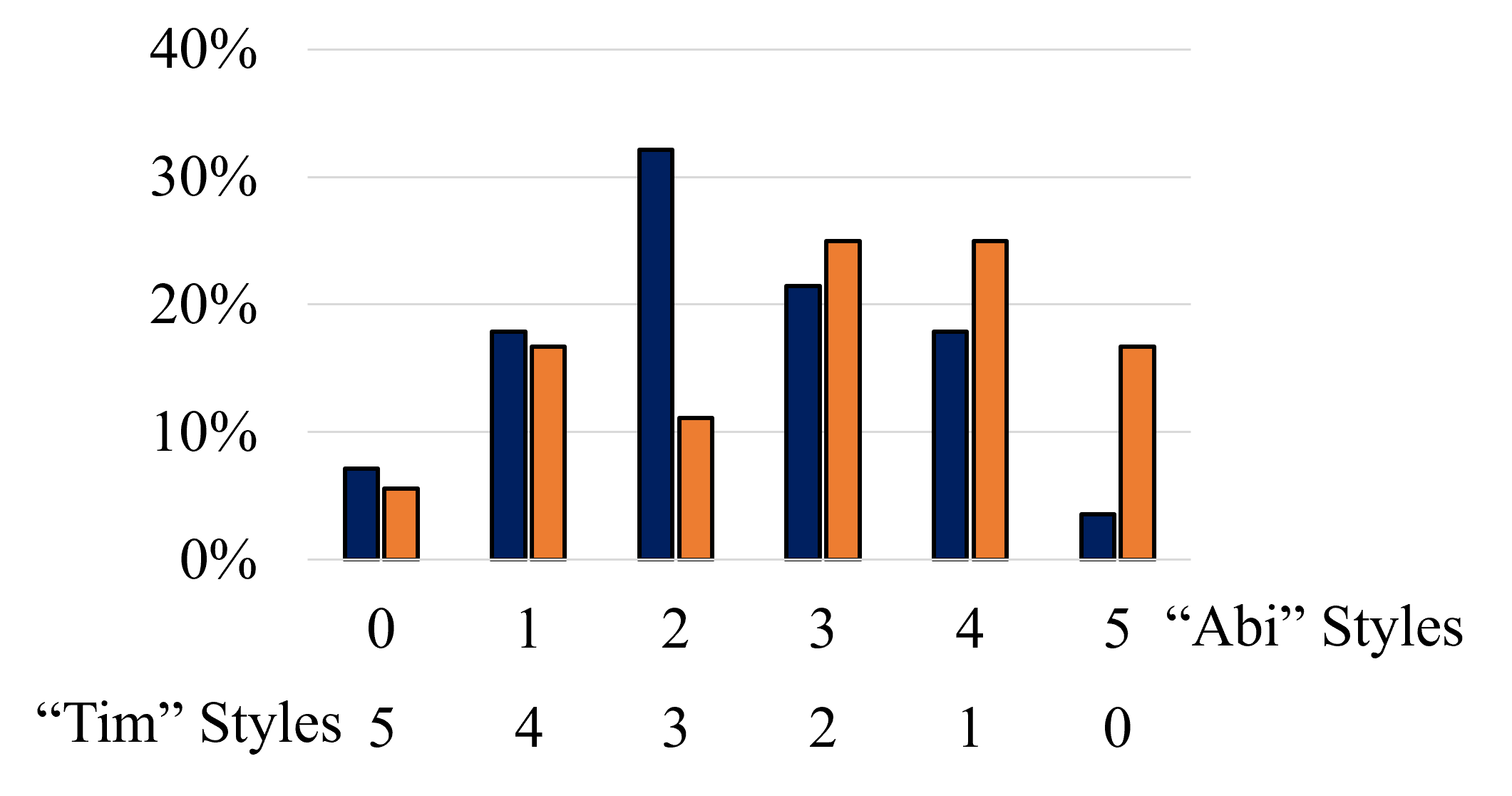}
    \caption{67\% of \womenCaption (n=36) had more Abi-like than Tim-like problem-solving styles, compared to 43\% of \menCaption (n=28); and 33\% of women had more Tim styles, compared to 57\% of men. }
    \label{figure:Gender_Facets}
\end{figure}

\boldification{Team Game considered Abi when finding and fixing the bugs, and the results support that the fixes worked for Abi and women had Abi's styles so it also worked for women.}
The relationship between problem-solving styles and gender also helps to explain the higher inclusivity for women.
Team Game aimed their fixes toward including Abi's problem-solving styles and, as Figure~\ref{figure:Gender_Facets} shows, women were particularly likely to report Abi-like problem-solving styles in their questionnaire responses.
Specifically, two-thirds ($67\%$) of the women had more Abi-like than Tim-like problem-solving styles (Figure~\ref{figure:Gender_Facets}).
In contrast, men tended more towards Tim ($57\%$) than Abi ($43\%$).
Thus, the fact that men's scores also increased slightly is again evidence of a curb-cut effect.

\begin{figure}[h]
    \includegraphics[scale=0.55]{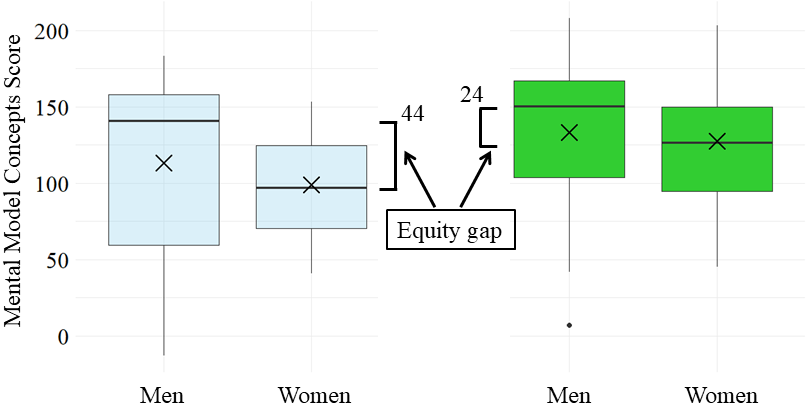}
    \caption{\PostGenderMag men and women show an improvement in equity compared to what \Original men and women had. 
    x: mean, ---: median.}
    \label{figure:equityGap}
\end{figure}

\boldification{What about equity? Team Game's fixes made Post-GM prototype more equitable as it reduced the gap between men and women.}
A related concept to inclusivity is equity. 
Equity is ``the quality of being fair and impartial''~\cite{lexico2019Equity}. 
In this study, equity would mean men and women achieving the same scores. 
As Figure~\ref{figure:equityGap} shows, although Team Game's inclusivity fixes did not fully achieve equity, they did improve it. 
With the Original prototype, women's median mental model concepts scores were 44 points lower than men's, but with the Post-GenderMag prototype, the gap narrowed to just 24 points---a 45\% improvement in equity.

\subsubsection{LGBTQIA* Results}

\boldification{For LGBTQIA* community followed what Andrew did--A table with their problem-solving style values}
Table~\ref{tab:LGBTQ_Facets} reports the problem-solving style values and results of the five participants from the LGBTQIA* community.
As with many other such reports of LGBTQIA* data (e.g.,~\cite{anderson2024measuring}), the dataset is too small for inferential statistics.
Still, we hope it contributes to the growing literature on LGBTQIA* participants (e.g.,~\cite{acena2021safe, freeman2022rediscovering, hardy2019participatory}) and helps future researchers expand the field's understanding of inclusive design across all gender identities.


\begin{table}[h]
\centering
\caption{LGBTQIA* problem-solving style values. Each row shows one LGBTQIA* participant’s GenderMag problem-solving style values. W = Woman, T = Transgender, NB = Non-Binary,
 PNS = Prefer Not to State, Select. = Selective, Comp. = Comprehensive, MMCS = Mental Model Concepts Score, PE = PredError.}
\label{tab:LGBTQ_Facets}
\begin{tabular}{l|c|c|c|c|c|c|c|c|c|c|c}
\hline
\textbf{PID} & \textbf{W} & \textbf{NB} & \textbf{T} & \textbf{PNS} & \textbf{SE} & \textbf{Motiv} & \textbf{Learn} & \textbf{Info} & \textbf{Risk} & \textbf{MMCS} & \textbf{PE} \\ \hline
P1   &  &  &  & \checkmark & High & Task & Tinker & Select. & Tolerant & 88 & 0.1224 \\ \hline
P11      &  &  &  & \checkmark  & High & Tech & Process & Select. & Averse & 207 & 0.142 \\ \hline
P13      &  & \checkmark &  &  & Low & Task & Process & Comp. & Tolerant & 142 & 0.0557  \\ \hline
P46      &  & \checkmark &  &  & Low & Task & Process & Comp. & Averse & 45 & 0.074 \\ \hline
P59      & \checkmark &  & \checkmark &  & High & Tech & Tinker & Comp. & Tolerant & 174 & 0.1176 \\ \hline
\end{tabular}
\end{table}

\section{Discussion}
\label{sec:discussion}


\subsection{When do predictions matter?}

\boldification{Do we need to consider prediction performance when evaluating mental models?}
Prediction accuracy is a common metric for measuring users' mental models of AI \cite{hoffman2019metrics, anderson2020mental}, but we believe the use of this metric needs careful consideration.
Predicting an AI agent's next move requires a user's mental model to be accurate enough to enable actually simulating an agent's behavior.
This raises the question of when this ability is needed.

\boldification{For example...}
Consider the example of self-driving cars. Potential buyers of such cars might need to be able to predict critical decisions like whether and when the car's AI will hit the brakes if a pedestrian starts crossing the road in front of it. 
In contrast, buyers may feel no need to predict decisions like the AI's route to reach some destination.

\boldification{This suggests that sometimes prediction is relevant and sometimes not.}
This suggests that evaluating users' ability to predict an AI's next action may be highly relevant for some XAI projects, but completely irrelevant in others.
In our study, we directed participants to ``understand'' and ``describe'' the agent and to rate how well explanations helped them do these things.
For this kind of explanation goal, predictions may not be a particularly relevant measure.  
On the other hand, suppose we had tasked participants to decide whether they would like the agent to ``play future games on their behalf'' (e.g.,~\cite{dodge2021AARAI}).
In that situation, users' ability to simulate what an agent will do in certain situations may be extremely relevant as a measure of the approach's effectiveness.

\boldification{So, we think this measure needs rethinking}
Thus, our call to the XAI research community is to rethink using predictions to measure a particular XAI project's effectiveness---using predictions as a measure when the explanations actually aim to support prediction, but not using predictions to measure effectiveness when the project's use-cases do not suggest the need for users to predict the AI. 


\subsection{Cognitive Lenses on Mental Model Concepts Scores vs. PredError Scores}
\DraftStatus{MMH: D2.5}

\boldification{When PredError does matter, cog theories can help us understand why certain expls won't help and what we should do instead.}

We have just pointed out that there are situations in which explanations do need to support users' ability to predict an AI's next action. 
Two theories of cognition offer interpretations of why Team Game's explanation fixes did not do so, even though they produced mental model improvements, and possible approaches to better enable explanations to support predicting.

\boldification{Dual-process theory of cognition might explain the contradictory effects of XAI in our study. Team Game's explanations might just supported general heuristic-based decision-making without encouraging analytical thinking.}

The ``dual-process theory of cognition'' suggests that humans mostly rely on heuristics and mental shortcuts for everyday decisions, and rarely engage in analytical thinking due to its cognitive demand~\cite{Buccinca21CogFunction, kahneman2011thinking, wason1974dual}.
Heuristic processing uses mental shortcuts such as simple decision rules to arrive quickly at a judgment, such as relying on currently held information or a perceived expert opinion~\cite{trumbo2019heuristicSystematic}.
In contrast, analytical thinking (also called systematic thinking) involves effortful scrutiny and comparison of information, or as Trumbo puts it 
``carefully examining, comparing, and relating arguments...and a scrutiny of arguments, maintaining higher standards for the quality of information''~\cite{trumbo2019heuristicSystematic}.  
Building on this theory, Bu\c{c}inca et al. posited that users tend to process explanations based on general heuristics rather than thinking analytically, which can lead to overreliance on AI decisions.
For this reason, they developed cognitive forcing functions for XAI to force users into more analytical thinking and thereby reduce AI overreliance~\cite{Buccinca21CogFunction}.
Their empirical results showed that the cognitive forcing did reduce AI overreliance.

The distinction between heuristic vs. analytical thinking may apply to our results as well.
We believe that in some cases, accurately predicting the AI's next move would require analytical thinking, because it tends to require to comparing information from multiple AI decision points.
It also seems likely that Team Game's explanation fixes instead encouraged participants to rely even more on heuristic thinking because it displayed easily noticed ``expert'' information they may have perceived to be sufficient, in the form of the Top 5 Moves display.

An example from our study is Game 1 Move 9 (the most difficult to predict, according to Figure~\ref{figure:PredError_All}), in which the AI's move turned out to be F3.
Post-GenderMag participants were more than twice as likely as Original participants to pick a displayed Top 5 Move here, but F3 was not one of the Top 5 Moves on display (recall Figure~\ref{figure:G1M8_GB_Top5_Foe}), so heuristic processing of the Top 5 Moves explanation would not help.
Rather, the information needed to accurately predict Game 1 Move 9  would require comparing the AI agent's past four moves' scores, win/loss/draw \%s, and rankings, while also considering the gameboard state so as to block the opponent.
This suggests the possibility of using cognitive forcing functionalities such as Bu\c{c}inca et al.'s~\cite{Buccinca21CogFunction} to force users into analytical thinking, for XAI applications in which ability to predict an AI is important.

\boldification{Another factor is "declarative" vs. "procedural" knowledge, and here's what these things are...}
Another factor that may have been at play is the difference between ``declarative knowledge'' vs. ``procedural knowledge.'' 
Declarative knowledge covers facts about something (e.g., New York is north of the Miami), whereas procedural knowledge covers steps and processes of \textit{how} to do something (e.g., to get to New York from Miami, first get on route..., then switch to...)~\cite{anderson2005cognitive,salaberry2018declarativeVsProcedural}.
Stated another way, declarative knowledge enables someone to describe something, whereas procedural knowledge enables them to actually do that thing~\cite{salaberry2018declarativeVsProcedural}.
Thus, declarative knowledge is often classified as a ``first stage'' of learning that, with practice and experience, can eventually lead to the next stage of learning, procedural knowledge~\cite{salaberry2018declarativeVsProcedural}.

\boldification{1. Team Game's fixes promoted declarative knowledge but not procedural knowledge (we know this from the MM scores). 2. Proc knowledge requires practice DOING the task -- education researchers know this.  
3. Wouldn't it be nice if we listened to them? 
4. So, expls that really want to support proc knowledge need to provide more proc knowledge of the AI, and then get participants some form of active DOING  simulations of that, to get their proc skills honed.}

The mental model concepts scores were based on declarative knowledge---i.e., facts about the AI such as ``Agent doesn’t consider opponent’s moves'' (recall Table~\ref{tab:exact_weights_4Criteria})---and the fact that the Post-GenderMag participants' mental model concepts scores were higher suggests that Team Game's explanation fixes were effective at promoting this declarative knowledge.
The fact that Post-GenderMag participants' PredError scores were not better despite their higher mental model concepts scores, could stem from a lack of procedural knowledge to be able to predict the next steps in a procedure~\cite{Johnson01Procedural, Braithwaite21Procedural}.
Education scholars have pointed out that procedural knowledge usually requires practice carrying out a task~\cite{salaberry2018declarativeVsProcedural}.
This suggests that, for explanations to support users' ability to actually simulate an AI, they need to be designed in ways that enable users to practice building exactly these skills.
How to harvest education research's approaches to creating this kind of user experience for applications where predictions matter is an open question.

\subsection{Threats to Validity}
\DraftStatus{MMH: D2.5}
Every study encounters potential challenges to its validity~\cite{Wohlin00SE}, and ours is no exception. Several factors could limit the generalizability and reliability of our findings.
\par 
\boldification{Most participants were students from the same university. This may limit the generalizability of our findings.}
The first issue is the demographic homogeneity. Although we had 69 participants, most of them were students and came from the same university. This may impact the generality of our results as our participants may not properly reflect the diversity of potential MNK or XAI users. 

\par 
\boldification{Another threat to generalizability is the MNK domain itself, which is not a common AI challenge domain. The agents also were not very good. G1 and G2 Blue agents are different.}
Moreover, the MNK domain itself presents a potential limitation. 
We decided to work with an MNK prototype as most people are familiar with Tic-Tac-Toe, and this familiarity gave  a pre-existing basis for judging the agents' decisions. 
For the same reason, MNK gave us the option to ask a wide range of questions related to predictions, flaws, changes, and after-action reviews. 
Still, MNK games might not be representative of other XAI applications.

\par 


Recall from Section~\ref{sec:method:subsec:experimentalDesign} that Agent Blue's noise levels differed in Game 2 from the other games.
This produced less determinism, which could be viewed as participants describing multiple agents rather than one.
To partially mitigate this threat, we eliminated participants' responses from the end-of-study question ``Please describe the steps in the agent’s approach or method...'', which would have encouraged them to somehow generalize all three games' behaviors into one description.
Still, the varying noise levels could have influenced participants' understanding of the AI agent during the games.
\par

\boldification{Another issue was: Our participants completed proxy tasks when predicting the AI and Bu\c{c}inca et al showed that proxy tasks could predict the results of actual decision-making tasks.}

Another concern is that our participants performed proxy tasks, i.e., they predicted which square the AI agent would move next, but their predictions did not impact the game. 
Such proxy tasks may not reflect how users would perform in actual decision-making tasks.
In Bu\c{c}inca et al.'s work, evaluations of XAI systems with proxy tasks did not predict the results of evaluations with actual decision-making tasks~\cite{Buccinca20Proxy}.
We used proxy tasks to reduce cognitive load and allow more focus on explanations, but they are not the same as immersive and high-stakes scenarios.
\par

\boldification{Q: Team Game focused on Abi's problem-solving needs---why not consider both Tim and Abi? A: they should but it's ok that they didn't}
Team Game focused on Abi-like problem-solving styles to find and fix inclusivity bugs, but did not repeat the process for Tim, because they did not have enough time available to do both.
When the time is too limited to do both, the GenderMag kit~\cite{burnett2020gendermagkit} recommends the ``Abi-first'' approach because Abi provides the strongest lens for inclusivity evaluation, as previous studies have shown~\cite{Vorvoreanu19Bias, Guizani22Inlcusivity}.
If Team Game had also run GenderMag sessions using Tim's perspective, their inclusivity bug finding and fixing efforts would likely have produced more and/or different changes to the Post-GenderMag explanations.

\boldification{We have not collected a click log for the explanations.}
Finally, we did not gather click log data for explanation interactions. Instead, we relied on participants' mentions of using the explanations as an engagement indicator. Maintaining a click log could have offered additional insights on explanation usage.
\par

Addressing such threats can only be achieved with additional studies using a variety of populations, AI agents, explanation types, and empirical methods.

\section{Conclusion}
\DraftStatus{MMH D2.5}


\begin{quote}
    \textit{``There’s an ingrained societal suspicion that intentionally supporting one group hurts another. That equity is a zero sum game.''} -Angela Glover Blackwell~\cite{Blackwell16Curb}
\end{quote}

Blackwell's concern is a powerful observation of societal misconceptions about inclusivity and equity-focused improvement efforts.
Yet, sidewalk curb cuts serve as a clear counterexample, showing that including a previously underserved group of users need not be a ``compromise'' that trades off benefits with those already well-served.

To investigate bringing curb cut effects to XAI, we analyzed the impacts of an AI product team's inclusivity fixes on their XAI prototype using GenderMag, an inclusive design approach that considers users' diverse problem-solving styles to generate actionable design directions.
The results revealed that, although the AI team had focused on improving inclusivity for ``Abi-like'' users, their fixes led to several curb-cut effects that particularly benefited not only Abi-like users, but also brought benefits to everyone:  

\begin{itemize}
    \item \textit{Mental model concepts scores improved:} Team Game's inclusivity fixes led to Post-GenderMag participants demonstrating significantly better conceptual mental model than the Original participants (Section~\ref{results_sec:mentalModelScores}). The differences revealed which of Team Game's inclusivity fixes brought about these changes, the inclusivity reasoning behind each of these fixes, and the AI agent-related concepts they successfully reinforced   (Section~\ref{results_sec:mentalModelScores}). These explicit ties between the inclusivity reasoning and the impacts on the entire group provides evidence of these fixes' curb-cut effects.
    
    \item \textit{Engaging with explanations helped:} Engagement with explanations significantly predicted participants' mental model concepts scores, and Post-GenderMag participants tended toward higher engagement than the Original participants (Sections~\ref{results_sec:predictionPerformance} and ~\ref{subsec:predDiffer_Why}). Since almost all of Team Game's changes were inclusivity fixes to explanations, this further evidences the presence of curb-cut effects in Team Game's inclusivity fixes.
    
    \item \textit{Inclusivity improved for Abi-like participants:} Team Game's underserved users were Abi-like users, and Team Game's inclusivity fixes were particularly effective for this group. Abi-like Post-GenderMag participants achieved significantly higher mental model concept scores than their Original counterparts (Section~\ref{sec:results:subsubsec:InclusivityforAbi}). Further, these gains for Abi-like users were accompanied by small (not significant) gains for Tim-like users, again consistent with curb-cut effects. 
    
    \item \textit{Inclusivity improved by gender:} Finally, the inclusivity fixes brought about significantly higher mental model concepts scores for an underserved gender, namely women, while also slightly increasing men's. The increase in inclusivity reduced the gender gap by 45\% (Section~\ref{subsec:inclusivitybyGender}).  

    \item \textit{A curb-fence effect with predictions:} However, there was also a ``curb-fence'' effect (opposite of the curb-cut effect): Participants' abilities to predict the AI agent's next move appeared to get worse.  While helpful for reinforcing conceptual mental model understanding at a declarative knowledge level, Team Game's fixes may have also encouraged participants to overrely on the explanations, affecting their abilities to predict the agent's next move   (Section~\ref{results_sec:predictionPerformance}).
\end{itemize}


In sum, these results suggest, first, that using an inclusive design to improve AI explanations can not only improve explanations' ability to inclusively serve users who are being left behind, but also can make AI explanations ``just plain better'' for everyone---in essence bringing curb-cut effects to XAI.
However, they also suggest the possibility of curb-fence effects, in which changes intended to increase inclusivity of explanations can be ``too effective,'' potentially leading to overreliance by everyone.
We hope other XAI researchers will join us in exploring the nuances of inclusive design approaches' potential to bring curb-cut effects into explanations while guarding against curb-fence effects, to address those who XAI currently underserves, while also benefiting everyone along the way.

\section{Acknowledgment}
We thank all the study participants for their time and effort.
We also thank Bhavika Madhwani, Jeramie Kim, Elizabeth Li, Sujay Koujalgi, Iyadunni Adenuga, Mitchell Hoesing, and Jeff Schulman for their help in this paper. This work has been supported in part by NSF \#1901031 and \#2042324, and by
USDA-NIFA/NSF \#2021-67021-35344.




\bibliographystyle{ACM-Reference-Format}
\bibliography{ms}


\begin{thebibliography}{66}


\ifx \showCODEN    \undefined \def \showCODEN     #1{\unskip}     \fi
\ifx \showDOI      \undefined \def \showDOI       #1{#1}\fi
\ifx \showISBNx    \undefined \def \showISBNx     #1{\unskip}     \fi
\ifx \showISBNxiii \undefined \def \showISBNxiii  #1{\unskip}     \fi
\ifx \showISSN     \undefined \def \showISSN      #1{\unskip}     \fi
\ifx \showLCCN     \undefined \def \showLCCN      #1{\unskip}     \fi
\ifx \shownote     \undefined \def \shownote      #1{#1}          \fi
\ifx \showarticletitle \undefined \def \showarticletitle #1{#1}   \fi
\ifx \showURL      \undefined \def \showURL       {\relax}        \fi
\providecommand\bibfield[2]{#2}
\providecommand\bibinfo[2]{#2}
\providecommand\natexlab[1]{#1}
\providecommand\showeprint[2][]{arXiv:#2}

\bibitem[Acena and Freeman(2021)]%
        {acena2021safe}
\bibfield{author}{\bibinfo{person}{Dane Acena} {and} \bibinfo{person}{Guo Freeman}.} \bibinfo{year}{2021}\natexlab{}.
\newblock \showarticletitle{"In My Safe Space": Social Support for LGBTQ Users in Social Virtual Reality}. In \bibinfo{booktitle}{\emph{Extended Abstracts of the 2021 CHI Conference on Human Factors in Computing Systems}}. \bibinfo{pages}{1--6}.
\newblock
\urldef\tempurl%
\url{https://doi.org/10.1145/3411763.3451611}
\showDOI{\tempurl}


\bibitem[Alipour et~al\mbox{.}(2021)]%
        {Alipour21Improving}
\bibfield{author}{\bibinfo{person}{Kamran Alipour}, \bibinfo{person}{Arijit Ray}, \bibinfo{person}{Xiao Lin}, \bibinfo{person}{Michael Cogswell}, \bibinfo{person}{Jurgen~P. Schulze}, \bibinfo{person}{Yi Yao}, {and} \bibinfo{person}{Giedrius~T. Burachas}.} \bibinfo{year}{2021}\natexlab{}.
\newblock \showarticletitle{Improving users' mental model with attention-directed counterfactual edits}.
\newblock \bibinfo{journal}{\emph{Appl. AI Lett.}} \bibinfo{volume}{2}, \bibinfo{number}{4} (\bibinfo{year}{2021}), \bibinfo{pages}{e47}.
\newblock
\urldef\tempurl%
\url{https://doi.org/10.1002/ail2.47}
\showDOI{\tempurl}


\bibitem[Amershi et~al\mbox{.}(2019)]%
        {Amershi19Guide}
\bibfield{author}{\bibinfo{person}{Saleema Amershi}, \bibinfo{person}{Dan Weld}, \bibinfo{person}{Mihaela Vorvoreanu}, \bibinfo{person}{Adam Fourney}, \bibinfo{person}{Besmira Nushi}, \bibinfo{person}{Penny Collisson}, \bibinfo{person}{Jina Suh}, \bibinfo{person}{Shamsi Iqbal}, \bibinfo{person}{Paul~N. Bennett}, \bibinfo{person}{Kori Inkpen}, \bibinfo{person}{Jaime Teevan}, \bibinfo{person}{Ruth Kikin-Gil}, {and} \bibinfo{person}{Eric Horvitz}.} \bibinfo{year}{2019}\natexlab{}.
\newblock \showarticletitle{Guidelines for Human-AI Interaction}. In \bibinfo{booktitle}{\emph{Proceedings of the 2019 CHI Conference on Human Factors in Computing Systems}} (Glasgow, Scotland Uk) \emph{(\bibinfo{series}{CHI '19})}. \bibinfo{publisher}{Association for Computing Machinery}, \bibinfo{address}{New York, NY, USA}, \bibinfo{pages}{1–13}.
\newblock
\showISBNx{9781450359702}
\urldef\tempurl%
\url{https://doi.org/10.1145/3290605.3300233}
\showDOI{\tempurl}


\bibitem[Anderson(2024)]%
        {Anderson24InclusiveHAI}
\bibfield{author}{\bibinfo{person}{Andrew Anderson}.} \bibinfo{year}{2024}\natexlab{}.
\newblock \bibinfo{title}{Inclusive Human-AI Interaction for Diverse Problem-solvers: The Problem and Two Solutions}.
\newblock
\newblock
\newblock
\shownote{PhD Thesis, Oregon State University}.


\bibitem[Anderson et~al\mbox{.}(2020)]%
        {anderson2020mental}
\bibfield{author}{\bibinfo{person}{Andrew Anderson}, \bibinfo{person}{Jonathan Dodge}, \bibinfo{person}{Amrita Sadarangani}, \bibinfo{person}{Zoe Juozapaitis}, \bibinfo{person}{Evan Newman}, \bibinfo{person}{Jed Irvine}, \bibinfo{person}{Souti Chattopadhyay}, \bibinfo{person}{Matthew Olson}, \bibinfo{person}{Alan Fern}, {and} \bibinfo{person}{Margaret Burnett}.} \bibinfo{year}{2020}\natexlab{}.
\newblock \showarticletitle{Mental models of mere mortals with explanations of reinforcement learning}.
\newblock \bibinfo{journal}{\emph{ACM Transactions on Interactive Intelligent Systems (TiiS)}} \bibinfo{volume}{10}, \bibinfo{number}{2} (\bibinfo{year}{2020}), \bibinfo{pages}{1--37}.
\newblock


\bibitem[Anderson et~al\mbox{.}(2024)]%
        {anderson2024measuring}
\bibfield{author}{\bibinfo{person}{Andrew Anderson}, \bibinfo{person}{Jimena Noa~Guevara}, \bibinfo{person}{Fatima Moussaoui}, \bibinfo{person}{Tianyi Li}, \bibinfo{person}{Mihaela Vorvoreanu}, {and} \bibinfo{person}{Margaret Burnett}.} \bibinfo{year}{2024}\natexlab{}.
\newblock \showarticletitle{Measuring User Experience Inclusivity in Human-AI Interaction via Five User Problem-Solving Styles}.
\newblock \bibinfo{journal}{\emph{ACM Transactions on Interactive Intelligent Systems}} (\bibinfo{year}{2024}).
\newblock


\bibitem[Anderson(2005)]%
        {anderson2005cognitive}
\bibfield{author}{\bibinfo{person}{John~R Anderson}.} \bibinfo{year}{2005}\natexlab{}.
\newblock \bibinfo{booktitle}{\emph{Cognitive psychology and its implications}}.
\newblock \bibinfo{publisher}{Macmillan}.
\newblock


\bibitem[Bansal et~al\mbox{.}(2019)]%
        {bansal2019beyond}
\bibfield{author}{\bibinfo{person}{Gagan Bansal}, \bibinfo{person}{Besmira Nushi}, \bibinfo{person}{Ece Kamar}, \bibinfo{person}{Walter~S Lasecki}, \bibinfo{person}{Daniel~S Weld}, {and} \bibinfo{person}{Eric Horvitz}.} \bibinfo{year}{2019}\natexlab{}.
\newblock \showarticletitle{Beyond accuracy: The role of mental models in human-AI team performance}. In \bibinfo{booktitle}{\emph{Proceedings of the AAAI conference on human computation and crowdsourcing}}, Vol.~\bibinfo{volume}{7}. \bibinfo{pages}{2--11}.
\newblock


\bibitem[Bhatt et~al\mbox{.}(2020)]%
        {bhatt2020explainable}
\bibfield{author}{\bibinfo{person}{Umang Bhatt}, \bibinfo{person}{Alice Xiang}, \bibinfo{person}{Shubham Sharma}, \bibinfo{person}{Adrian Weller}, \bibinfo{person}{Ankur Taly}, \bibinfo{person}{Yunhan Jia}, \bibinfo{person}{Joydeep Ghosh}, \bibinfo{person}{Ruchir Puri}, \bibinfo{person}{Jos{\'e}~MF Moura}, {and} \bibinfo{person}{Peter Eckersley}.} \bibinfo{year}{2020}\natexlab{}.
\newblock \showarticletitle{Explainable machine learning in deployment}. In \bibinfo{booktitle}{\emph{Proceedings of the 2020 conference on fairness, accountability, and transparency}}. \bibinfo{pages}{648--657}.
\newblock


\bibitem[Blackmon et~al\mbox{.}(2003)]%
        {Blackmon03CWW}
\bibfield{author}{\bibinfo{person}{Marilyn~Hughes Blackmon}, \bibinfo{person}{Muneo Kitajima}, {and} \bibinfo{person}{Peter~G. Polson}.} \bibinfo{year}{2003}\natexlab{}.
\newblock \showarticletitle{Repairing Usability Problems Identified by the Cognitive Walkthrough for the Web}. In \bibinfo{booktitle}{\emph{Proceedings of the SIGCHI Conference on Human Factors in Computing Systems}} (Ft. Lauderdale, Florida, USA) \emph{(\bibinfo{series}{CHI '03})}. \bibinfo{publisher}{Association for Computing Machinery}, \bibinfo{address}{New York, NY, USA}, \bibinfo{pages}{497–504}.
\newblock
\showISBNx{1581136307}
\urldef\tempurl%
\url{https://doi.org/10.1145/642611.642698}
\showDOI{\tempurl}


\bibitem[Blackwell(2016)]%
        {Blackwell16Curb}
\bibfield{author}{\bibinfo{person}{A.~G. Blackwell}.} \bibinfo{year}{2016}\natexlab{}.
\newblock \showarticletitle{The Curb-Cut Effect}.
\newblock \bibinfo{journal}{\emph{Stanford Social Innovation Review}} \bibinfo{volume}{15}, \bibinfo{number}{1} (\bibinfo{year}{2016}), \bibinfo{pages}{28--33}.
\newblock
\urldef\tempurl%
\url{https://doi.org/10.48558/YVMS-CC96}
\showDOI{\tempurl}


\bibitem[Brachman et~al\mbox{.}(2023)]%
        {Brachman23Following}
\bibfield{author}{\bibinfo{person}{Michelle Brachman}, \bibinfo{person}{Qian Pan}, \bibinfo{person}{Hyo~Jin Do}, \bibinfo{person}{Casey Dugan}, \bibinfo{person}{Arunima Chaudhary}, \bibinfo{person}{James~M. Johnson}, \bibinfo{person}{Priyanshu Rai}, \bibinfo{person}{Tathagata Chakraborti}, \bibinfo{person}{Thomas Gschwind}, \bibinfo{person}{Jim~A Laredo}, \bibinfo{person}{Christoph Miksovic}, \bibinfo{person}{Paolo Scotton}, \bibinfo{person}{Kartik Talamadupula}, {and} \bibinfo{person}{Gegi Thomas}.} \bibinfo{year}{2023}\natexlab{}.
\newblock \showarticletitle{Follow the Successful Herd: Towards Explanations for Improved Use and Mental Models of Natural Language Systems}. In \bibinfo{booktitle}{\emph{28th International Conference on Intelligent User Interfaces}} (Sydney, NSW, Australia) \emph{(\bibinfo{series}{IUI '23})}. \bibinfo{publisher}{Association for Computing Machinery}, \bibinfo{address}{New York, NY, USA}, \bibinfo{pages}{220–239}.
\newblock
\showISBNx{9798400701061}
\urldef\tempurl%
\url{https://doi.org/10.1145/3581641.3584088}
\showDOI{\tempurl}


\bibitem[Braithwaite and Sprague(2021)]%
        {Braithwaite21Procedural}
\bibfield{author}{\bibinfo{person}{David Braithwaite} {and} \bibinfo{person}{Lauren Sprague}.} \bibinfo{year}{2021}\natexlab{}.
\newblock \showarticletitle{Conceptual Knowledge, Procedural Knowledge, and Metacognition in Routine and Nonroutine Problem Solving}.
\newblock \bibinfo{journal}{\emph{Cognitive Science A Multidisciplinary Journal}}  \bibinfo{volume}{45} (\bibinfo{date}{10} \bibinfo{year}{2021}), \bibinfo{pages}{e13048}.
\newblock
\urldef\tempurl%
\url{https://doi.org/10.1111/cogs.13048}
\showDOI{\tempurl}


\bibitem[Bu\c{c}inca et~al\mbox{.}(2020)]%
        {Buccinca20Proxy}
\bibfield{author}{\bibinfo{person}{Zana Bu\c{c}inca}, \bibinfo{person}{Phoebe Lin}, \bibinfo{person}{Krzysztof~Z. Gajos}, {and} \bibinfo{person}{Elena~L. Glassman}.} \bibinfo{year}{2020}\natexlab{}.
\newblock \showarticletitle{Proxy tasks and subjective measures can be misleading in evaluating explainable AI systems}. In \bibinfo{booktitle}{\emph{Proceedings of the 25th International Conference on Intelligent User Interfaces}} (Cagliari, Italy) \emph{(\bibinfo{series}{IUI '20})}. \bibinfo{publisher}{Association for Computing Machinery}, \bibinfo{address}{New York, NY, USA}, \bibinfo{pages}{454–464}.
\newblock
\showISBNx{9781450371186}
\urldef\tempurl%
\url{https://doi.org/10.1145/3377325.3377498}
\showDOI{\tempurl}


\bibitem[Bu\c{c}inca et~al\mbox{.}(2021)]%
        {Buccinca21CogFunction}
\bibfield{author}{\bibinfo{person}{Zana Bu\c{c}inca}, \bibinfo{person}{Maja~Barbara Malaya}, {and} \bibinfo{person}{Krzysztof~Z. Gajos}.} \bibinfo{year}{2021}\natexlab{}.
\newblock \showarticletitle{To Trust or to Think: Cognitive Forcing Functions Can Reduce Overreliance on AI in AI-assisted Decision-making}.
\newblock \bibinfo{journal}{\emph{Proc. ACM Hum.-Comput. Interact.}} \bibinfo{volume}{5}, \bibinfo{number}{CSCW1}, Article \bibinfo{articleno}{188} (\bibinfo{date}{April} \bibinfo{year}{2021}), \bibinfo{numpages}{21}~pages.
\newblock
\urldef\tempurl%
\url{https://doi.org/10.1145/3449287}
\showDOI{\tempurl}


\bibitem[Burnett et~al\mbox{.}(2017)]%
        {Burnett17MS}
\bibfield{author}{\bibinfo{person}{Margaret Burnett}, \bibinfo{person}{Robin Counts}, \bibinfo{person}{Ronette Lawrence}, {and} \bibinfo{person}{Hannah Hanson}.} \bibinfo{year}{2017}\natexlab{}.
\newblock \showarticletitle{Gender HCl and microsoft: Highlights from a longitudinal study}. In \bibinfo{booktitle}{\emph{2017 IEEE Symposium on Visual Languages and Human-Centric Computing (VL/HCC)}}. \bibinfo{pages}{139--143}.
\newblock
\urldef\tempurl%
\url{https://doi.org/10.1109/VLHCC.2017.8103461}
\showDOI{\tempurl}


\bibitem[Burnett et~al\mbox{.}(2020)]%
        {burnett2020gendermagkit}
\bibfield{author}{\bibinfo{person}{Margaret Burnett}, \bibinfo{person}{Simone Stumpf}, \bibinfo{person}{Laura Beckwith}, {and} \bibinfo{person}{Anicia Peters}.} \bibinfo{year}{2020}\natexlab{}.
\newblock \bibinfo{title}{The gendermag kit: How to use the gendermag method to find inclusiveness issues through a gender lens}.
\newblock
\newblock


\bibitem[Burnett et~al\mbox{.}(2016)]%
        {Burnett16GM}
\bibfield{author}{\bibinfo{person}{Margaret Burnett}, \bibinfo{person}{Simone Stumpf}, \bibinfo{person}{Jamie Macbeth}, \bibinfo{person}{Stephann Makri}, \bibinfo{person}{Laura Beckwith}, \bibinfo{person}{Irwin Kwan}, \bibinfo{person}{Anicia Peters}, {and} \bibinfo{person}{William Jernigan}.} \bibinfo{year}{2016}\natexlab{}.
\newblock \showarticletitle{GenderMag: A Method for Evaluating Software’s Gender Inclusiveness}.
\newblock \bibinfo{journal}{\emph{Interacting with Computers}}  \bibinfo{volume}{forthcoming} (\bibinfo{date}{01} \bibinfo{year}{2016}).
\newblock
\urldef\tempurl%
\url{https://doi.org/10.1093/iwc/iwv046}
\showDOI{\tempurl}


\bibitem[Cai et~al\mbox{.}(2019)]%
        {cai2019effects}
\bibfield{author}{\bibinfo{person}{Carrie~J Cai}, \bibinfo{person}{Jonas Jongejan}, {and} \bibinfo{person}{Jess Holbrook}.} \bibinfo{year}{2019}\natexlab{}.
\newblock \showarticletitle{The effects of example-based explanations in a machine learning interface}. In \bibinfo{booktitle}{\emph{Proceedings of the 24th international conference on intelligent user interfaces}}. \bibinfo{pages}{258--262}.
\newblock


\bibitem[Chatterjee et~al\mbox{.}(2022)]%
        {Chatterjee22Inclusivity}
\bibfield{author}{\bibinfo{person}{Amreeta Chatterjee}, \bibinfo{person}{Lara Letaw}, \bibinfo{person}{Rosalinda Garcia}, \bibinfo{person}{Doshna~Umma Reddy}, \bibinfo{person}{Rudrajit Choudhuri}, \bibinfo{person}{Sabyatha~Sathish Kumar}, \bibinfo{person}{Patricia Morreale}, \bibinfo{person}{Anita Sarma}, {and} \bibinfo{person}{Margaret Burnett}.} \bibinfo{year}{2022}\natexlab{}.
\newblock \showarticletitle{Inclusivity Bugs in Online Courseware: A Field Study}. In \bibinfo{booktitle}{\emph{Proceedings of the 2022 ACM Conference on International Computing Education Research - Volume 1}} (Lugano and Virtual Event, Switzerland) \emph{(\bibinfo{series}{ICER '22})}. \bibinfo{publisher}{Association for Computing Machinery}, \bibinfo{address}{New York, NY, USA}, \bibinfo{pages}{356–372}.
\newblock
\showISBNx{9781450391948}
\urldef\tempurl%
\url{https://doi.org/10.1145/3501385.3543973}
\showDOI{\tempurl}


\bibitem[Cunningham et~al\mbox{.}(2016)]%
        {cunningham2016supporting}
\bibfield{author}{\bibinfo{person}{Sally~Jo Cunningham}, \bibinfo{person}{Annika Hinze}, {and} \bibinfo{person}{David~M Nichols}.} \bibinfo{year}{2016}\natexlab{}.
\newblock \showarticletitle{Supporting gender-neutral digital library creation: A case study using the GenderMag Toolkit}. In \bibinfo{booktitle}{\emph{Digital Libraries: Knowledge, Information, and Data in an Open Access Society: 18th International Conference on Asia-Pacific Digital Libraries, ICADL 2016, Tsukuba, Japan, December 7--9, 2016, Proceedings 18}}. Springer, \bibinfo{pages}{45--50}.
\newblock


\bibitem[Dodge et~al\mbox{.}(2022)]%
        {Dodge22Rank}
\bibfield{author}{\bibinfo{person}{Jonathan Dodge}, \bibinfo{person}{Andrew~A. Anderson}, \bibinfo{person}{Matthew Olson}, \bibinfo{person}{Rupika Dikkala}, {and} \bibinfo{person}{Margaret Burnett}.} \bibinfo{year}{2022}\natexlab{}.
\newblock \showarticletitle{How Do People Rank Multiple Mutant Agents?}. In \bibinfo{booktitle}{\emph{27th International Conference on Intelligent User Interfaces}} (Helsinki, Finland) \emph{(\bibinfo{series}{IUI '22})}. \bibinfo{publisher}{Association for Computing Machinery}, \bibinfo{address}{New York, NY, USA}, \bibinfo{pages}{191–211}.
\newblock
\showISBNx{9781450391443}
\urldef\tempurl%
\url{https://doi.org/10.1145/3490099.3511115}
\showDOI{\tempurl}


\bibitem[Dodge et~al\mbox{.}(2021a)]%
        {dodge2021AARAI}
\bibfield{author}{\bibinfo{person}{Jonathan Dodge}, \bibinfo{person}{Roli Khanna}, \bibinfo{person}{Jed Irvine}, \bibinfo{person}{Kin-Ho Lam}, \bibinfo{person}{Theresa Mai}, \bibinfo{person}{Zhengxian Lin}, \bibinfo{person}{Nicholas Kiddle}, \bibinfo{person}{Evan Newman}, \bibinfo{person}{Andrew Anderson}, \bibinfo{person}{Sai Raja}, {et~al\mbox{.}}} \bibinfo{year}{2021}\natexlab{a}.
\newblock \showarticletitle{After-action review for AI (AAR/AI)}.
\newblock \bibinfo{journal}{\emph{ACM Transactions on Interactive Intelligent Systems (TiiS)}} \bibinfo{volume}{11}, \bibinfo{number}{3-4} (\bibinfo{year}{2021}), \bibinfo{pages}{1--35}.
\newblock


\bibitem[Dodge et~al\mbox{.}(2021b)]%
        {Dodge21AAR}
\bibfield{author}{\bibinfo{person}{Jonathan Dodge}, \bibinfo{person}{Roli Khanna}, \bibinfo{person}{Jed Irvine}, \bibinfo{person}{Kin-ho Lam}, \bibinfo{person}{Theresa Mai}, \bibinfo{person}{Zhengxian Lin}, \bibinfo{person}{Nicholas Kiddle}, \bibinfo{person}{Evan Newman}, \bibinfo{person}{Andrew Anderson}, \bibinfo{person}{Sai Raja}, \bibinfo{person}{Caleb Matthews}, \bibinfo{person}{Christopher Perdriau}, \bibinfo{person}{Margaret Burnett}, {and} \bibinfo{person}{Alan Fern}.} \bibinfo{year}{2021}\natexlab{b}.
\newblock \showarticletitle{After-Action Review for AI (AAR/AI)}.
\newblock \bibinfo{journal}{\emph{ACM Trans. Interact. Intell. Syst.}} \bibinfo{volume}{11}, \bibinfo{number}{3–4}, Article \bibinfo{articleno}{29} (\bibinfo{date}{sep} \bibinfo{year}{2021}), \bibinfo{numpages}{35}~pages.
\newblock
\showISSN{2160-6455}
\urldef\tempurl%
\url{https://doi.org/10.1145/3453173}
\showDOI{\tempurl}


\bibitem[Dodge et~al\mbox{.}(2019)]%
        {Dodge19Fairness}
\bibfield{author}{\bibinfo{person}{Jonathan Dodge}, \bibinfo{person}{Q.~Vera Liao}, \bibinfo{person}{Yunfeng Zhang}, \bibinfo{person}{Rachel K.~E. Bellamy}, {and} \bibinfo{person}{Casey Dugan}.} \bibinfo{year}{2019}\natexlab{}.
\newblock \showarticletitle{Explaining Models: An Empirical Study of How Explanations Impact Fairness Judgment}. In \bibinfo{booktitle}{\emph{Proceedings of the 24th International Conference on Intelligent User Interfaces}} (Marina del Ray, California) \emph{(\bibinfo{series}{IUI '19})}. \bibinfo{publisher}{Association for Computing Machinery}, \bibinfo{address}{New York, NY, USA}, \bibinfo{pages}{275–285}.
\newblock
\showISBNx{9781450362726}
\urldef\tempurl%
\url{https://doi.org/10.1145/3301275.3302310}
\showDOI{\tempurl}


\bibitem[Ehsan et~al\mbox{.}(2021)]%
        {Ehsan21Expanding}
\bibfield{author}{\bibinfo{person}{Upol Ehsan}, \bibinfo{person}{Q.~Vera Liao}, \bibinfo{person}{Michael Muller}, \bibinfo{person}{Mark~O. Riedl}, {and} \bibinfo{person}{Justin~D. Weisz}.} \bibinfo{year}{2021}\natexlab{}.
\newblock \showarticletitle{Expanding Explainability: Towards Social Transparency in AI systems}. In \bibinfo{booktitle}{\emph{Proceedings of the 2021 CHI Conference on Human Factors in Computing Systems}} (Yokohama, Japan) \emph{(\bibinfo{series}{CHI '21})}. \bibinfo{publisher}{Association for Computing Machinery}, \bibinfo{address}{New York, NY, USA}, Article \bibinfo{articleno}{82}, \bibinfo{numpages}{19}~pages.
\newblock
\showISBNx{9781450380966}
\urldef\tempurl%
\url{https://doi.org/10.1145/3411764.3445188}
\showDOI{\tempurl}


\bibitem[Freeman et~al\mbox{.}(2022)]%
        {freeman2022rediscovering}
\bibfield{author}{\bibinfo{person}{Guo Freeman}, \bibinfo{person}{Divine Maloney}, \bibinfo{person}{Dane Acena}, {and} \bibinfo{person}{Catherine Barwulor}.} \bibinfo{year}{2022}\natexlab{}.
\newblock \showarticletitle{(Re)discovering the Physical Body Online: Strategies and Challenges to Approach Non-Cisgender Identity in Social Virtual Reality}. In \bibinfo{booktitle}{\emph{CHI Conference on Human Factors in Computing Systems}}. \bibinfo{pages}{1--15}.
\newblock
\urldef\tempurl%
\url{https://doi.org/10.1145/3491102.3517436}
\showDOI{\tempurl}


\bibitem[Frost(2021)]%
        {Frost21tail}
\bibfield{author}{\bibinfo{person}{Jim Frost}.} \bibinfo{year}{2021}\natexlab{}.
\newblock \bibinfo{booktitle}{\emph{When Can I Use One-Tailed Hypothesis Tests?}}
\newblock
\urldef\tempurl%
\url{https://statisticsbyjim.com/hypothesis-testing/use-one-tailed-tests/}
\showURL{%
\tempurl}
\newblock
\shownote{Accessed on November 21, 2023}.


\bibitem[Gajos and Mamykina(2022)]%
        {Gajos22CognitiveAI}
\bibfield{author}{\bibinfo{person}{Krzysztof~Z. Gajos} {and} \bibinfo{person}{Lena Mamykina}.} \bibinfo{year}{2022}\natexlab{}.
\newblock \showarticletitle{Do People Engage Cognitively with AI? Impact of AI Assistance on Incidental Learning}. In \bibinfo{booktitle}{\emph{Proceedings of the 27th International Conference on Intelligent User Interfaces}} (Helsinki, Finland) \emph{(\bibinfo{series}{IUI '22})}. \bibinfo{publisher}{Association for Computing Machinery}, \bibinfo{address}{New York, NY, USA}, \bibinfo{pages}{794–806}.
\newblock
\showISBNx{9781450391443}
\urldef\tempurl%
\url{https://doi.org/10.1145/3490099.3511138}
\showDOI{\tempurl}


\bibitem[Gerlings et~al\mbox{.}(2022)]%
        {gerlings2022explainable}
\bibfield{author}{\bibinfo{person}{Julie Gerlings}, \bibinfo{person}{Millie~S{\o}ndergaard Jensen}, {and} \bibinfo{person}{Arisa Shollo}.} \bibinfo{year}{2022}\natexlab{}.
\newblock \showarticletitle{Explainable AI, but explainable to whom? An exploratory case study of xAI in healthcare}.
\newblock \bibinfo{journal}{\emph{Handbook of Artificial Intelligence in Healthcare: Vol 2: Practicalities and Prospects}} (\bibinfo{year}{2022}), \bibinfo{pages}{169--198}.
\newblock


\bibitem[Gong et~al\mbox{.}(2024)]%
        {Gong24SocialExplanation}
\bibfield{author}{\bibinfo{person}{Yeaeun Gong}, \bibinfo{person}{Lanyu Shang}, {and} \bibinfo{person}{Dong Wang}.} \bibinfo{year}{2024}\natexlab{}.
\newblock \showarticletitle{Integrating Social Explanations Into Explainable Artificial Intelligence (XAI) for Combating Misinformation: Vision and Challenges}.
\newblock \bibinfo{journal}{\emph{IEEE Transactions on Computational Social Systems}} \bibinfo{volume}{11}, \bibinfo{number}{5} (\bibinfo{year}{2024}), \bibinfo{pages}{6705--6726}.
\newblock
\urldef\tempurl%
\url{https://doi.org/10.1109/TCSS.2024.3404236}
\showDOI{\tempurl}


\bibitem[Green et~al\mbox{.}(2000)]%
        {Green00CW}
\bibfield{author}{\bibinfo{person}{T.R.G. Green}, \bibinfo{person}{M.M. Burnett}, \bibinfo{person}{A.J. Ko}, \bibinfo{person}{K.J. Rothermel}, \bibinfo{person}{C.R. Cook}, {and} \bibinfo{person}{J. Schonfeld}.} \bibinfo{year}{2000}\natexlab{}.
\newblock \showarticletitle{Using the cognitive walkthrough to improve the design of a visual programming experiment}. In \bibinfo{booktitle}{\emph{Proceeding 2000 IEEE International Symposium on Visual Languages}}. \bibinfo{pages}{172--179}.
\newblock
\urldef\tempurl%
\url{https://doi.org/10.1109/VL.2000.874381}
\showDOI{\tempurl}


\bibitem[Guizani et~al\mbox{.}(2022)]%
        {Guizani22Inlcusivity}
\bibfield{author}{\bibinfo{person}{Mariam Guizani}, \bibinfo{person}{Igor Steinmacher}, \bibinfo{person}{Jillian Emard}, \bibinfo{person}{Abrar Fallatah}, \bibinfo{person}{Margaret Burnett}, {and} \bibinfo{person}{Anita Sarma}.} \bibinfo{year}{2022}\natexlab{}.
\newblock \showarticletitle{How to Debug Inclusivity Bugs? A Debugging Process with Information Architecture}. In \bibinfo{booktitle}{\emph{2022 IEEE/ACM 44th International Conference on Software Engineering: Software Engineering in Society (ICSE-SEIS)}}. \bibinfo{pages}{90--101}.
\newblock
\urldef\tempurl%
\url{https://doi.org/10.1145/3510458.3513009}
\showDOI{\tempurl}


\bibitem[Hamid et~al\mbox{.}(2024)]%
        {Hamid23GM}
\bibfield{author}{\bibinfo{person}{Md~Montaser Hamid}, \bibinfo{person}{Amreeta Chatterjee}, \bibinfo{person}{Mariam Guizani}, \bibinfo{person}{Andrew Anderson}, \bibinfo{person}{Fatima Moussaoui}, \bibinfo{person}{Sarah Yang}, \bibinfo{person}{Isaac Escobar}, \bibinfo{person}{Anita Sarma}, {and} \bibinfo{person}{Margaret Burnett}.} \bibinfo{year}{2024}\natexlab{}.
\newblock \bibinfo{booktitle}{\emph{Equity, Diversity, and Inclusion in Software Engineering: Best Practices and Insights}}.
\newblock \bibinfo{publisher}{Apress}, Chapter : How to Measure Diversity Actionably in Technology.
\newblock


\bibitem[Hardy and Vargas(2019)]%
        {hardy2019participatory}
\bibfield{author}{\bibinfo{person}{Jean Hardy} {and} \bibinfo{person}{Stefani Vargas}.} \bibinfo{year}{2019}\natexlab{}.
\newblock \showarticletitle{Participatory Design and the Future of Rural LGBTQ Communities}. In \bibinfo{booktitle}{\emph{Companion Publication of the 2019 on Designing Interactive Systems Conference 2019 Companion}}. \bibinfo{pages}{195--199}.
\newblock
\urldef\tempurl%
\url{https://doi.org/10.1145/3301019.3323894}
\showDOI{\tempurl}


\bibitem[Hilderbrand et~al\mbox{.}(2020)]%
        {Hilderbrand20Trench}
\bibfield{author}{\bibinfo{person}{Claudia Hilderbrand}, \bibinfo{person}{Christopher Perdriau}, \bibinfo{person}{Lara Letaw}, \bibinfo{person}{Jillian Emard}, \bibinfo{person}{Zoe Steine-Hanson}, \bibinfo{person}{Margaret Burnett}, {and} \bibinfo{person}{Anita Sarma}.} \bibinfo{year}{2020}\natexlab{}.
\newblock \showarticletitle{Engineering Gender-Inclusivity into Software: Ten Teams' Tales from the Trenches}. In \bibinfo{booktitle}{\emph{Proceedings of the ACM/IEEE 42nd International Conference on Software Engineering}} (Seoul, South Korea) \emph{(\bibinfo{series}{ICSE '20})}. \bibinfo{publisher}{Association for Computing Machinery}, \bibinfo{address}{New York, NY, USA}, \bibinfo{pages}{433–444}.
\newblock
\showISBNx{9781450371216}
\urldef\tempurl%
\url{https://doi.org/10.1145/3377811.3380371}
\showDOI{\tempurl}


\bibitem[Hoffman et~al\mbox{.}(2019)]%
        {hoffman2019metrics}
\bibfield{author}{\bibinfo{person}{Robert~R. Hoffman}, \bibinfo{person}{Shane~T. Mueller}, \bibinfo{person}{Gary Klein}, {and} \bibinfo{person}{Jordan Litman}.} \bibinfo{year}{2019}\natexlab{}.
\newblock \bibinfo{title}{Metrics for Explainable AI: Challenges and Prospects}.
\newblock
\newblock
\showeprint[arxiv]{1812.04608}


\bibitem[Kahneman(2011)]%
        {kahneman2011thinking}
\bibfield{author}{\bibinfo{person}{Daniel Kahneman}.} \bibinfo{year}{2011}\natexlab{}.
\newblock \bibinfo{booktitle}{\emph{Thinking, Fast and Slow}}.
\newblock \bibinfo{publisher}{Macmillan}.
\newblock


\bibitem[Kirillova and Malikh(2017)]%
        {kirillova2017gender}
\bibfield{author}{\bibinfo{person}{YS Kirillova} {and} \bibinfo{person}{DA Malikh}.} \bibinfo{year}{2017}\natexlab{}.
\newblock \showarticletitle{Gender accessories recognition of user web-applications by classifiers}.
\newblock \bibinfo{journal}{\emph{Alley of Science}} \bibinfo{volume}{4}, \bibinfo{number}{9} (\bibinfo{year}{2017}), \bibinfo{pages}{854--857}.
\newblock


\bibitem[Koujalgi et~al\mbox{.}(2024)]%
        {koujalgi2024PredictionAccuracy}
\bibfield{author}{\bibinfo{person}{Sujay Koujalgi}, \bibinfo{person}{Andrew Anderson}, \bibinfo{person}{Iyadunni Adenuga}, \bibinfo{person}{Shikha Soneji}, \bibinfo{person}{Rupika Dikkala}, \bibinfo{person}{Teresita~Guzman Nader}, \bibinfo{person}{Leo Soccio}, \bibinfo{person}{Sourav Panda}, \bibinfo{person}{Rupak~Kumar Das}, \bibinfo{person}{Margaret Burnett}, {and} \bibinfo{person}{Jonathan Dodge}.} \bibinfo{year}{2024}\natexlab{}.
\newblock \bibinfo{title}{How to Measure Human-AI Prediction Accuracy in Explainable AI Systems}.
\newblock
\newblock
\showeprint[arxiv]{2409.00069}~[cs.HC]
\urldef\tempurl%
\url{https://arxiv.org/abs/2409.00069}
\showURL{%
\tempurl}


\bibitem[Kulesza et~al\mbox{.}(2013)]%
        {kulesza2013too}
\bibfield{author}{\bibinfo{person}{Todd Kulesza}, \bibinfo{person}{Simone Stumpf}, \bibinfo{person}{Margaret Burnett}, \bibinfo{person}{Sherry Yang}, \bibinfo{person}{Irwin Kwan}, {and} \bibinfo{person}{Weng-Keen Wong}.} \bibinfo{year}{2013}\natexlab{}.
\newblock \showarticletitle{Too much, too little, or just right? Ways explanations impact end users' mental models}. In \bibinfo{booktitle}{\emph{2013 IEEE Symposium on Visual Languages and Human Centric Computing}}. IEEE, \bibinfo{pages}{3--10}.
\newblock


\bibitem[Liao and Varshney(2021)]%
        {Liao21Human}
\bibfield{author}{\bibinfo{person}{Q.~Vera Liao} {and} \bibinfo{person}{Kush~R. Varshney}.} \bibinfo{year}{2021}\natexlab{}.
\newblock \bibinfo{title}{Human-Centered Explainable AI (XAI): From Algorithms to User Experiences}.
\newblock
\newblock
\showeprint[arxiv]{2110.10790}


\bibitem[Markman and Gentner(1996)]%
        {markman1996differences}
\bibfield{author}{\bibinfo{person}{Arthur~B Markman} {and} \bibinfo{person}{Dedre Gentner}.} \bibinfo{year}{1996}\natexlab{}.
\newblock \showarticletitle{Commonalities and differences in similarity comparisons}.
\newblock \bibinfo{journal}{\emph{Memory \& cognition}} \bibinfo{volume}{24}, \bibinfo{number}{2} (\bibinfo{year}{1996}), \bibinfo{pages}{235--249}.
\newblock


\bibitem[Millecamp et~al\mbox{.}(2020)]%
        {Millecamp20Cognitive}
\bibfield{author}{\bibinfo{person}{Martijn Millecamp}, \bibinfo{person}{Robin Haveneers}, {and} \bibinfo{person}{Katrien Verbert}.} \bibinfo{year}{2020}\natexlab{}.
\newblock \showarticletitle{Cogito Ergo Quid? The Effect of Cognitive Style in a Transparent Mobile Music Recommender System}. In \bibinfo{booktitle}{\emph{Proceedings of the 28th ACM Conference on User Modeling, Adaptation and Personalization}} (Genoa, Italy) \emph{(\bibinfo{series}{UMAP '20})}. \bibinfo{publisher}{Association for Computing Machinery}, \bibinfo{address}{New York, NY, USA}, \bibinfo{pages}{323–327}.
\newblock
\showISBNx{9781450368612}
\urldef\tempurl%
\url{https://doi.org/10.1145/3340631.3394871}
\showDOI{\tempurl}


\bibitem[Murphy-Hill et~al\mbox{.}(2024)]%
        {murphy2024gendermag}
\bibfield{author}{\bibinfo{person}{Emerson Murphy-Hill}, \bibinfo{person}{Alberto Elizondo}, \bibinfo{person}{Ambar Murillo}, \bibinfo{person}{Marian Harbach}, \bibinfo{person}{Bogdan Vasilescu}, \bibinfo{person}{Delphine Carlson}, {and} \bibinfo{person}{Florian Dessloch}.} \bibinfo{year}{2024}\natexlab{}.
\newblock \showarticletitle{GenderMag Improves Discoverability in the Field, Especially for Women}. In \bibinfo{booktitle}{\emph{2024 IEEE/ACM 46th International Conference on Software Engineering (ICSE)}}. IEEE Computer Society, \bibinfo{pages}{973--973}.
\newblock


\bibitem[Nam et~al\mbox{.}(2024)]%
        {nam2024using}
\bibfield{author}{\bibinfo{person}{Daye Nam}, \bibinfo{person}{Andrew Macvean}, \bibinfo{person}{Vincent Hellendoorn}, \bibinfo{person}{Bogdan Vasilescu}, {and} \bibinfo{person}{Brad Myers}.} \bibinfo{year}{2024}\natexlab{}.
\newblock \showarticletitle{Using an LLM to Help With Code Understanding}. In \bibinfo{booktitle}{\emph{Proceedings of the IEEE/ACM 46th International Conference on Software Engineering}} (Lisbon, Portugal) \emph{(\bibinfo{series}{ICSE '24})}. \bibinfo{publisher}{Association for Computing Machinery}, \bibinfo{address}{New York, NY, USA}, Article \bibinfo{articleno}{97}, \bibinfo{numpages}{13}~pages.
\newblock
\showISBNx{9798400702174}
\urldef\tempurl%
\url{https://doi.org/10.1145/3597503.3639187}
\showDOI{\tempurl}


\bibitem[NASA({[n.\,d.]})]%
        {NASA}
\bibfield{author}{\bibinfo{person}{NASA}.} \bibinfo{year}{[n.\,d.]}\natexlab{}.
\newblock \bibinfo{booktitle}{\emph{NASA TLX: Task Load Index}}.
\newblock
\urldef\tempurl%
\url{https://humansystems.arc.nasa.gov/groups/TLX/}
\showURL{%
Retrieved December 4, 2023 from \tempurl}


\bibitem[Nourani et~al\mbox{.}(2021)]%
        {nourani2021anchoring}
\bibfield{author}{\bibinfo{person}{Mahsan Nourani}, \bibinfo{person}{Chiradeep Roy}, \bibinfo{person}{Jeremy~E Block}, \bibinfo{person}{Donald~R Honeycutt}, \bibinfo{person}{Tahrima Rahman}, \bibinfo{person}{Eric Ragan}, {and} \bibinfo{person}{Vibhav Gogate}.} \bibinfo{year}{2021}\natexlab{}.
\newblock \showarticletitle{Anchoring bias affects mental model formation and user reliance in explainable ai systems}. In \bibinfo{booktitle}{\emph{26th International Conference on Intelligent User Interfaces}}. \bibinfo{pages}{340--350}.
\newblock


\bibitem[Press(2019)]%
        {lexico2019Equity}
\bibfield{author}{\bibinfo{person}{Oxford~University Press}.} \bibinfo{year}{2019}\natexlab{}.
\newblock \bibinfo{title}{Lexico US Dictionary}.
\newblock
\newblock
\urldef\tempurl%
\url{https://www.lexico.com/}
\showURL{%
\tempurl}


\bibitem[Rittle-Johnson et~al\mbox{.}(2001)]%
        {Johnson01Procedural}
\bibfield{author}{\bibinfo{person}{Bethany Rittle-Johnson}, \bibinfo{person}{Robert Siegler}, {and} \bibinfo{person}{Martha Alibali}.} \bibinfo{year}{2001}\natexlab{}.
\newblock \showarticletitle{Developing conceptual understanding and procedural skill in mathematics: An iterative process.}
\newblock \bibinfo{journal}{\emph{Journal of Educational Psychology}}  \bibinfo{volume}{93} (\bibinfo{date}{06} \bibinfo{year}{2001}), \bibinfo{pages}{346--362}.
\newblock
\urldef\tempurl%
\url{https://doi.org/10.1037//0022-0663.93.2.346}
\showDOI{\tempurl}


\bibitem[Rosenfeld and Richardson(2019)]%
        {rosenfeld2019explainability}
\bibfield{author}{\bibinfo{person}{Avi Rosenfeld} {and} \bibinfo{person}{Ariella Richardson}.} \bibinfo{year}{2019}\natexlab{}.
\newblock \showarticletitle{Explainability in human--agent systems}.
\newblock \bibinfo{journal}{\emph{Autonomous agents and multi-agent systems}}  \bibinfo{volume}{33} (\bibinfo{year}{2019}), \bibinfo{pages}{673--705}.
\newblock


\bibitem[Salaberry(2018)]%
        {salaberry2018declarativeVsProcedural}
\bibfield{author}{\bibinfo{person}{M~Rafael Salaberry}.} \bibinfo{year}{2018}\natexlab{}.
\newblock \showarticletitle{Declarative versus procedural knowledge}.
\newblock \bibinfo{journal}{\emph{The TESOL Encyclopedia of English language teaching}} (\bibinfo{year}{2018}), \bibinfo{pages}{1--7}.
\newblock


\bibitem[Santos et~al\mbox{.}(2023)]%
        {santos2023designing}
\bibfield{author}{\bibinfo{person}{Italo Santos}, \bibinfo{person}{Jo{\~a}o~Felipe Pimentel}, \bibinfo{person}{Igor Wiese}, \bibinfo{person}{Igor Steinmacher}, \bibinfo{person}{Anita Sarma}, {and} \bibinfo{person}{Marco~A Gerosa}.} \bibinfo{year}{2023}\natexlab{}.
\newblock \showarticletitle{Designing for cognitive diversity: Improving the github experience for newcomers}. In \bibinfo{booktitle}{\emph{2023 IEEE/ACM 45th International Conference on Software Engineering: Software Engineering in Society (ICSE-SEIS)}}. IEEE, \bibinfo{pages}{1--12}.
\newblock


\bibitem[Schmidt and Biessmann(2020)]%
        {schmidt2020calibrating}
\bibfield{author}{\bibinfo{person}{Philipp Schmidt} {and} \bibinfo{person}{Felix Biessmann}.} \bibinfo{year}{2020}\natexlab{}.
\newblock \showarticletitle{Calibrating human-AI collaboration: Impact of risk, ambiguity and transparency on algorithmic bias}. In \bibinfo{booktitle}{\emph{Machine Learning and Knowledge Extraction: 4th IFIP TC 5, TC 12, WG 8.4, WG 8.9, WG 12.9 International Cross-Domain Conference, CD-MAKE 2020, Dublin, Ireland, August 25–28, 2020, Proceedings}}, \bibfield{editor}{\bibinfo{person}{Andreas Holzinger}, \bibinfo{person}{Peter Kieseberg}, \bibinfo{person}{A~Min Tjoa}, {and} \bibinfo{person}{Edgar Weippl}} (Eds.). \bibinfo{publisher}{Springer}, \bibinfo{pages}{431--449}.
\newblock
\urldef\tempurl%
\url{https://doi.org/10.1007/978-3-030-57321-8_25}
\showDOI{\tempurl}


\bibitem[Schraagen et~al\mbox{.}(2020)]%
        {Schraagen20Trusting}
\bibfield{author}{\bibinfo{person}{Jan~Maarten Schraagen}, \bibinfo{person}{Pia Elsasser}, \bibinfo{person}{Hanna Fricke}, \bibinfo{person}{Marleen Hof}, {and} \bibinfo{person}{Fabyen Ragalmuto}.} \bibinfo{year}{2020}\natexlab{}.
\newblock \showarticletitle{Trusting the X in XAI: Effects of different types of explanations by a self-driving car on trust, explanation satisfaction and mental models}.
\newblock \bibinfo{journal}{\emph{Human Factors and Ergonomics Society Annual Meeting}} \bibinfo{volume}{64}, \bibinfo{number}{1} (\bibinfo{year}{2020}), \bibinfo{pages}{339--343}.
\newblock
\urldef\tempurl%
\url{https://doi.org/10.1177/1071181320641077}
\showDOI{\tempurl}


\bibitem[Shekhar and Marsden(2018)]%
        {shekhar2018cognitive}
\bibfield{author}{\bibinfo{person}{Arun Shekhar} {and} \bibinfo{person}{Nicola Marsden}.} \bibinfo{year}{2018}\natexlab{}.
\newblock \showarticletitle{Cognitive Walkthrough of a learning management system with gendered personas}. In \bibinfo{booktitle}{\emph{Proceedings of the 4th Conference on Gender \& IT}}. \bibinfo{pages}{191--198}.
\newblock


\bibitem[Stemler(2004)]%
        {Stemler04IRR}
\bibfield{author}{\bibinfo{person}{Steven Stemler}.} \bibinfo{year}{2004}\natexlab{}.
\newblock \showarticletitle{A Comparison of Consensus, Consistency, and Measurement Approaches to Estimating Interrater Reliability}.
\newblock \bibinfo{journal}{\emph{Practical Assessment, Research, and Evaluation}}  \bibinfo{volume}{9} (\bibinfo{date}{01} \bibinfo{year}{2004}), \bibinfo{pages}{1--19}.
\newblock


\bibitem[Sun et~al\mbox{.}(2022)]%
        {Sun22Investigating}
\bibfield{author}{\bibinfo{person}{Jiao Sun}, \bibinfo{person}{Q.~Vera Liao}, \bibinfo{person}{Michael Muller}, \bibinfo{person}{Mayank Agarwal}, \bibinfo{person}{Stephanie Houde}, \bibinfo{person}{Kartik Talamadupula}, {and} \bibinfo{person}{Justin~D. Weisz}.} \bibinfo{year}{2022}\natexlab{}.
\newblock \showarticletitle{Investigating Explainability of Generative AI for Code through Scenario-based Design}. In \bibinfo{booktitle}{\emph{Proceedings of the 27th International Conference on Intelligent User Interfaces}} (Helsinki, Finland) \emph{(\bibinfo{series}{IUI '22})}. \bibinfo{publisher}{Association for Computing Machinery}, \bibinfo{address}{New York, NY, USA}, \bibinfo{pages}{212–228}.
\newblock
\showISBNx{9781450391443}
\urldef\tempurl%
\url{https://doi.org/10.1145/3490099.3511119}
\showDOI{\tempurl}


\bibitem[Trumbo(1999)]%
        {trumbo2019heuristicSystematic}
\bibfield{author}{\bibinfo{person}{C.~W. Trumbo}.} \bibinfo{year}{1999}\natexlab{}.
\newblock \showarticletitle{Heuristic-systematic information processing and risk judgment}.
\newblock \bibinfo{journal}{\emph{Risk Anal.}} \bibinfo{volume}{19}, \bibinfo{number}{3} (\bibinfo{date}{June} \bibinfo{year}{1999}), \bibinfo{pages}{391--400}.
\newblock


\bibitem[Tsai et~al\mbox{.}(2021)]%
        {Tsai21Exploring}
\bibfield{author}{\bibinfo{person}{Chun-Hua Tsai}, \bibinfo{person}{Yue You}, \bibinfo{person}{Xinning Gui}, \bibinfo{person}{Yubo Kou}, {and} \bibinfo{person}{John~M. Carroll}.} \bibinfo{year}{2021}\natexlab{}.
\newblock \showarticletitle{Exploring and Promoting Diagnostic Transparency and Explainability in Online Symptom Checkers}. In \bibinfo{booktitle}{\emph{2021 CHI Conference on Human Factors in Computing Systems}} (Yokohama, Japan) \emph{(\bibinfo{series}{CHI '21})}. \bibinfo{publisher}{Association for Computing Machinery}, \bibinfo{address}{New York, NY, USA}, Article \bibinfo{articleno}{152}, \bibinfo{numpages}{17}~pages.
\newblock
\showISBNx{9781450380966}
\urldef\tempurl%
\url{https://doi.org/10.1145/3411764.3445101}
\showDOI{\tempurl}


\bibitem[Vorvoreanu et~al\mbox{.}(2019)]%
        {Vorvoreanu19Bias}
\bibfield{author}{\bibinfo{person}{Mihaela Vorvoreanu}, \bibinfo{person}{Lingyi Zhang}, \bibinfo{person}{Yun-Han Huang}, \bibinfo{person}{Claudia Hilderbrand}, \bibinfo{person}{Zoe Steine-Hanson}, {and} \bibinfo{person}{Margaret Burnett}.} \bibinfo{year}{2019}\natexlab{}.
\newblock \showarticletitle{From Gender Biases to Gender-Inclusive Design: An Empirical Investigation}. In \bibinfo{booktitle}{\emph{Proceedings of the 2019 CHI Conference on Human Factors in Computing Systems}} (Glasgow, Scotland Uk) \emph{(\bibinfo{series}{CHI '19})}. \bibinfo{publisher}{Association for Computing Machinery}, \bibinfo{address}{New York, NY, USA}, \bibinfo{pages}{1–14}.
\newblock
\showISBNx{9781450359702}
\urldef\tempurl%
\url{https://doi.org/10.1145/3290605.3300283}
\showDOI{\tempurl}


\bibitem[Wang and Yin(2021)]%
        {wang21explanations}
\bibfield{author}{\bibinfo{person}{Xinru Wang} {and} \bibinfo{person}{Ming Yin}.} \bibinfo{year}{2021}\natexlab{}.
\newblock \showarticletitle{Are Explanations Helpful? A Comparative Study of the Effects of Explanations in AI-Assisted Decision-Making}. In \bibinfo{booktitle}{\emph{Proceedings of the 26th International Conference on Intelligent User Interfaces}} (College Station, TX, USA) \emph{(\bibinfo{series}{IUI '21})}. \bibinfo{publisher}{Association for Computing Machinery}, \bibinfo{address}{New York, NY, USA}, \bibinfo{pages}{318–328}.
\newblock
\showISBNx{9781450380171}
\urldef\tempurl%
\url{https://doi.org/10.1145/3397481.3450650}
\showDOI{\tempurl}


\bibitem[Wang et~al\mbox{.}(2024)]%
        {wang2024roadmap}
\bibfield{author}{\bibinfo{person}{Ziming Wang}, \bibinfo{person}{Changwu Huang}, {and} \bibinfo{person}{Xin Yao}.} \bibinfo{year}{2024}\natexlab{}.
\newblock \showarticletitle{A Roadmap of Explainable Artificial Intelligence: Explain to Whom, When, What and How?}
\newblock \bibinfo{journal}{\emph{ACM Transactions on Autonomous and Adaptive Systems}} (\bibinfo{year}{2024}).
\newblock


\bibitem[Wason and Evans(1974)]%
        {wason1974dual}
\bibfield{author}{\bibinfo{person}{Peter~C. Wason} {and} \bibinfo{person}{J.~St. B.~T. Evans}.} \bibinfo{year}{1974}\natexlab{}.
\newblock \showarticletitle{Dual Processes in Reasoning?}
\newblock \bibinfo{journal}{\emph{Cognition}} \bibinfo{volume}{3}, \bibinfo{number}{2} (\bibinfo{year}{1974}), \bibinfo{pages}{141--154}.
\newblock


\bibitem[Wharton et~al\mbox{.}(1994)]%
        {Wharton94CW}
\bibfield{author}{\bibinfo{person}{Cathleen Wharton}, \bibinfo{person}{John Rieman}, \bibinfo{person}{Clayton~H. Lewis}, {and} \bibinfo{person}{Peter~G. Polson}.} \bibinfo{year}{1994}\natexlab{}.
\newblock \showarticletitle{The cognitive walkthrough method: a practitioner's guide}.
\newblock
\urldef\tempurl%
\url{https://api.semanticscholar.org/CorpusID:53925427}
\showURL{%
\tempurl}


\bibitem[Wohlin et~al\mbox{.}(2000)]%
        {Wohlin00SE}
\bibfield{author}{\bibinfo{person}{Claes Wohlin}, \bibinfo{person}{Per Runeson}, \bibinfo{person}{Martin H\"{o}st}, \bibinfo{person}{Magnus~C. Ohlsson}, \bibinfo{person}{Bj\"{o}orn Regnell}, {and} \bibinfo{person}{Anders Wessl\'{e}n}.} \bibinfo{year}{2000}\natexlab{}.
\newblock \bibinfo{booktitle}{\emph{Experimentation in Software Engineering: An Introduction}}.
\newblock \bibinfo{publisher}{Kluwer Academic Publishers}, \bibinfo{address}{USA}.
\newblock
\showISBNx{0792386825}


\end{thebibliography}

\appendix

\newpage
\section*{Appendix}

\begin{longtable}{p{5cm}|p{8cm}}
\caption{Concepts and definitions}\label{tab:codes_definitions} \\
\hline
\textbf{Concept} & \textbf{Definition} \\
\hline
\endfirsthead

\multicolumn{2}{c}%
{{\tablename\ \thetable{} -- Concepts and definitions (continued from previous page)}} \\
\hline
\textbf{Concept} & \textbf{Definition} \\
\hline
\endhead

\hline
\multicolumn{2}{r}{{Continued on next page}} \\
\endfoot

\hline
\endlastfoot

\textbf{Prediction Why} & \\
\hline
Participant used current game state & Participant drew on aspects of the current game state, such as gameboard, move placement, sequence orientation, win/loss position/state (condition of gameboard). \\
\hline
Participant used prior actions & Participants drew on something they'd seen the agent do (games/moves). \\
\hline
Participant used prior explanation & Participants drew on something they'd seen in the previous explanation(s) including the explanations from a previous game. \\
\hline
Participant predicted next explanation & Participants hypothesized what the explanation would look like after the agent took the action they predicted. \\
\hline
Participant said agent was random or baseless & Participant stated something about the agent that seemed random/wasn't made for a ``good'' reason. \\
\hline
Participant stated what they'd do & Participants placed themselves as a pilot, directing what they want to attack. \\
\hline
Participant is randomly guessing & Participant gives a random guess, particularly for an empty gameboard or early game. \\
\hline
\textbf{Flaws} & \\
\hline
Yes failure & Participant detected an issue, manifested an incorrect output i.e., wrong move, value too high/low, missing opportunity, missing an opportunity to prevent a loss. \\
\hline
Yes fault & Participant identified why the agent performed poorly or something wrong about how the agent processed its input, or agent is focusing too much on its own moves. \\
\hline
General miscalculation & Participant says or perceives the score is wrong/agent miscalculated the score/there are prediction flaws but they did not specify which particular square they were referring to. \\
\hline
Square ``X'' is miscalculated & Participant says the score is wrong/agent miscalculated the score/there are prediction flaws and they specified which particular square they were referring to. \\
\hline
Agent misselection & Participant says the agent selected the wrong square because it is not the high/highest score (not following its reasoning). \\
\hline
Agent chose the wrong action & Participant says there is a lost win opportunity, missed a critical move, or missed the opportunity to block. \\
\hline
General flaw & Participant only mentioned that there is a flaw but did not specify why it's a flaw. \\
\hline
\textbf{Strategy Should be} & 
Participant says some general idea about how the agent SHOULD play the game or suggestions that the agent should follow. \\
\hline
Agent should pick high score & Participant says the agent should use the highest / high score or top 5. \\
\hline
Agent should do location-based Strategy & Participant says the agent should pick based on the gameboard, past moves, orientation, or should continue the sequence to 4 in a row, prioritize established sequence unless under threat, take winning move, block opponent's move, follow offensive or defensive strategy. \\
\hline
Agent should create multiple options & Participant says the agent should create multiple options, e.g., 2 2 X in a row. \\
\hline
Agent should follow other strategy & Participant says the agent should follow another strategy such as changing how it calculates, reduce the number of moves, and/or consider more options/data (e.g., consider opponent's moves). \\
\hline
\textbf{How Agent Behaves} &
Participant's perception of the agent's behavior, such as how the agent plays the game, how it behaves during the game, what its priority, strengths and weaknesses are. \\
\hline
Agent picks high score & Participant says the agent is picking the square with the highest score or within the top 5. \\
\hline
Agent doesn't pick high score & Participant says the agent is not picking the square with the highest score or within the top 5. \\
\hline
Agent does location-based Strategy & Participant says the agent picks based on the gameboard, past moves, orientation, or will continue the sequence to 4 in a row, prioritize established sequence unless under threat, take winning move, block opponent's move, follow offensive or defensive strategy. \\
\hline
Agent does not do location-based Strategy & Participant says the agent does not pick based on the gameboard, past moves, orientation, continue the sequence to 4 in a row, prioritize established sequence unless under threat, take winning move, block opponent's move, follow offensive or defensive strategy. \\
\hline
Agent doesn't consider opponent's moves & Participant says the agent does not consider the opponent or consider their effect. \\
\hline
Agent does make multiple options & Participant says the agent makes multiple options, e.g., 2 2 X in a row. \\
\hline
Agent thinks ahead & Participant says explicitly that the agent can think ahead. \\
\hline
Agent doesn't think ahead & Participant says that the agent does not think ahead. \\
\hline
Opponent is random/illogical & Participant says red is dumb, not smart, making weird moves, illogical, random. \\
\hline
Agent is random/illogical & Participant says generally the agent is random, behaving illogically, making weird moves, or doesn't know what it is doing. \\
\hline
Agent prolongs game & Participant says the agent prolongs/stalls/lengthens the game (e.g., not doing finishing or winning move). \\
\hline
Agent is/can learn & Participant says the agent maybe/is learning. \\
\hline
Participant has uncertainty/confusion & Participant said they are unsure/confused about something with respect to how the agent works (possibly speculatively). \\
\hline
Participant has a question & Participant explicitly asks a question about the agent, game, or explanations and may want more information. \\
\hline
Starting first has advantages & Participant says that going first has advantages or going second has fewer advantages towards winning the game. \\
\hline
Start doesn't matter & Participant says that starting moves do not matter. \\
\hline
Score should be symmetrical & Participant says scores should be symmetrical with respect to the gameboard. \\
\hline
\end{longtable}


\begin{longtable}{p{2.5cm}|p{4.5cm}|p{6cm}}
\caption{All of Team Game's Fixes, why they fixed them (problem-solving styles impacted), and the changes made for the fixes}\label{tab:fixes_all} \\
\hline 
\textbf{Fix-ID \& Fix Name} & \textbf{Fixed-Why + Problem Solving Styles(PSS)} & \textbf{Changes Made for the Fixes} \\
\hline
\endfirsthead

\multicolumn{3}{c}%
{{\tablename\ \thetable{} -- Fixes (continued from previous page)}} \\
\hline
\textbf{Fix-ID \& Fix Name} & \textbf{Fixed-Why + Problem Solving Styles(PSS)} & \textbf{Changes Made for the Fixes} \\
\hline
\endhead

\hline
\multicolumn{3}{r}{{Continued on next page}} \\
\endfoot

\hline
\endlastfoot

Fix-1: Blurb of Instructions & ``Attitude toward technology is by process,...look for documentation..'' PSS: Info, Risk. &
\hspace{-2mm} -- Added instructions on getting started.

\hspace{-2mm} -- Added summary of each explanation.

\hspace{-2mm} -- Added minimize button for instructions.\\
\hline
Fix-2: Exact Values and Exact Actions  & ``She is not clear on what she should do.....will not take this step.'' PSS: Risk, Learn, Info, SE, Motiv. & 
\hspace{-2mm} -- Showed exact win\%, loss\%, and draw\% for every rectangle in the BTW, STT, and OTB explanations using tooltips.

\hspace{-2mm} -- Added tooltips that informed about the exact actions offered by the next and the prior move buttons.

 \\
\hline
Fix-3: Buttons Design and Colors & ``The game info...gives the hint....needs further indications.'' PSS: Motiv, Info. & 
\hspace{-2mm} -- Made buttons look more like buttons.

\hspace{-2mm} -- Increased color contrast between the button and the instructions background.

\hspace{-2mm} -- Used undo icon with the prior move button.\\
\hline
Fix-4: Info Button &  ``Not sure if Abi knows....what type of info it would provide.'' PSS: SE, Learn. & 
\hspace{-2mm} -- Added circular info button with each explanation that gives a short summary of the explanation.\\
\hline
Fix-5: Colors & ``Hovering does....This falls into her learning style..'' PSS: SE, Learn. & 
\hspace{-2mm} -- Changed the color of the interactive elements.

\hspace{-2mm} -- Increased the contrast between the interactive element and the interface background.\\
\hline
Fix-6: Labels and Titles & ``She sees new information...needs to understand what....trying to provide.''
PSS: Motiv, Info. & 
\hspace{-2mm} -- Changed labels for STT x-axis: Move Number instead of Time.

\hspace{-2mm} -- Changed the BTW and STT y-axis: 100\% win and 100\% loss instead of just win and loss.

\hspace{-2mm} -- Added a title bar ``Spot the Broken AI''.\\
\hline
Fix-7: Top 5 Moves & ``...but they may not be sure how to reason with this information.'' PSS: Risk, Learn, Info, Motiv. & 
\hspace{-2mm} -- Added ``Top 5'' moves with their score, win, loss, and draw percentages.\\
\hline
Fix-8: Tying Arrows &``..no update for the other player...she may consider....something wrong.'' PSS: Info, Risk, SE. & 
\hspace{-2mm} -- Added arrows on BTW, STT, OTB  that point toward the rectangles representing the scores of the most recent move.

\hspace{-2mm} -- These arrows appeared for a few seconds when the user clicked the next/prior move button.\\
\hline
Fix-9: Game Log & ``...nothing has changed....so Abi might be confused.'' PSS: Risk. &
\hspace{-2mm} -- Added a game log beside the move rank list.

\hspace{-2mm} -- Added info on all the moves in the log.

\hspace{-2mm} -- Added a scroll bar and a separator between the game log and the top 5 moves.\\
\hline
Fix-10: BTW Legend & ``The relationship between...board and the BTW...isn’t clear.'' PSS: SE, Risk. & 
\hspace{-2mm} -- Added legend to the top right for btw.

\hspace{-2mm} -- Changed the coloring of the move lines. Darker color for more recent moves.

\hspace{-2mm} -- Made the legend interactive.\\
\hline
Fix-11: Text and Prompts & ``Nothing is clearly stating that the game is over...'' PSS: SE. & 
\hspace{-2mm} -- Added a text message at the beginning shows the game is starting.

\hspace{-2mm} -- Added a text message at the end that shows the game has ended.

\hspace{-2mm} -- Alongside the last move's number added who won.

\hspace{-2mm} -- Removed who moves next for the last move.

\hspace{-2mm} -- Mentioned the game number with the move, e.g., Game 1 Move 1.\\
\hline
Fix-12: Chart Highlight Changes & ``When highlighting..., there are multiple....This causes confusion for Abi..likes to have a complete understanding...'' PSS: Info, SE, Learn. & 
\hspace{-2mm} --  Added the same highlighting on OTB for the winning sequence.

\hspace{-2mm} -- Added a thick vertical bar on STT indicates that this is the last move.

\hspace{-2mm} -- Changed BTW interaction to just highlight the rectangle from the current data series.\\
\hline
Fix-13: Game History & ``Because of IP/Risk, ...rethink what they’re doing before moving on.'' PSS: Info, SE, Risk. & 
\hspace{-2mm} --   Provided a drop-down game history at the top.

\hspace{-2mm} --  Provided a summary of what happened in the earlier games inside the game history drop-down.\\
\hline
Fix-14: Move History & ``Abi will want to spend more time, because with the prior move...what it’s telling her.'' PSS: Info, SE, Risk. & 
\hspace{-2mm} -- Team Game wanted to have the feature to go back to any move from an older game. We have not implemented it.\\
\hline 
Fix-15: Zoom in/out & ``Not sure that Abi would get the information...given her information processing style. PSS: Info, SE. & 
\hspace{-2mm} -- Team Game wanted to have a zoom-in/out feature for the explanation charts. We have not implemented it.\\
\hline
\end{longtable}


\end{document}